\documentclass[usenatbib]{mn2e} \voffset=-0.5in

\usepackage{graphicx} 
\usepackage{amsmath} 
\usepackage{amssymb}
\usepackage{exscale, relsize}
\usepackage{amscd}
\usepackage{natbib}
\usepackage{lscape, rotating}      
\usepackage{afterpage}

\usepackage{times}

\usepackage{hyperref}
\hypersetup{colorlinks=true, linkcolor=blue, bookmarks=true, citecolor=black}

\newcommand{\eg}{\textit{e.g.}} 
\newcommand{\ie}{\textit{i.e.}} 
\newcommand{\viz}{\textit{viz.}} 
\newcommand{\cf}{\textit{cf.}}

\newcommand{\Ell}{\mathcal{L}}
\newcommand{\prob}{\mathrm{Pr}}

\renewcommand{\vec}[1]{\underline{#1}}
\newcommand{\mat}[1]{\underline{\underline{\mathrm{#1}}} }
\newcommand{\set}[1]{ \mathbf{ #1 } }

\newcommand{\parset}{\mathbf{P}}

\renewcommand{\eqref}[1]{\hyperref[eq:#1]{Eq.\ (\ref*{eq:#1})}}
\newcommand{\secref}[1]{\hyperref[ch:#1]{\textsection \ref*{ch:#1}}}
\newcommand{\figref}[1]{Fig.\ \ref{fig:#1}}
\newcommand{\appref}[1]{Appendix \ref*{app:#1}}

\renewcommand{\sun}{_\odot}

\newcommand{\red}{{}_\mathrm{R}}
\newcommand{\blue}{{}_\mathrm{B}}
\newcommand{\redsel}{{}_\mathrm{red}}
\newcommand{\bluesel}{{}_\mathrm{blue}}
\newcommand{\bad}{{}_\mathrm{bad}}
\newcommand{\good}{{}_\mathrm{good}}

\newcommand{\redi}{{}_{\mathrm{R},i}}
\newcommand{\bluei}{{}_{\mathrm{B},i}}
\newcommand{\subi}{ {}_i }
\newcommand{\commai}{ {}_{,i} }

\renewcommand{\max}{_\mathrm{max}}
\newcommand{\helio}{_\mathrm{helio}}
\newcommand{\maxi}{_{\mathrm{max},i}}

\newcommand{\Sersic}{S\'ersic}
\newcommand{\autott}{\textsc{auto}}
\newcommand{\petrott}{\textsc{petro}}
\newcommand{\modeltt}{\textsc{model}}

\newcommand{\pet}{_\mathrm{petro}}

\newcommand{\eff}{_\mathrm{eff}}

\newcommand{\starcol}{(g^*\!- i^*)}
\newcommand{\dust}{A_{V}}
\newcommand{\LWage}{ \left< t_* \right>}
\newcommand{\gauss}{\mathcal{G}}
\newcommand{\emcee}{\textsc{emcee}}

\begin{document}


\title[Galaxy And Mass Assembly: Deconstructing Bimodality --- I.\ Red ones
and blue ones] {Galaxy And Mass Assembly (GAMA): Deconstructing Bimodality---
I.\ Red ones and blue ones}


\author[Edward N. Taylor et al.]
{Edward N.~Taylor,$^{1,2,}$\thanks{ent@ph.unimelb.edu.au}
Andrew M.~Hopkins,$^3$
Ivan K.~Baldry,$^4$
Joss Bland-Hawthorn,$^2$\newauthor
Michael J.I.~Brown,$^{5}$ 
Matthew Colless,$^{3,6}$
Simon Driver,$^{7,8}$
Peder Norberg,$^{9}$
Aaron S.G.~\newauthor Robotham,$^{7,8}$
Mehmet Alpaslan,$^7$
Sarah Brough,$^3$
Michelle E.~Cluver,$^{10}$
Madusha\newauthor Gunawhardhana,$^3$
Lee S.~Kelvin,$^{8,11}$
Jochen Liske,$^{12}$
Christopher J.~Conselice,$^{13}$
Scott\newauthor Croom,$^2$
Caroline Foster,$^{3}$
Thomas H.~Jarrett,$^{10}$
Maritza Lara Lopez,$^3$
 Jon Loveday$^{14}$ \\
$^{1}$School of Physics, the University of Melbourne, Parkville,
  VIC 3010, Australia \\
$^{2}$Sydney Institute for Astronomy, School of Physics, University of
Sydney, NSW 2006, Australia \\
$^{3}$Australian Astronomical Observatory, PO Box 915, North Ryde, NSW 1670, Australia \\
$^{4}$Astrophysics Research Institute, Liverpool John Moores University, Twelve Quays House, Egerton Wharf, Birkenhead, CH41 1LD, UK \\
$^{5}$School of Physics, Monash University, Clayton, Victoria 3800, Australia \\
$^6$Research School of Astronomy \& Astrophysics, Australian National University, Weston Creek, ACT 2611, Australia \\
$^7$ICRAR, The University of Western Australia, 35 Stirling Highway, Crawley, WA 6009, Australia \\
$^{8}$School of Physics \& Astronomy, University of St Andrews, North
Haugh, St Andrews, KY16 9SS, UK \\
$^{9}$Institute for Computational Cosmology, Department of Physics,
Durham University, Durham DH1 3LE, UK \\
$^{10}$University of Cape Town, Astronomy Dept, Private Bag X3, Rondebosch 7701, South Africa \\
$^{11}$ Institut f\" ur Astro- und Teilchenphysik, Universit\" at Innsbruck, 
Technikerstra\ss e 25, 6020 Innsbruck, Austria \\
$^{12}$European Southern Observatory, Karl-Schwarzschild-Str.~2, 85748
Garching, Germany \\
$^{13}$Centre for Astronomy and Particle Theory, University of
Nottingham, University Park, Nottingham NG7 2RD, UK \\
$^{14}$Astronomy Centre, University of Sussex, Falmer, Brighton BN1 9QH, UK
}

\maketitle 

\begin{abstract} 
    We measure the mass functions for generically red and blue
galaxies, using a $z < 0.12$ sample of $\log M_* > 8.7$ field galaxies from
the Galaxy And Mass Assembly (GAMA) survey. Our motivation is that, as we
show, the dominant uncertainty in existing measurements stems from how `red'
and `blue' galaxies have been selected/defined. Accordingly, we model our data
as two naturally overlapping populations, each with their own mass function
and colour--mass relation, which enables us characterise the two populations
without having to specify {\em a priori} which galaxies are `red' and `blue'.
Our results then provide the means to {\em derive} objective operational
definitions for the terms `red' and `blue', which are based on the
phenomenology of the colour--mass diagrams. 

Informed by this descriptive modelling, we show that: 1.) after accounting for
dust, the stellar colours of `blue' galaxies do not depend strongly on mass;
2.) the tight, flat `dead sequence' does not extend much below $\log M_* \sim
10.5$; instead, 3.) the stellar colours of `red' galaxies vary rather strongly
with mass, such that lower mass `red' galaxies have bluer stellar populations;
4.) below $\log M_* \sim 9.3$, the `red' population dissolves into obscurity,
and it becomes problematic to talk about two distinct populations; as a
consequence, 5.) it is hard to meaningfully constrain the shape, including the
existence of an upturn, of the `red' galaxy mass function below $\log M_* \sim
9.3$. Points 1--4 provide meaningful targets for models of galaxy formation
and evolution to aim for. \end{abstract}



\section*{} 

\begin{center}
{\em ``All nature is perverse \& will not do as I wish it''} \\
 --- Charles Darwin (1855) \hspace{3cm}
\end{center}

\section{Introduction, aims, and overview}

\subsection{Introduction}

Quantitative studies of galaxy demographics --- that is, of the multivariate
distribution functions that connect global galaxy properties --- provide the
empirical bedrock on which theoretical models of galaxy formation and
evolution are founded. The quality of a cosmological model of galaxy formation
\citep[\eg ][]{Croton2006, Bower2006, Bower2008, Naab2007, Somerville2008,
Schaye2010} is judged by its ability to reproduce the most basic demographics
of real galaxy samples. This includes univariate distributions like the mass
or luminosity functions, and/or bivariate distributions like the size--mass,
colour--mass or mass--density relations. The field of galaxy formation and
evolution is thus largely data-driven, and is likely to remain so for the
foreseeable future.

It has long since been established that there exist a number of empirical
`laws' that describe the remarkably tight correlations between most, if not
all, of the global parameters used to describe galaxies: \eg, luminosity,
mass, colour, mean stellar age, star formation rate, size, shape, structure,
morphology, dynamics, etc.\ \citep[\eg][]{Freeman1970, TullyFisher,
FaberJackson, FP1, FP2}. Since 2000, an explosion in the volume and quality of
galaxy survey data at low- and high-redshifts has helped to consolidate and
make concrete these insights, at least for high mass galaxies.

One of the most important and influential insights has been the idea that
these scaling relations can be best understood as a unidimensional sequence in
stellar mass \citep[\eg,][]{Shen2003, Kauffmann2003, Tremonti2004,
Blanton2005, Gallazzi2005, Gallazzi2006, Baldry2006, deRijcke2007}---but see
also, \eg, \citet{BelldeJong, Bernardi2005, Kauffmann2006, Franx2008,
Graves2009, Williams2010, Wake2012}, who argue that stellar surface density or
central velocity dispersion may be the more fundamental parameter. In this
picture, once mass is controlled for, local environment potentially plays an
important but secondary role \citep[\eg][]{Hogg2004, Baldry2006,
VanDerWel2008, vandenBosch2008, Bamford2009, PengLilly, Geha2012,
Wijesinghe2012}.

There is also the longstanding idea that the galaxy population can be
naturally subdivided into two (and only two) broad classes. Even before
galaxies came to be called `galaxies', \citet{Hubble} recognised that the
`extragalactic nebulae' could be qualitatively separated into two distinct
phenomenological classes, based on their morphologies. Broadly speaking, at
fixed mass, the elliptical `early type' galaxies are older, redder, less
likely to be star forming, and smaller than the `late type' spirals
\citep[\eg][]{Strateva2001, Shen2003, Blanton2003a, Bell2004-Morph, Ellis2005,
Driver2006, Papovich2012}. In this way, there appear to be two (and only two)
distinct sets of relations that describe how galaxies' properties scale with
mass; one set for each of the early- and late-type populations. Further, early
types are, on average, more massive and more likely to be found in denser
environments \citep[\eg,][]{Dressler1980, Kauffmann2003, Blanton2005,
Baldry2006, VanDerWel2008}. The idea has thus been that these two populations
correspond to two (and only two) distinct evolutionary states.

One aspect of this `bimodality'---or, better, the dichotomy between the
`developing' and `developed' galaxy populations---has been particularly
influential, inasmuch as it has received a huge amount of attention from
observers and modellers alike. In order to reproduce the distributions of
galaxy {\em colours} \citep[\eg,][]{Bell2003, Baldry2004, Balogh2004}, and in
particular the evolving mass functions (MFs) of red/blue galaxies
\citep[\eg,][]{Bell2004, Tanaka2005, Borch2006, Arnouts2007, Faber2007,
Brown2008, Drory2009, PengLilly, Ilbert2010, Brammer2011}, cosmological models
have had to introduce an {\em ad hoc} `quenching' mechanism (or mechanisms) to
disrupt or prevent star formation. Within the models, these inclusions act on
more massive galaxies and/or galaxies in denser environments, either by the
removal/expulsion of the existing gas reservoir, or by preventing the
accretion of new material.

The physical nature of the quenching process remains controversial. The most
popular candidates are energetic `feedback' from an AGN
\citep[\eg][]{Croton2006, Menci2006, Bower2006, Bower2008, Somerville2008}, or
a distinction between `hot-' and `cold-mode' accretion
\citep[\eg][]{Keres2005, DekelBirnboim2006, Cattaneo2008, vandenBosch2008}
resulting from the presence or absence of persistent shock-fronts in infalling
gas. The quenching mechanism is usually taken to be linked to halo mass, and
may or may not have an environmental component \citep[\eg][]{PengLilly}.

\subsection{Aims}

With the above as background, our immediate goal in this paper is to derive a
quantitative, phenomenological description of the bivariate colour--mass
distribution function for field galaxies in the local universe, with
particular emphasis on the colour--mass relations (CMRs) and mass functions
(MFs) for the redder and bluer of the galaxy subpopulations.

In essence, our goals are similar to those of \citet{Baldry2004}, who set out
to quantitatively model the two apparently distinct but overlapping `red' and
`blue' populations seen in the colour--magnitude plane. The colour--magnitude
diagram is astronomy's most basic diagnostic plot. For galaxies, as a measure
of the integrated starlight, magnitude is typically taken as a proxy for total
stellar content; \ie, stellar mass. Colour is a simple observable diagnostic
parameter that characterises galaxies' stellar populations. In particular,
modulo dust, colour acts as a proxy for the luminosity-weighted mean stellar
age, which can also be taken as an average specific star formation rate (SFR)
over long ($\sim$ Gyr) timescales.

Our analysis improves on that of \citet{Baldry2004} in two
ways. First, we use the results of stellar population synthesis (SPS)
modelling of broadband spectral energy distributions (SEDs), rather than
simple restframe luminosities and colours. Specifically, we use SPS-derived
stellar mass estimates as our proxy for total stellar content, and we use
dust-corrected intrinsic stellar colour as (at least in principle) a more
direct tracer of galaxies' stellar populations. Second, we extend the
\citet{Baldry2004} analysis by developing and applying a statistically
rigorous mixture modelling formalism to derive a quantitative,
phenomenological description of the bimodality in galaxies' stellar
populations.

The crux of the problem is that the (optical) colour distributions of the
apparently distinct red and blue populations are seen to overlap. In the first
instance, this presents an operational problem: how best to disentangle these
two populations. In the second instance, the fact of overlap makes it
difficult to interpret the terms `red' and `blue' in concrete, astrophysical
terms. Given the role that these kinds of observations have in guiding
theories of galaxy formation and evolution, a secondary goal of this work is
to elucidate some of the important conceptual subtleties and difficulties
inherent to this kind of analysis, which are too often glossed over---if not
ignored altogether.

\subsection{Overview}

Our discussion proceeds in four parts, as follows. 

The GAMA data and our basic analysis of them are laid out in \secref{data} and
\secref{complete}. We discuss our ability to meaningfully constrain dust
obscurations and intrinsic stellar colours in \secref{colours} and
\figref{colours}. The (limited) role of incompleteness and selection effects
in our results are discussed in \secref{complete}, as well as \secref{state}.

In the second part, we motivate and describe our approach to the problem. In
\secref{redness}, we show how and why previous studies have found
qualitatively and quantitatively different results for the red/blue MFs:
namely, the different---and almost always arbitrary---ways that the `red' and
`blue' galaxy samples have been selected or defined. The extent to which these
results provide meaningful constraints on the process of galaxy formation and
evolution is therefore limited by the extent to which the terms `red' and
`blue' can be shown to be astrophysically meaningful.

This is why we have set out to derive objective and phenomenological,
operational definitions for the terms `red' and `blue'. In addition to the
description of our modelling procedure given in \secref{method}, we provide a
more pedagogical discussion of our approach in \appref{themodel}, in which we
develop our analysis starting from a simple $\chi^2$-minimisation fit. This
material is intended to help fast-track researchers intending to apply a
similar mixture-modelling analysis to their own data. The reader that is
concerned about how we have decided to parameterise our fits to the bivariate
colour--mass distributions should focus on \secref{modelsel}.

In the third part, we present the results of our descriptive modelling. The
quality of our fits is illustrated in \figref{rfdist} and \figref{stardist},
and discussed in \secref{dist}. The specific question of how to interpret our
results at very low masses is discussed in \secref{wrong}. In
\secref{results}, we present and discuss our characterisations of the CMRs for
the two galaxy populations (\secref{cmrs}), the objective classification
scheme that we derive from our modelling of the CMDs (\secref{classes}), and
the MFs for the two populations (\secref{mfs}). Our most important
astrophysical results and conclusions can be found in \secref{observations},
in which we describe the essential characteristics of the bimodal (or, better,
two-population) distribution of galaxies' stellar populations.

The fourth and final part comprises a discussion of our results and methods
(\secref{discuss}). We revisit the results of earlier studies in light of our
analysis and results in \secref{others}, including illustrating how our
objective, phenomenological red/blue classifications compare to those used
previously; this effectively closes the loop opened in \secref{redness}. In
\secref{howelse}, we show how our objective classification scheme maps onto
two commonly used diagnostic diagrams. \figref{nircols} and \figref{HdD4}
provide important validation and illustration of how our objective
classifications discriminate between galaxies with qualitatively different
stellar populations. Finally, in \secref{fallacies} we discuss potential
objections to our analysis and results.

This paper is long. Given the increasing awareness of the need for more
detailed and rigorous statistical analysis of large galaxy catalogues, our
hope is that this paper will serve as a useful pedagogical resource for
researchers working on similar problems in the future. Therefore, some of the
more technical description and discussion of our statistical formalism may not
be of interest to some readers; or, for those researchers familiar with
Bayesian MCMC fitting techniques, for example, they may seem {\em overly}
detailed. In recognition of this, we have made efforts to make the structure
of the paper as modular as possible, so that the reader can choose which
sections to read closely, and which to skip altogether.

For the casual or first time reader, we make the following recommendations.
Start with \figref{colours} and its caption. Then, read \secref{redness} and
the opening of \secref{method} for the motivation for our analysis, and an
outline of the basic assumptions that underpin our approach. \secref{cuts} and
\secref{overlap} are particularly important, in that they provide our
rationale for favouring the more neutral designations `B' and `R', in place of
the more laden terms `blue' and `red'. Next, move to \secref{quality} and
\secref{done}, which offer an intuitive way of understanding how a mixture
modelling approach can be used to characterise the two populations without
ever specifying which galaxies belong to which population. After reviewing
Fig.s \ref{fig:rfcmr}--\ref{fig:starmf} and their captions, move to
\secref{observations}, in which we discuss our main results and conclusions in
astrophysical terms. \secref{howelse}, in which we show that our R-type
galaxies really do have different and much older stellar populations than
B-type galaxies, is very important. \secref{others}, and especially
\secref{peng}, in which we compare our results to those of \citet{PengLilly},
is also important for readers interested in the problem of quenching. Readers
that remain concerned about the validity of our methods and results---as well
as those of previous studies---should read \secref{fallacies} carefully.

A summary of our analysis, results, and conclusions is given in
\secref{summary}. \figref{mcmc} serves as a table of results for the various
parameters that we have fit for. Machine-readable tables of the results shown
in Fig.s \ref{fig:rfcmr}---\ref{fig:starmf} are made available as additional
online content. We are happy to provide the source code for our modelling on
request.

Throughout, we adopt the concordance cosmology: $(\Omega_m, \, \Omega_\Lambda,
\, H_0)$ = $(0.3, \, 0.7, \, 70 \, \mathrm{km/s/Mpc})$. All stellar mass
estimates have been derived assuming, or have been approximately scaled to
match, a \citet{Chabrier2003} stellar initial mass function (IMF). All
magnitudes are expressed in the AB system. Finally, a note on notation: in the
more technical sections of this work, which describe the formal basis and
justification for our modelling, we will represent vectors as $\vec{v}$,
matrices as $\mat{M}$, and sets as $\set{S}$, as distinct from scalar
quantities like $x$, $Q$, $\zeta$, $\Phi$, $\ell$, or $\Ell$.


\section{Data --- The Galaxy And Mass Assembly (GAMA) survey} \label{ch:data}

\subsection{Spectroscopic redshifts and flow-corrected distances}
\label{ch:zdist}


As an optical spectroscopic survey, the Galaxy and Mass Assembly
\citep[GAMA;][]{Driver2009,Driver2011} survey has now completed its
observations of three separate equatorial fields of 60 $\square^\circ$ each.
The spectroscopic target selection is described by \citet{Baldry2010}. Targets
have been selected on the basis of dust-corrected \petrott\ $r$-band
magnitudes from the SDSS DR7  \citep{Abazajian2009}. For GAMA-II, all three
fields have been surveyed to a depth of $r\pet < 19.8$ mag. In GAMA-II
nomenclature, these define the \texttt{SURVEY\_CLASS} $\ge 4$ sample selection
limits.

The GAMA survey strategy \citep{Robotham2010} has been optimised for uniform
and near total spectroscopic completeness ($\gtrsim 98$\,\%), even in regions
with high target density. Targeting completeness is better than 99.9\,\%, with
only 160/189059 main survey targets not having been observed. As a function of
the SDSS $r$-band \texttt{fiber} magnitudes, redshift success is 99 and 95\,\%
for $r_\mathtt{fiber} = 19.5$ and 20.5 respectively, where success is defined
as $> 98$\,\% confidence that the given redshift is correct.

Whereas previously, the GAMA spectroscopic redshifts were based on by-eye
determinations done by observers at the telescope, the spectroscopic redshifts
given in the GAMA-II catalogues have been derived using an automated pipeline,
as described by \citet{Baldry2014}. This has reduced the standard redshift
error from $\sim 100$ to $\sim 33$ km/s, and reduced the redshift blunder rate
for high confidence redshifts from $\lesssim 5$ to $\lesssim 0.1$\,\%, as
determined through comparisons between repeat observations of GAMA targets,
and through comparisons between GAMA and SDSS observations of common targets.

For the purposes of calculating luminosity and comoving distances, these
heliocentric redshifts have been corrected for local bulk flows using the
model of \citet{Tonry2000} for the very lowest redshifts ($z\helio < 0.02$),
and then tapering to a Cosmic Microwave Background (CMB)-centric frame for $z
> 0.03$. The details of this conversion are given by \citet{Baldry2011}. It is
these flow-corrected redshifts that we will use as the basis of our analysis,
including sample selection. 


\subsection{Imaging and photometry} \label{ch:photometry}

The photometric backbone of the GAMA-II dataset comprises optical {\em ugriz}
imaging from SDSS (DR7) and near infrared {\em ZYJHK} imaging from the VIKING
survey. The SDSS data have been extensively described \citep[see,
\eg,][]{Strauss2002, Abazajian2009}, and have been obtained from the SDSS Data
Archive Server\footnote{DAS for DR7: \texttt{das.sdss.org}}. The VIKING data
reduction has been done by the Cambridge Astronomical Data Unit (CASU)
pipeline for VISTA\footnote{See
\texttt{http://casu.ast.cam.ac.uk/surveys-projects/vista} for online
documentation.}, and have been obtained from the VISTA Science
Archive\footnote{\texttt{http://surveys.roe.ac.uk/vsa/}}
\citep[VSA;][]{Cross2012}.   The GAMA-II photometric catalogue is based on an
independent reanalysis of these imaging data \citep[see][]{Hill2010,
Driver2011, Kelvin2012, Driver2014}.

For the purpose of constructing multi-band Spectral Energy Distributions
(SEDs), a set of PSF-matched mosaics ($2''$ FWHM) have been made. These have
been fed to SExtractor \citep{SExtractor}, which has been run in dual-image
mode, using the $r$-band mosaics as the detection images, to yield seeing- and
aperture-matched {\em ugriZYJHK} SEDs. Comparisons between earlier versions of
this photometry and the SDSS \modeltt\ and \petrott\ photometry are presented
by \citet{Hill2010} and \citet{Taylor2011}. For this work, we have used the
latest GAMA-II photometric catalogue (internal designation
\texttt{ApMatchedCatv05}), which will be described by Driver et al.\ (in
prep.). In comparison to the earlier \texttt{ApMatchedCatv01} catalogue
presented by \citet{Hill2010}, the most significant change is the supersession
of the UKIDSS LAS NIR data with {\em ZYJHK} data from VST VIKING.

As is well known, the finite \autott\ aperture is prone to miss a significant
amount of flux for galaxies that are faint and/or have significant low-surface
brightness wings. To account for this, we characterise the total, observers'
frame $r$-band flux by fitting a \Sersic\ profile to the observed 2D light
distributions for each galaxy. As described by \citet{Kelvin2012}, this has
been done using \textsc{galfit3} \citep{galfit, galfit3}, incorporating a
galaxy-specific model for the PSF, and taking care to isolate and deblend the
target from any and all nearby galaxies. In the fits, the \Sersic\ profile has
been truncated at 10 $R\eff$, which typically corresponds to a surface
brightness limit of $\mu\eff \approx 30$ mag / $\square''$. For this work, we
have used these \Sersic -fit estimates of total $r$-band flux, taken from the
\texttt{SersicCatv09} catalogue, to normalise the \autott\ SEDs described
above. At fixed \Sersic\ index, the RMS in the values of these
corrections-to-total is of order 0.05 mag or less, even for the faintest
apparent magnitudes.


\subsection{Stellar Population Synthesis (SPS) modelling}
\label{ch:sedfits}

The redshifts and multiband photometry described above have been combined to
estimate stellar population parameters including masses, restframe photometry,
luminosity-weighted mean stellar ages, dust obscurations, metalliticies,
specific star formation rates, etc. The basic procedure is the same for the
GAMA-I masses presented by \citet{Taylor2011}, but with one significant
improvement. For GAMA-II, each band is weighted such that the SPS fits are
done to a fixed {\em restframe} wavelength range of 3000--11000 \AA, which
corresponds roughly to restframe $u$--$Y$. Between this change, and the change
from the UKIDSS to the VIKING NIR data, the large systematic errors in the SPS
fits discussed at length in \citet{Taylor2011} have been reduced significantly
in the $Z$, $Y$, and $J$ bands. This suggests that at least part of these
problems were due to calibration errors in the UKIDSS data. The issues with
the $H$ and $K$ band data persist, but at a lower level, which is why we have
not pushed further into the restframe NIR for the SPS fits.

Following standard practice, in the course of these fits, an arbitrary error
floor is imposed on each photometric point by adding an additional uncertainty
of 0.05 mag in quadrature to the catalogued photometric uncertainties. This
decision is typically justified as providing protection against both errors
and uncertainties in the relative or cross calibration of the photometry in
different bands, as well as against template mismatch and/or aliasing errors
in the SPS fitting. Given that these uncertainties are treated as being both
random and independent, however, neither of these justifications are really
well-founded. With the exception of the $u$-band photometry, the catalogued
errors are almost always comparable to, or even less than, 0.05 mag. This
imposed error floor is thus the limiting factor in setting the formal
uncertainties on our SPS-derived results including, in order of increasing
importance, $(g-i)$, $\log M_*$, $\dust$, and $\starcol$. In fact, as we will
argue in \secref{errors} below, this decision leads to drastic overestimates
of the `true' errors on the intrinsic (\ie, dust corrected) stellar colours
$\starcol$.

\subsubsection{Stellar mass and stellar population parameter estimates}

The SED fitting process involves comparing the observed photometry to a
library of synthetic stellar population spectra. This stellar population
library (SPL) was constructed using the \citep{BC03} stellar evolution models
for a \citet{Chabrier2003} IMF, and making the following common simplifying
assumptions: 1.)\ exponentially declining star formation histories, 2.)\
uniform, single screen dust attenuation, and 3.)\ uniform stellar
metallicities. The estimates for both the values of and uncertainties in the
SP parameters for individual galaxies have been made in a Bayesian way. The
RMS difference between our values for $M_*/L_i$ and those from the MPA/JHU
catalogues for SDSS DR7 is $\lesssim 0.07$ dex, with no appreciable systematic
differences as a function of colour, structure, mass, or apparent magnitude.

We note that the quantitative values of the mass estimates can be recovered to
high precision ($1 \sigma$ error of $0.06$ dex) using the following simple,
empirical relation \citep{Taylor2011}:
\begin{equation}
	\log M_* / [\mathrm{M}\sun] = 1.15 + 0.70 \, (g - i) - 0.4 ~ M_i ~ .
\end{equation} 
In this way, the $(g-i)$--$M_*$ colour--stellar mass diagram can be
transparently viewed as simple, linear shearing of the $(g-i)$--$M_i$
colour--magnitude diagram. Further, for the reader wishing to compare their
data to our results, this relation offers a simple and transparent basis for
comparison between our (and by extension, the SDSS) stellar mass estimates. 

\begin{figure} \centering \includegraphics[width=8.7cm]{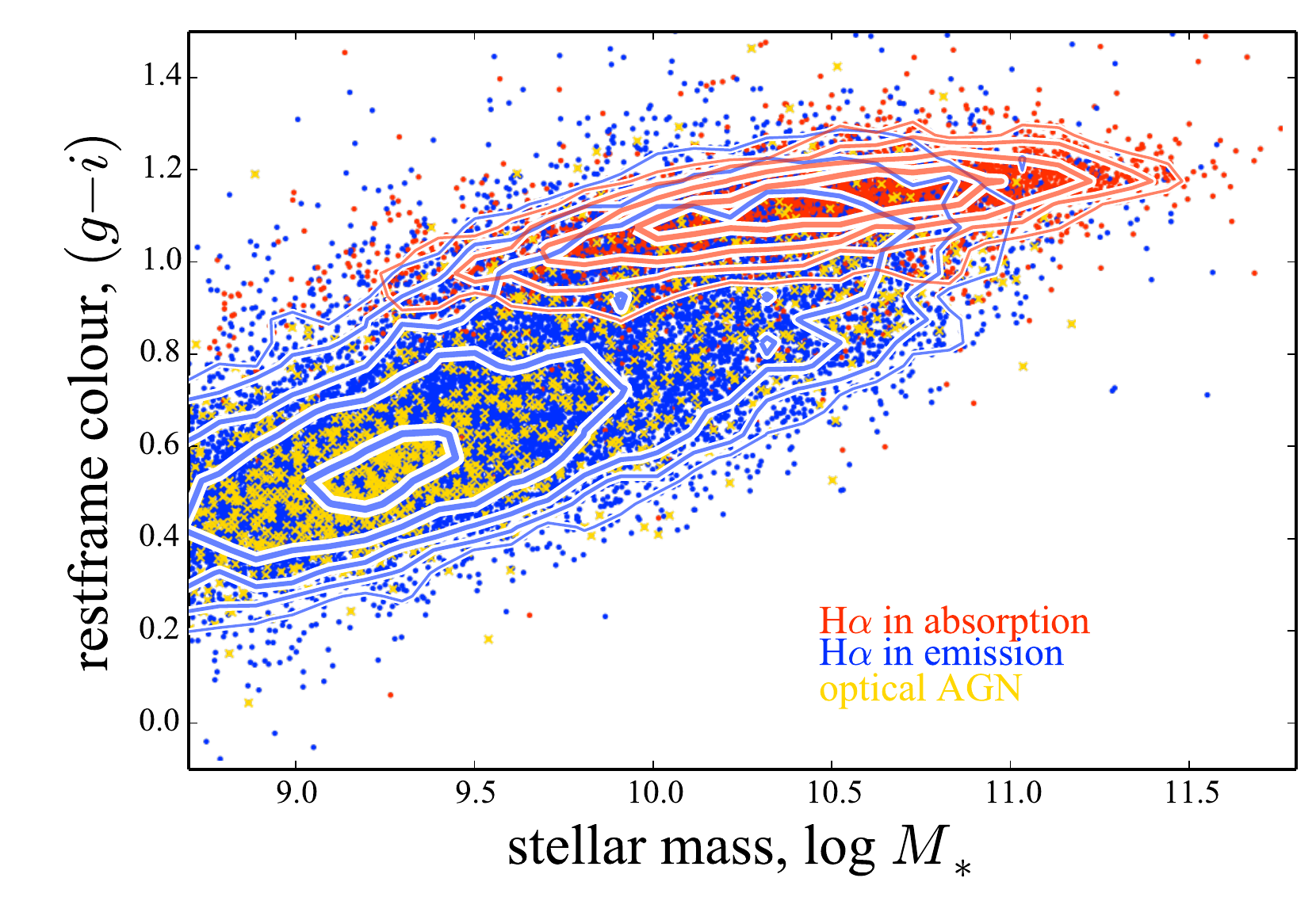}
\vspace{-0.2cm} \includegraphics[width=8.7cm]{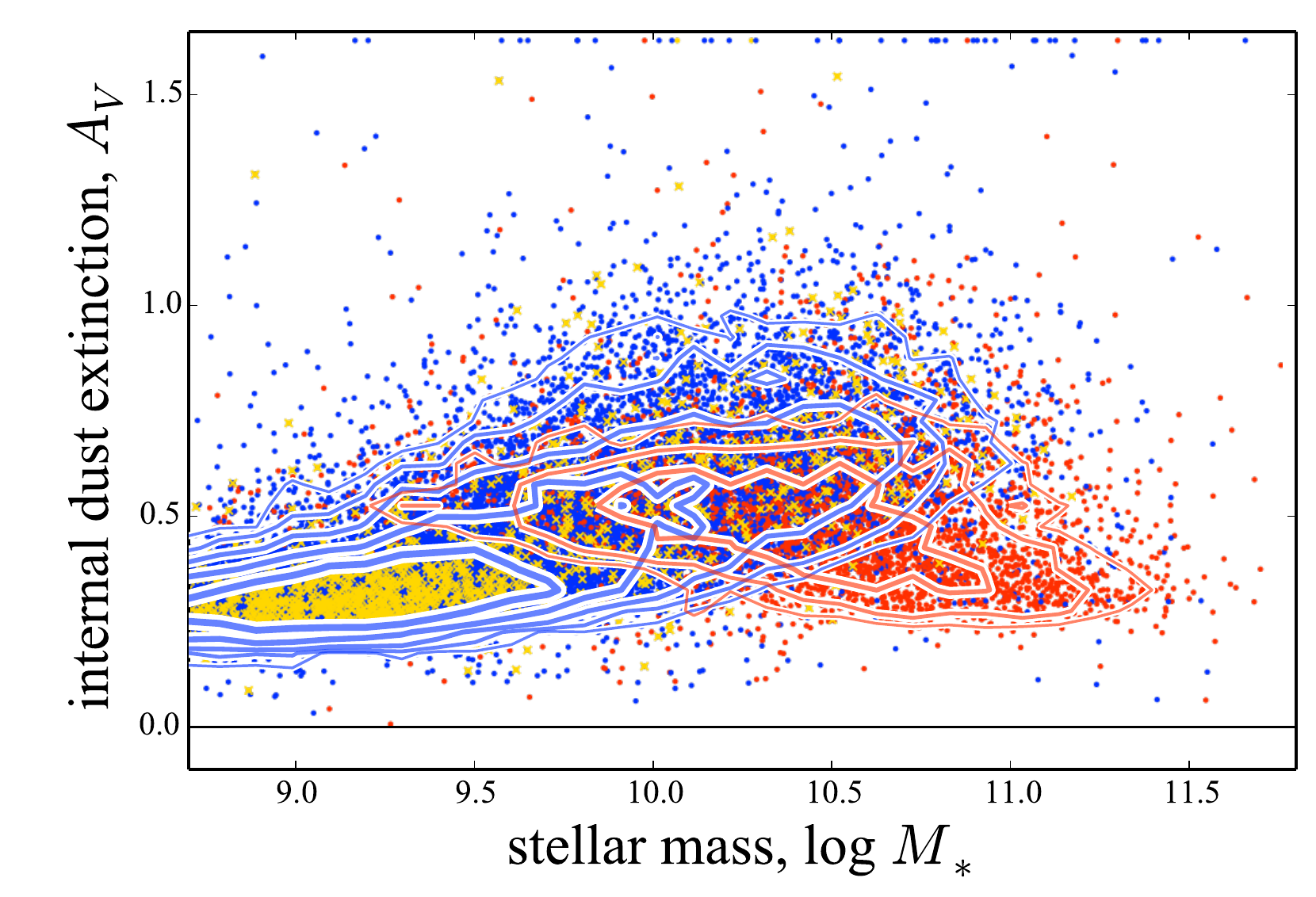}
\vspace{-0.2cm} \includegraphics[width=8.7cm]{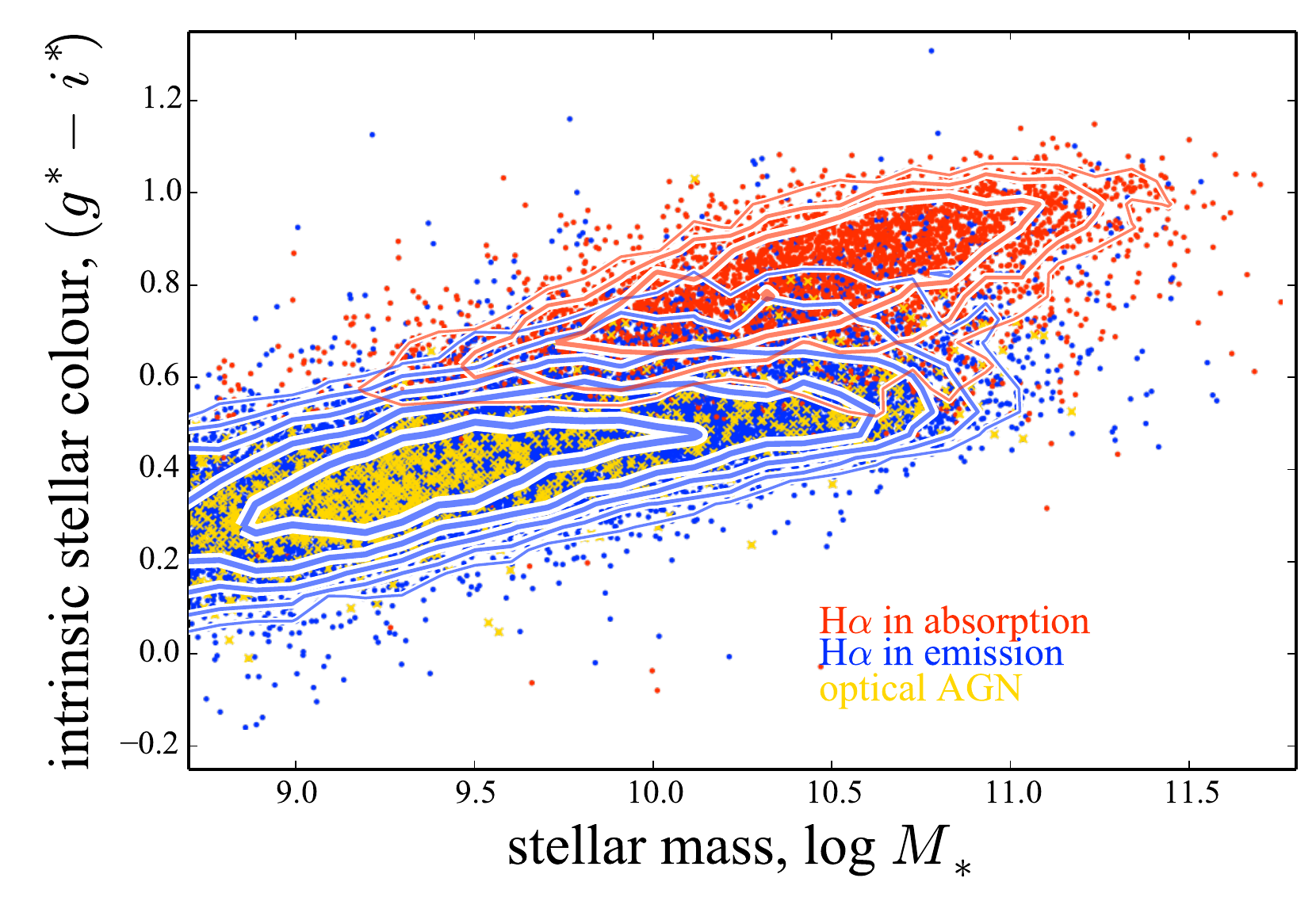}
\caption{Stellar population parameters, as derived from SPS modelling of
broadband SEDs.--- The upper and lower panels show the distribution of $z <
0.12$ galaxies in the colour--stellar mass plane using restframe $(g-i)$
colour, and using the intrinsic (\ie, corrected for internal dust extinction)
stellar colour $\starcol$, respectively. The middle panel shows the dust
extinction, $\dust$, as a function of mass, and can be thought of as linking
the other two. Within each panel, the blue and red points distinguish galaxies
that do or do not show clear H$\alpha$ line emission; optically identified AGN
are highlighted yellow. These spectral classifications are entirely
independent of the SED-fit results. Note that the AGN seem to be almost
exclusively associated with the blue sequence in the $\starcol$ CMD. As
expected, there is a strong correspondence between galaxies with old, red
stellar populations, galaxies with little to no dust, and galaxies with little
or no H$\alpha$ emission. That is, there is good consistency between the
wholly independent photometric and spectroscopic classifications of blue/red,
young/old, and active/passive galaxies. (But this does not imply that all
`blue' galaxies are star-forming, or vice versa.) This should give some
confidence in the SPS-derived modelling used to derive all of $M_*$, $(g-i)$,
$\dust$, and $\starcol$. \label{fig:colours}} \end{figure}

\subsubsection{Effective restframe and intrinsic stellar colours}
\label{ch:colours}

Restframe luminosities and colours are derived for each galaxy in the course
of the SPS fitting process, in the same way as for $M_*$ or $M_*/L_i$.
Naturally, these values reflect the galaxies' constituent stellar population,
modulated by interstellar dust within each galaxy. In order to more directly
trace the stellar populations, we will therefore also consider dust corrected
or intrinsic stellar colours. To our knowledge, this approach was first
pursued by \citet{CowieBarger}. It has also been described by, \eg,
\citet{Brammer2009} and \citet{Cardamone2010}, on the basis of 32-band
photometric redshifts from the MUSYC Narrow-Medium Band Survey (NMBS). Here,
we demonstrate the feasibility and applicability of this kind of analysis for
broadband SEDs, given spectroscopic redshifts.

We will focus on the intrinsic stellar $(g-i)$ colour, which we will represent
as $\starcol$. This parameter is a very good proxy for luminosity weighted
mean stellar age\footnote{We note that optical colour is also frequently
treated as a proxy for specific star formation rate. For the (smooth)
exponentially declining SFHs used for the SPL, there is naturally a close
connection between SSFR and $\LWage$ for these models. However, the
correlation between the SPS-inferred and H$\alpha$-derived SSFR for real
galaxies is weak at best. It seems to us that SPS fits are good at
constraining $\LWage$, which is very closely related to $M_*/L$, but much less
so at constraining the instantaneous SSFR. Thus, we consider $\starcol$ to be
a better proxy for $\LWage$ than for SSFR.}, $\LWage$. Quantitatively, at
fixed $\starcol$, the range in $\LWage$s is everywhere $\lesssim 0.1$ dex, and
$\lesssim 0.05$ dex for $\starcol \gtrsim 0.5$.

The values of $\starcol$ have also been derived in the course of the SPS fits,
but can be derived to within $\lesssim 0.01$ mag from the values of $(g-i)$
and $\dust$ directly:
\begin{equation}
	\starcol \cong (g-i) - 6.0 ~ \dust ~ .
\end{equation}
The coefficient of 6.0 in this relation reflects \citet{Calzetti} dust
extinction applied to the average (SED-fit) galaxy spectrum within our sample.
The first order effect of adopting a different dust obscuration law would be a
scaling of this coefficient. The choice of dust obscuration law is thus a
potential source of unaccounted-for random and systematic error in our
characterisations of the $\starcol$ CMRs (but much less so for the MFs).

The formal uncertainties in the derived values of $\dust$ ranges from $\approx
0.15$ mag for very blue galaxies to $\approx 0.3$ for very red ones. Again,
formally, the dominant factor in these uncertainties is the imposed 0.05 mag
error floor on each point in the SED (rather than the catalogued photometric
uncertainties), and thus they are not strongly magnitude dependent. Given that
the parameter $\dust$ is bounded---the amount of dust cannot be
negative---these random uncertainties will lead to a systematic overestimate
of $\dust$ in the case that the true value of $\dust$ is close to zero. This
may in turn induce a differential systematic bias in the inferred $\starcol$
colours of galaxies with little to no dust (compared to those with some dust),
such that the value of $\starcol$ may be too blue by $\sim 0.1$ mag, but not
more than 0.2 mag. This limits our ability to accurately determine the locus
of the CMR for red sequence galaxies, at least where such galaxies have little
to no dust. Without diminishing this point, what is more important for our
analysis---and especially when it comes to determining the MFs for the red and
blue populations---is that we are able to make the {\em qualitative}
distinction between an old stellar population and a younger one with some
dust.

With this in mind, \textbf{\figref{colours}} shows a simple sanity check on
these values. In this Figure, we distinguish between galaxies with and without
strong H$\alpha$ line emission. Specifically, those galaxies with an
equivalent width of 1 \AA\ or greater are plotted as blue; AGN-host galaxies
(see \secref{sample}) are plotted as yellow crosses; the remainder of the
population with H$\alpha$ seen in absorption are plotted as red.

The top panel of \figref{colours} shows the effective, restframe $(g-i)$ CMD
for our $z < 0.12$ sample. As expected, galaxies with H$\alpha$ seen in
absorption can be seen to form a tight red sequence in $(g-i)$ colours.
However, there are also many galaxies with strong H$\alpha$ emission that lie
embedded in or very near to this red sequence. In general terms, these are the
dusty star-formers.

The central panel of \figref{colours} shows the SED-fit values of $\dust$ as a
function of $\log M_*$, using the same plotting scheme to distinguish `active
star-formers' from `quiescent' galaxies. The emission- and absorption-line
galaxies can be seen to follow different $\log M_*$--$\dust$ relations: in
general, the galaxies without H$\alpha$ emission have low (but non-zero;
$\dust \sim 0.2$--0.35) dust extinctions. While this is as expected, it is
crucial to realise that the spectral classifications are independent of the
SED-fit values for $\log M_*$ and $\dust$. These results thus demonstrate that
our SPS fits are indeed able to reliably distinguish between old SPs with
little or no dust, and dusty star-forming galaxies. (See also the discussions
based on optical--NIR colours or stellar spectral diagnostics described in
\secref{howelse}, below.)

For the galaxies with H$\alpha$ emission, there is a trend towards higher
values of $\dust$ with increasing $M_*$. The obvious implication is that the
star-forming population will be observed to become redder in $(g-i)$ towards
higher $M_*$ by virtue of their higher dust content, independently of any
variation in their stellar populations. This complicates any attempt to
disentangle the young/star forming and old/passive populations based on
$(g-i)$ alone.

As can be seen in the lower panel of \figref{colours}, however, the active and
quiescent populations are much better separated using the dust-corrected,
intrinsic stellar colour, $\starcol$. Again, we stress that the determination
of $\starcol$ is independent of the spectral classification---the fact that
galaxies that show H$\alpha$ in absorption are almost all red in $\starcol$
thus demonstrates that we are in fact able to distinguish between `red and
dead' galaxies from dusty star-forming galaxies on the basis of their
broad-band SEDs. (See \secref{nircols} for further discussion of this point.)

Note that our immediate goal in this paper is {\em not} to distinguish between
galaxies based on their instantaneous SFRs. (We will do this in another paper,
using the H$\alpha$ measurements.) Here, our goal is to characterise galaxies'
stellar populations, using the intrinsic stellar colour, $\starcol$, which is
a close proxy for luminosity weighted mean stellar age, $\LWage$. This relies
on our ability to meaningfully constrain the dust obscuration, which is what
\figref{colours} is intended to show. Taken together, the three panels of
\figref{colours} should thus give some confidence in the reliability of our
estimates of all of $M_*$, $(g-i)$, $\dust$, and $\starcol$.

\subsubsection{Covariant errors in $M_*$, $(g-i)$, and $\starcol$}
\label{ch:errors}

When we come to fitting the galaxy distributions in colour--mass space in
\secref{method}, we will want to account for the fact that the
measurement errors/uncertainties in $M_*$ and $(g-i)$ are correlated.
The strength of this correlation is characterised by the (Pearson) correlation
coefficient,
\begin{equation} \label{eq:rho}
	\rho_{xy} 
	\equiv \left< \frac{\left( x - \left< x \right> \right)
				\left( y - \left< y \right> \right) }
		{ \sigma_x ~ \sigma_y } \right> ~ .
\end{equation}
Here, $x$ and $y$ can be taken to be $\log M_*$ and either $(g-i)$ or
$\starcol$; $\sigma_x$ and $\sigma_y$ are the uncertainties in these values;
and $\left<Q\right>$ represents the expectation value for a generic quantity
$Q$. In the parlance of \citet{Taylor2011}, $\left<Q\right>$ is the Bayesian
`most likely' value, which is computed as the probability-weighted integral
over the posterior distribution function (PDF) for that quantity \citep[see
Eq.\ 5 of][]{Taylor2011}. By definition, the value of $\rho$ is constrained to
be $-1 \le 0 \le 1$, with the cases $\rho = -1, ~ 0,~ 1$ corresponding to
total anti-correlation, total independence, and total correlation,
respectively. The values of $\rho_{xy}$ have been computed with the formal
uncertainties $\sigma_x$ and $\sigma_y$ on a per galaxy basis in the course of
the SPS fitting process. For the galaxies in our sample, the covariance
between $\log M_*$ and $(g-i)$ is typically $\sim 0.4$; the $\log
M_*$--$\starcol$ covariance is typically in the range $0.1 \lesssim \rho
\lesssim 0.9$.

With this definition, the error/uncertainty ellipse for any individual galaxy
can then be expressed in the usual way for a bivariate Gaussian distribution:
\begin{equation} \label{eq:errorellipse}
	p(\vec{x}\subi-\vec{x}) = \frac{1}{ 2 \pi \, | \mat{S}\subi |^{1/2} } ~ 
		\exp \left[ \frac{-1}{2} (\vec{x}\subi - \vec{x} )^T 
				\mat{S}\subi^{-1} (\vec{x}\subi - \vec{x} ) \right] ,
\end{equation}
where the vector $\vec{x}\subi = (x_i, y_i)$ represents the observed data point and the associated error/uncertainty matrix, $\mat{S}\subi$, is:
\begin{equation} \label{eq:covar}
	\mat{S}\subi \equiv \left( \begin{matrix} 
	\sigma_{x,i}^2      &     \rho_{xy,i} ~ \sigma_{x,i} ~ \sigma_{y,i} \\
	\rho_{xy,i} ~ \sigma_{x,i} ~ \sigma_{y,i}      &     \sigma_{y,i}^2  
	\end{matrix} \right) ~ .
\end{equation}
Note that if $\rho_{xy,i} = 0$, then the matrix $\mat{S}\subi^{-1}$ is
diagonal with entries $\sigma_{x,i}^{-2}$ and $\sigma_{y,i}^{-2}$, and
\eqref{errorellipse} reduces to the familiar form for a 2D Gaussian with
$p(x,~y) \propto \exp [-\frac{1}{2}( x^2/\sigma_x^2 + y^2/\sigma_y^2 ) ]$.

As mentioned at the beginning of this Section, the SPS fits to the SEDs
includes an error floor of 0.05 mag, and it is this decision that largely
determines the formal uncertainties in $\starcol$. The median formal
uncertainty in $\starcol$ within our sample is 0.18 mag; 99\,\% of our sample
have uncertainties greater than 0.10 mag. By comparison, the {\em observed}
width of the blue and red sequences in the $\starcol$ CMD are on the order of
0.10 mag (see \figref{stardist}, below); \ie, significantly smaller than the
formal uncertainties.

This indicates that the formal (random) errors in $\starcol$ are badly
overestimated. For this reason, when we model the CMD, we rescale the formal
error estimates using a multiplicative factor $A_y$. The value of this scaling
factor is fit for as a nuisance parameter along with the rest of the model.
From our modelling of $\starcol$ CMD, we find $A_y \approx 0.24$; in effect,
we are ultimately using nearly uniform uncertainties in $\starcol$ of
$\approx$ 0.05 mag. Note that we do not rescale the formal uncertainties for
$\log M_*$, nor do we adjust the correlation coefficients $\rho_{xy}$. For
comparison, fitting to the $(g-i)$ CMD, the inferred value is $A_y \approx 1$;
\ie, we see no signs that the formal uncertainties on $(g-i)$ ought to be
rescaled.


\subsection{Sample Definition} \label{ch:sample}

Our analysis is based on a subset of the full GAMA database. Specifically, we
limit our analysis to those GAMA galaxies with $\log M_* > 8.7$ (\ie, $M_*
\gtrsim 5 \times 10^8$ M$\sun$) and $z < 0.12$. These mass and redshift limits
are motivated and justified in \secref{complete}, below. To ensure the
reliability and robustness of the spectroscopic redshift measurements, we will
only consider those galaxies with \texttt{nQ} $\ge 3$. We only consider the
$r$-band selected sample; that is, we ignore 12 H-ATLAS selected galaxies, and
355 filler targets with $19.80 < r\pet < 19.85$. With these selections, we
have a sample of 26368 galaxies.

98.5\,\% of our sample has effective surface brightness $\mu\eff < 23$. Based
on the completeness curves shown in \citet{Loveday2012}, we expect there to be
no significant surface brightness selection effects inherited from the (SDSS)
photometric parent catalogues, at least for $\log M_* \gtrsim 9$. We have
explored the impact of surface brightness-dependent redshift failure rates, by
applying completeness corrections as a function of the SDSS $r_\mathtt{fiber}$
magnitude. The effect on the MFs is negligible: only 1\,\% for $\log M_* =
9.5$, and still just 3\,\% for $\log M_* = 8.7$.

We do not explicitly exclude AGN from our analysis. In \figref{colours}, we
highlight the 1522 galaxies that are identified as AGN hosts, based on their
position in the BPT \citep{BPT} diagram, coupled with an H$\alpha$ equivalent
width $>$ 6 \AA\ selection. This is similar in spirit to the WHaN selection
described by \citet{Cid2012}, and was chosen to approximately reproduce the
by-eye spectral classifications by \citet{Robotham2013}. We note that the vast
majority of these AGN hosts are inferred to have `normal' B-type $\starcol$
colours. We have verified that none of our main results or conclusions
(including the shape of the B and R MFs) change if we choose to exclude these
galaxies.


\section{Quantifying and accounting for incompleteness as a function of mass,
colour, and redshift} \label{ch:complete}

The upper panels of \figref{colours} show the basic data for our analysis;
namely the $(g-i)$ and $\starcol$ CMDs. In both cases, the relative number of
red sequence galaxies in both the $(g-i)$-- and the $\starcol$--$M_*$ diagrams
peaks somewhere around $\log M_* \sim 10.5$. There is a drop-off in the
fraction of red galaxies below this mass, such that there is little to no
clear evidence for a continuation of the $\starcol$ red sequence below $\log
M_* \sim 9.5$. The principal difficulty in interpreting this result is the
extent to which our $z < 0.12$ sample is sensitive to truly `red and dead'
galaxies at these relatively low masses of $\log M_* \lesssim 10$. We explore
this issue in two complementary ways in this Section.

\boldmath
\subsection{Incompleteness and $\mathbf{1/V\max}$ corrections}
\unboldmath
\label{ch:vmax}

We use the standard $1/V\max$ technique \citep{Schmidt1968} to account and
correct for incompleteness as a function of both stellar mass and stellar
population. The essential idea behind $1/V\max$ corrections is to estimate
that maximal volume, $V\max$, over which any given galaxy would satisfy our
($r$-band) selection criteria. If we can estimate or predict the apparent
$r$-band magnitude for a given galaxy if it were to be placed at some generic
redshift $z'$ as $r(z')$, then $V\max$ can be derived by integrating over the
survey volume in which $r(z')$ is brighter than our selection limits. This has
been done in the course of the SPS SED fits, using the single best-fit SPL
template, as described in \citet{Taylor2011}.

There are two things that need to be accounted for when estimating the values
of $V\max$ for galaxies in our sample. First, there is the fact that the GAMA
target selection has been done on the basis of SDSS \petrott\ magnitudes. We
can account for this by calculating $r(z')$ as $r\pet + \Delta r(z')$, where
$r\pet$ is the (foreground extinction corrected) SDSS \petrott\ magnitude, and
$\Delta r(z')$ can be thought of as the $z'$-dependent $k$-correction implied
by the SPS fit. Second, there is the difference between the cosmological
redshift, which maps directly to comoving distance, $D$, as $c z \approx H_0 D
$, and the heliocentric redshift, which includes Doppler shifting from
peculiar motions due to local bulk flows. This is done by recognising that $(1
+ z'_\mathrm{helio}) = (1 + z') (1 + v_\mathrm{flow}/c)$, where
$v_\mathrm{flow}$ is the peculiar velocity arising from local bulk flows (see
\secref{zdist}). The value of $V\max$ is then defined via the maximum
(flow-corrected) redshift, $z'$, for which both the $r(z') < 19.8$ and $z' <
0.12$ selection criteria are satisfied.

We have experimented with using a density corrected $1/V\max$ weighting to
account for large scale structure at the lowest redshifts. \citet{Baldry2011}
have shown first that large scale structure in the $z < 0.06$ GAMA volume can
have a significant impact on the recovered mass functions, and second that
these effects can be largely mitigated by using a Density Defining Population.
Our case is rather different, however: even for $\log M_* \sim 9$, most of our
galaxies lie at $z > 0.06$. (Plus, the GAMA-II survey area is 25\,\% larger, as
well as 0.4 mag deeper in two of three fields.) Using the \citet{Baldry2011}
scheme, the corrections to the MFs are at the level of $\sim 5$\,\% for $\log
M_* \lesssim 9.5$. The problem is that using different DDPs yield different
corrections. The difference in the recovered $\log M_* \lesssim 9.5$ MF when
defining the DDP to be $\log M_* > 10.5$ or $\log M_* > 11$ galaxies is on the
order of $\sim 3$\,\%; that is, comparable to the size of the corrections
themselves. For this reason, we do not apply these negligible corrections.

\begin{figure} \includegraphics[width=8.5cm]{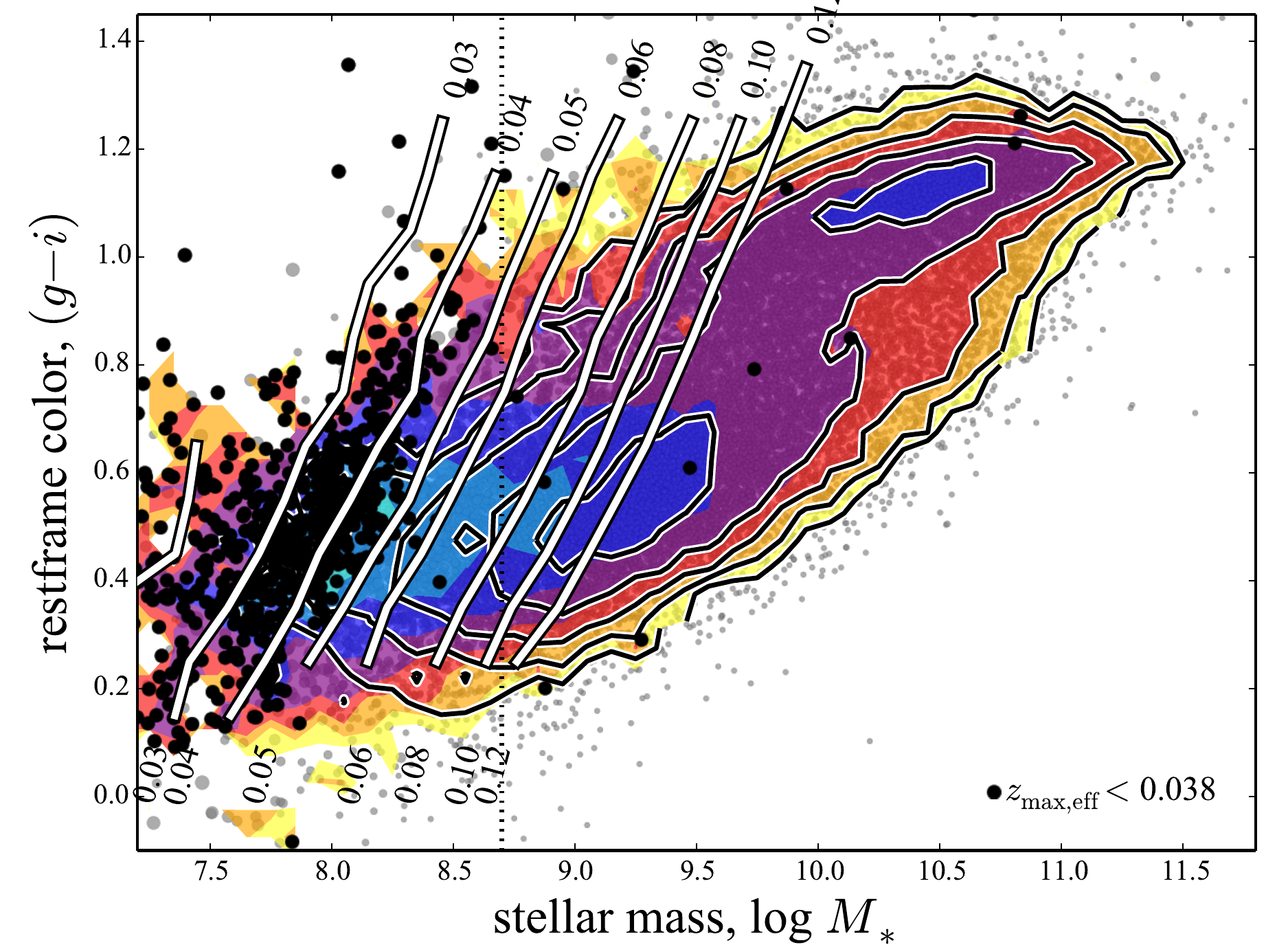}
\includegraphics[width=8.5cm]{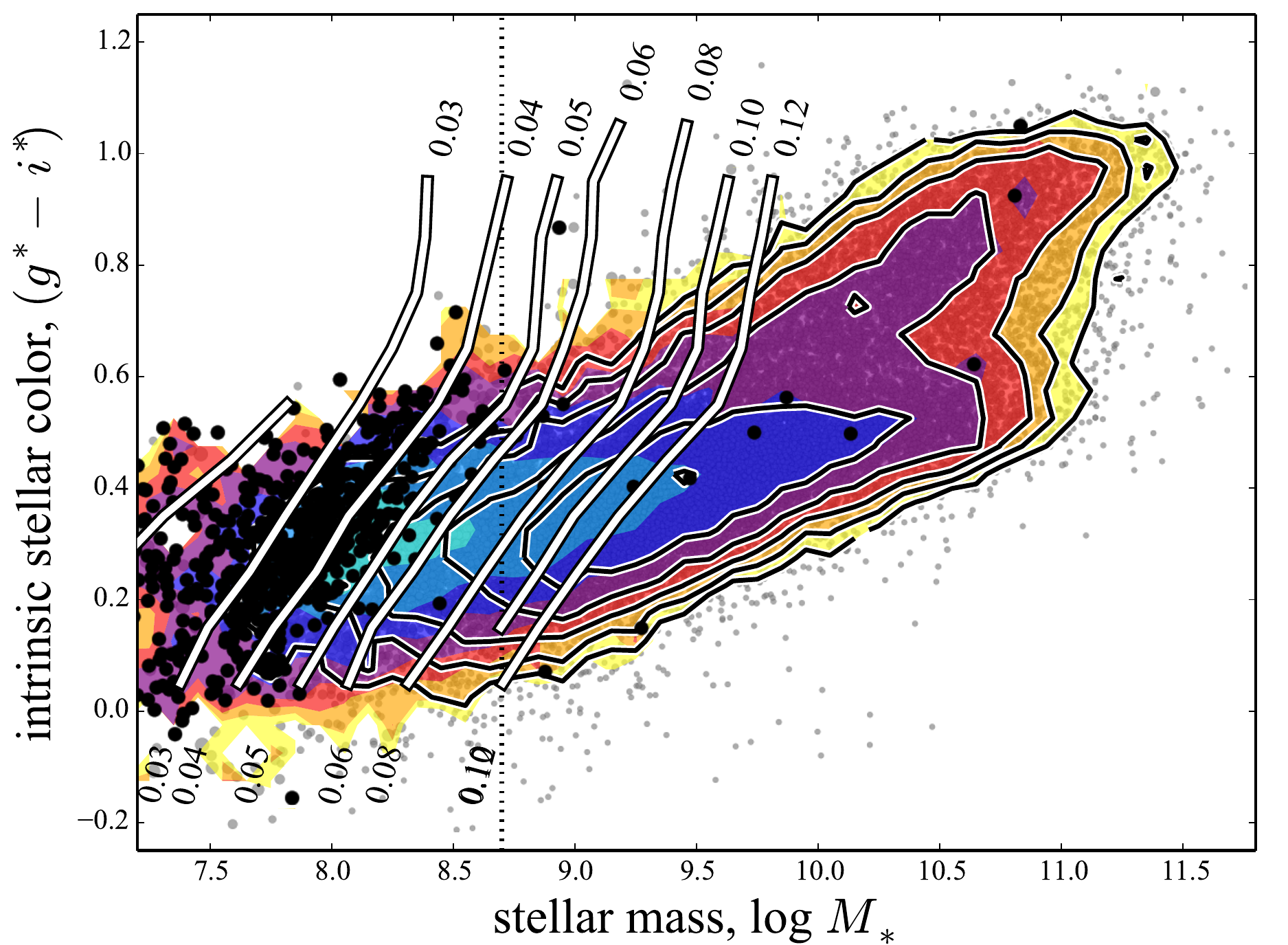} \caption{Completeness
limits as a function of mass and colour.--- This figure is discussed at length
in \secref{complete}. The upper and lower panels of this Figure show the
$(g-i)$ and $\starcol$ CMDs, respectively, for all $z < 0.12$ galaxies in the
GAMA catalogue. In both panels, the points show individual galaxies, with the
sizes of each point chosen to reflect the implied $1/V\max$ incompleteness
corrections (see \secref{vmax}). The line contours show the observed data
density without corrections for incompleteness; the filled coloured contours
show the $1/V\max$-weighted results. The white lines show empirical estimates
of how the $r\pet < 19.8$ selection limit maps onto the CMD at different
redshifts (see \secref{vmaxcheck}). The $z \approx 0.12$ curve shows that we
are complete (volume limited) for all galaxy colours for $\log M_* \gtrsim
10$. Our decision to limit our $V_\mathrm{survey}/V\max$ weightings to $\le
30$ effectively means that we will be under-correcting for incompleteness for
$z < 0.038$; those galaxies with relative weightings $> 30$, or $z\max <
0.038$, are plotted in black. The extent to which the empirical completeness
curve for $z \approx 0.04$ approximately bounds the black
$V_\mathrm{survey}/V\max > 30$ points thus shows the consistency between these
two complementary means of estimating our selection limits as a function of
colour, mass, and redshift. Our analysis is thus {\em conservatively} limited
to $\log M_* > 8.7$ (\ie, $M_* \gtrsim 5 \times 10^8$ M$\sun$); this limit is
shown as the vertical dotted line. \label{fig:cmds} } \end{figure}

Note that to protect against catastrophic errors in the $V\max$ estimates, we
limit the maximum relative weighting of any individual galaxy to be
$V_\mathrm{survey}/V\max < 30$. In effect this means that we will be
under-correcting for any galaxies that have $z\max < 0.038$. As can be seen in
\figref{cmds}, this decision affects only 16 galaxies in our $\log M_* > 8.7$
sample. Limiting our sample to being $z > 0.035$ excludes all of these
`problem' objects. With this $z > 0.035$ limit, the inferred mass functions
are depressed by $\lesssim 0.1$ dex for $\log M_* \lesssim 9$ (due to
incompleteness), but none of our qualitative results or conclusions change.

We have done the usual consistency tests \citep{Schmidt1968} to check the
reasonableness of these incompleteness corrections. We have verified that
where the values of $V\max$ imply that we are properly volume limited (\ie,
mass complete), the median $z$ is approximately equal to the volumetric centre
of the $z < 0.12$ survey window. For the bluest galaxies ($0.25 < (g-i) <
0.50$), this is true for $\log M_* \gtrsim 9.5$; for the reddest galaxies,
this is true for $\log M_* \gtrsim 10$. We have also verified that the median
value of $V(z)/V\max \approx 1/2$; even after binning by colour, this is true
for all masses $\log M_* \gtrsim 8.7$.

The problem is that there are too few low mass red galaxies in our sample for
us to look at our completeness for $(g-i) \gtrsim 1$ galaxies with $\log M_*
\lesssim 9$ in this way. We have only 12 galaxies with $(g-i) > 1.0$ and $\log
M_* < 9$ in our sample, all of which are at $z \lesssim 0.04$. Leaving
aside the question of field-to-field variance, the concern is whether the
apparent dearth of such red, low-mass galaxies in the GAMA catalogues is a
fair characterisation of the GAMA $z \lesssim 0.04$ survey volume, or if
instead we have over-estimated our sensitivity to these very faint galaxies.

\subsection{An empirical characterisation of our completeness limits as a
function of mass, colour, and redshift} \label{ch:vmaxcheck}

In \textbf{\figref{cmds}}, we again show the distribution of $z < 0.12$
galaxies in the $(g-i)$ and $\starcol$ colour--mass diagrams. In both panels,
the filled, coloured contours show the inferred bivariate colour--mass
distribution function after applying our $1/V\max$ weightings to account for
incompleteness. These contours should be compared to the black and white
line-contours, which show the raw, observed data-density in the colour--mass
diagrams; \ie, without incompleteness corrections.

We have plotted the individual galaxies in our sample as the gray points; the
size of each point directly reflects the magnitude of the $1/V\max$ factor
used to account for incompleteness. The black points in these panels highlight
those few galaxies with relative weightings $w = V_\mathrm{survey}/V\max >
30$. Since we have chosen to limit our weightings to be $\le 30$, these are
the galaxies for which (formally) we would be under-correcting for
incompleteness. It is entirely possible, however, that these points reflect
somehow catastrophic errors in our $V\max$ and/or $M_*$ estimations: there are
many more galaxies with similar masses and colours for which the implied
values of $w$ are considerably smaller.\footnote{In fact, eye-balling these
galaxies most are badly blended with a nearby galaxy or bright star, and the
redshift for the one clearly isolated galaxy is suspect.}

In order to investigate our sensitivity to low mass, red sequence galaxies
further, we have therefore sought to quantify our sample completeness limits
in a way that is independent of our $V\max$ calculations. We have done so by
taking all observed galaxies and simply scaling their total
luminosities/masses down to match the $r\pet = 19.8$ selection limit. Then, by
dividing our sample in narrow redshift intervals of width $\Delta z = 0.01$
and centred on $z = 0.01$, 0.02, ..., 0.12, we take the median value of this
limiting mass in narrow bins of restframe or intrinsic colour. 

This analysis thus provides an empirical description of our 50\,\% mass
completeness limits, as a function of redshift and colour, but in a way that
is independent of the SPS fits that have been used to derive $M_*$, $(g-i)$,
and $\starcol$; the results are shown as the heavy white-and-black lines in
\figref{cmds}. The $z = 0.03$, 0.04, 0.06, 0.08, and 0.10 curves can be taken
as corresponding to relative volume completenesses of $V\max/V_\mathrm{survey}
\approx$ 0.02, 0.04, 0.13, 0.30, and 0.59, respectively.

The first point to make is that this independent, empirical characterisation
of our mass-completeness limits agrees very well with the results of our
$1/V\max$ calculations. The fact that the distribution of the black points in
each panel of \figref{cmds} is approximately bounded by the empirical
completeness limit for $z = 0.04$ should thus give some confidence in our
$V\max$ estimates. In the same way, the $z = 0.12$ curve can be taken as
indicative of where we are truly volume limited. Taken together, these two
curves thus bound the region of the colour-mass diagrams in which our
incompleteness corrections are important and reasonable. Since the black and
colour-filled contours in these panels show the data density without and with
$1/V\max$ corrections, where these contours coincide shows where
incompleteness corrections are unnecessary. Again, the fact that the $z =
0.12$ curve very accurately bounds the regions in both the $(g-i)$--$M_*$ and
$\starcol$--$M_*$ diagrams over which this is true should give confidence in
our $V\max$ estimates.

\subsection{Are we seeing the low mass end of the red population?}
\label{ch:missing}

Given all of the above, are we (or are we even capable of) seeing the low-mass
end of the red sequence? We can be all but certain that there are very few red
galaxies with $8.7 < \log M_* < 9.0$ galaxies in the $z < 0.06$ GAMA survey
volume ($\approx 3 \times 10^{-3}$ Gpc$^3$). Addressing this question any
further is made problematic, however, by the effects of large scale structure,
and particularly by the degree to which low-mass and red galaxies are biased
towards rich-group and cluster environments.

\citet{Geha2012} have looked at the fraction of low mass galaxies in SDSS that
are (spectroscopically) identified as having {\em both} old stellar
populations {\em and} no ongoing star formation, and found that all such
galaxies with $\log M_* \sim 7$--9 are satellites within 1.5 Mpc (comoving,
projected) of a $\log M_* \gtrsim 10.4$ `host' galaxy. For our sample, the
number density of $\log M_* > 10.4$ galaxies at $z < 0.06$ is actually 10\,\%
higher than for $0.06 < z < 0.12$: that is, we may in fact be biased {\em
towards} old, low mass galaxies. Based on the group catalogue described by
\citet {Robotham2011}, our sample contains 580 $\log M_* > 10.4$ and $z <
0.06$ galaxies, in 232 separate groups, 22 of which have multiplicities of 10
or more. Included in these groups are 194/2895 (6.7\,\%) of the $\log M_* < 9$
and $z < 0.06$ galaxies in our sample, all of which are within 0.5 Mpc (\cf\
the \citet{Geha2012} limit of 1.5 Gpc) of their hosts. These numbers give some
sense of the environments we are probing---\ie, from isolation up to low-
and-moderate sized groups.

At the same time, we point out that based on the derived values of $V\max$,
the results shown in \figref{cmds} suggest that our completeness may still be
$\gtrsim 95$\,\% even for $\log M_* \sim 8.5$. Further, we stress that our
relative volume completeness is greater than 50\,\% ($z_\mathrm{max,eff}
\gtrsim 0.095$) for even the reddest galaxies with $\log M_* > 9.5$; we
consider it highly unlikely that our results above this mass scale are
strongly affected by incompleteness. We have also verified that none of our
results change significantly if we limit our analysis to $z < 0.06$, or to
$\log M_* > 9.5$. 

In light of all this, and with the above caveats, we continue our analysis
with a nominal mass limit of $\log M_* > 8.7$.


\begin{figure*} \centering
\includegraphics[width=17.2cm]{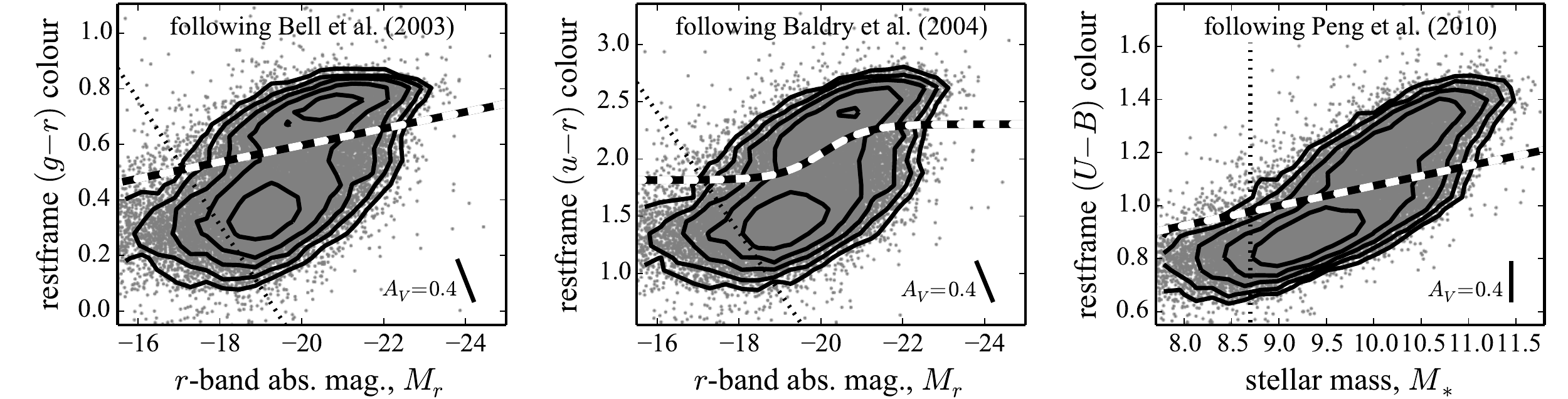}
\includegraphics[width=17.2cm]{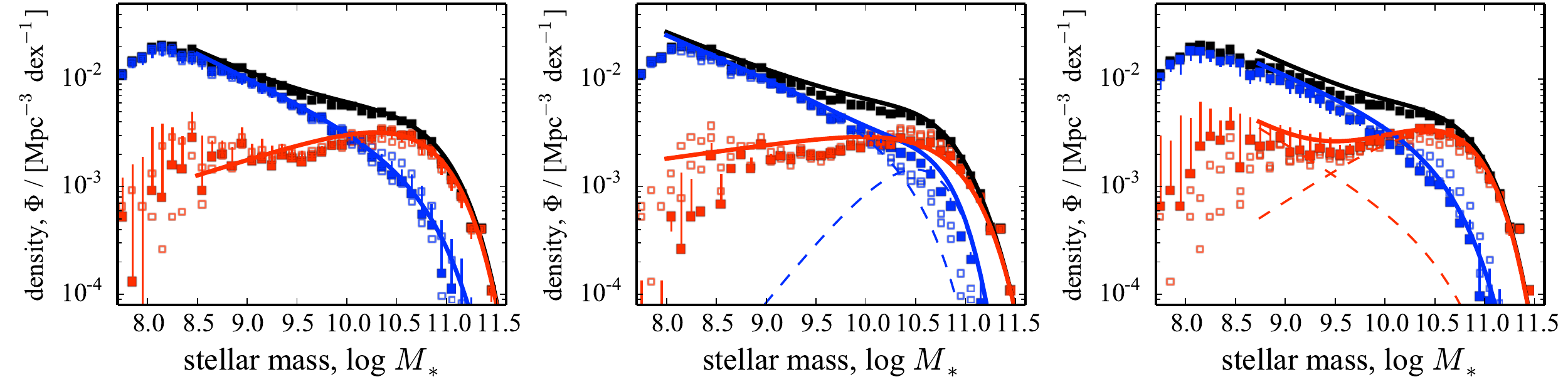} \caption{Reproducing
previous analyses using our GAMA sample, and illustrating the crucial
importance of how the terms `red' and `blue' are defined.--- The upper panels
show the selection used to separate red from blue galaxies by
\citet{Bell2003}, \citet{Baldry2004}, and \citet{PengLilly}, in either a
colour-magnitude or colour-mass diagram. In these panels, the points show all
$z < 0.12$ galaxies from GAMA, and the contours show the (logarithmic) data
density, without corrections for incompleteness. The lighter dotted line shows
our sample selection limit of $\log M_* = 8.7$, and the heavy black-and-white
lines show the binary red/blue cuts used or advocated by each set of authors.
In the lower panels, we show the inferred mass functions for red and blue
galaxies when applying these different authors' cuts. In these panels, the
smooth curves show these authors' fits, which have all been derived using SDSS
data. Where these fits are the sum of two Schechter functions, the separate
components are shown as the thin dashed lines. These should be compared to the
points, which show what we find when we apply each of these red/blue
selections to our GAMA sample. We are able to reproduce each set of results
with our data. However, the different authors' fits should also be compared to
one another. To help with this, in each of these panels, the results from the
other two analyses are reproduced as the smaller, open squares. There are
major qualitative and quantitative differences between the results of the
different analyses, which are entirely due to the different ways of defining
`red' and `blue'. \label{fig:others} } \end{figure*}

\section{What --- if anything --- do you mean by `red'?}
\label{ch:redness}

\subsection{The state of play} \label{ch:state}

There are a number of ways of discriminating between `developed' and
`developing' galaxies, based on, \eg, restframe colour, spectral
classification, Hubble type (\ie, morphology), or \Sersic\ index (\ie,
structure). There is considerable, but by no means total, overlap between
these different kinds of selections \citep[see, \eg,][]{Robotham2013}.
However, as we will show in future papers in this series, inappropriate
conflation of the terms red/blue, early-/late-type, and quiescent/active has
the potential to be dangerously misleading.

Our overarching goal in this work is to look at the bimodality as seen in the
optical CMD---in other words, we are specifically interested in the bimodality
that exists in galaxies' {\em stellar populations}. With this in mind, our
specific goal is to derive a quantitative description of phenomenology of the
joint colour--mass distribution of galaxies, in terms of both the CMRs and
the MFs for the apparently distinct `red' and `blue' populations.

As a motivating introduction to our method for attacking this problem,
consider \textbf{\figref{others}}. In this figure, we show our best attempts
at reproducing the SDSS-based analyses of \citet{Bell2003},
\citet{Baldry2004}, and \citet{PengLilly} using our $\log M_* > 8.7$ and $z <
0.12$ GAMA sample. In the upper panels of this Figure, we show the different
ways that each set of authors have separated the red and blue galaxy
populations, based on either a colour-magnitude, or a colour--mass diagram. In
rough terms, the \citet{Bell2003} cut can be seen as a relatively conservative
means of selecting `red' galaxies: the selection line appears to hug the lower
limits of the red sequence. By contrast, the \citet{PengLilly} cut is rather
aggressive: it falls closer to the upper edge of the blue cloud. The cut
advocated by \cite{Baldry2004} is in a sense intermediate: it can be seen to
be aggressive at lower luminosities, and conservative at higher luminosities.
In the lower panels, the filled squares show the inferred red/blue galaxy mass
functions, when applying each of the different selections to our GAMA dataset.
In general, the agreement between each set of SDSS- and GAMA-derived results
is very good.

Further to our discussion of incompleteness in the previous section, we also
highlight the fact that the GAMA MFs---including the red MFs---are continuous
for $\log M_* \gtrsim 8.5$. This is despite the distracting and unfortunate
downtick in the number of galaxies with $\log M_* \approx 8.7$ (our mass
selection limit). We are not obviously incomplete for $8.7 \lesssim \log M_*
\lesssim 9$.

There are some obvious systematic differences in the inferred number densities
for $\log M_* \lesssim 10$. As a result, the integrated number density of
galaxies with $\log M_* > 8.7$ from GAMA is 7\,\%, 12\,\%, and 13\,\% lower
than that from \citet{Bell2003}, \citet{Baldry2004}, and \citet{PengLilly},
respectively. \citep[Not surprisingly, however, we agree almost exactly
with][not shown]{Baldry2011}. These differences come down to the different
means of estimating stellar masses.

The role of various kinds of systematic errors/uncertainties in determining
the net MF (the black curves and points in \figref{others}) has been explored
by \citet{Baldry2011}; that is not our main purpose here. For our purposes, it
is sufficient to note that having controlled for everything we can (\eg,
taking SDSS \textsc{model} fluxes as total; matching IMFs and cosmologies) we
match the inferred integrated stellar mass density for $\log M_* > 9.5$ to
within 3--4\,\% in each case.

Instead, we are specifically concerned with sources of systematic error or
uncertainty on the MFs for the `red' or for the `blue' galaxy populations.
That is, we are particularly interested in the red and blue lines/points shown
in \figref{others}.

In this regard, the most noticeable discrepancy is our failure to reproduce
the apparent upturn in the red mass function seen by \citet{PengLilly} for
$\log M_* \lesssim 9.5$. We suggest that at least part of this discrepancy is
due to differences in how we have derived our $(U-B)$ colours. Comparing our
$(U-B)$ colours, derived in the course of the SPS fits, to those from
$\textsc{kcorrect}$ \citep{kcorrect}, we find that there is considerable
scatter (at the level of 0.15 mag), even when analysing the same SDSS
\textsc{petro} photometry. We have tried simply perturbing our $(U-B)$ colours
by 0.15 mag. This has the net effect of scattering a small fraction of `blue'
galaxies into the `red' sample, which leads to a significant increase in the
inferred numbers of red galaxies, as illustrated by the thin vertical lines in
the lower-right panel of \figref{others}. Note that the other MFs are more
robust to photometric scatter at this level, as is shown.

We also note that, in the middle panel, the agreement between our GAMA-derived
results and the \citet{Baldry2004} fits is imperfect, particularly for the red
MF. We will defer detailed discussion of this discrepancy to \secref{others}.
For now, we simply note that the \citet{Baldry2004} cut is based on their fits
to the MF, rather than the other way around; we therefore expect some small
quantitative differences between the \citet{Baldry2004} fits and the MFs
derived using the \citet{Baldry2004} cut. At this stage, the important point
is that we see the same qualitative results.

In order to facilitate easy comparison between the results of these different
analyses, in each of the lower panels of \figref{others}, there are two sets
of small open squares, which re-plot the results of the other two analyses.
The range spanned by these points thus reflects the systematic uncertainty on
the blue and red mass functions, arising from the different ways that the
`blue' and `red' galaxy samples have been selected/defined. Note in particular
the size of these uncertainties at and around the knee of the mass function,
as well as at low masses, for both the red and blue mass functions.

Based on \figref{others}, we make the following four observations:
\begin{itemize}

\item Comparing the different authors' fit MFs {\em to one another}, there are
important discrepancies in the shapes of both the red and the blue mass
functions for $10 \lesssim \log M_* \lesssim 11.3$. This is unfortunate,
because this leads to large uncertainties ($\gtrsim 0.3$ dex) in the
mass-scale at which galaxies transition from one population to the other.

\item There are also large discrepancies in the values of the low-mass slope
of the red mass function: it might be slowly declining ($\alpha \approx
-0.7$), or nearly constant ($\alpha \approx -0.9$), or has a sharp upturn
($\alpha \approx -1.5$). This is unfortunate, as it leaves the behaviour of
the low mass red population largely unconstrained.

\item Further, there is not even consensus as to how the two MFs ought to be
described and understood {\em qualitatively}. The \citet{Bell2003} MFs are
each well described by a single Schechter function; \cite{Baldry2004} finds a
need for a second Schechter component to describe the blue MF;
\citet{PengLilly} find instead that it is the red MF that needs a second
Schechter component.

\item That we can reproduce each set of results using our dataset shows that
these discrepancies come from differences in how the data are analysed, rather
than differences in the data themselves.

\end{itemize}

In other words, {\em current understanding of the MFs for the red and blue
galaxy populations is limited by systematic errors}. As mentioned in the
Introduction, measurements of the MFs for the red and blue galaxy populations
have played a pivotal role in informing our understanding of galaxy formation
and evolution. It is therefore critically important to understand how and why
there can be such large discrepancies between the results of these different
analyses. Only then will we be able to formulate an analytical approach that
will allow us to robustly measure these quantities.

\begin{figure*} \centering
\includegraphics[width=17.2cm]{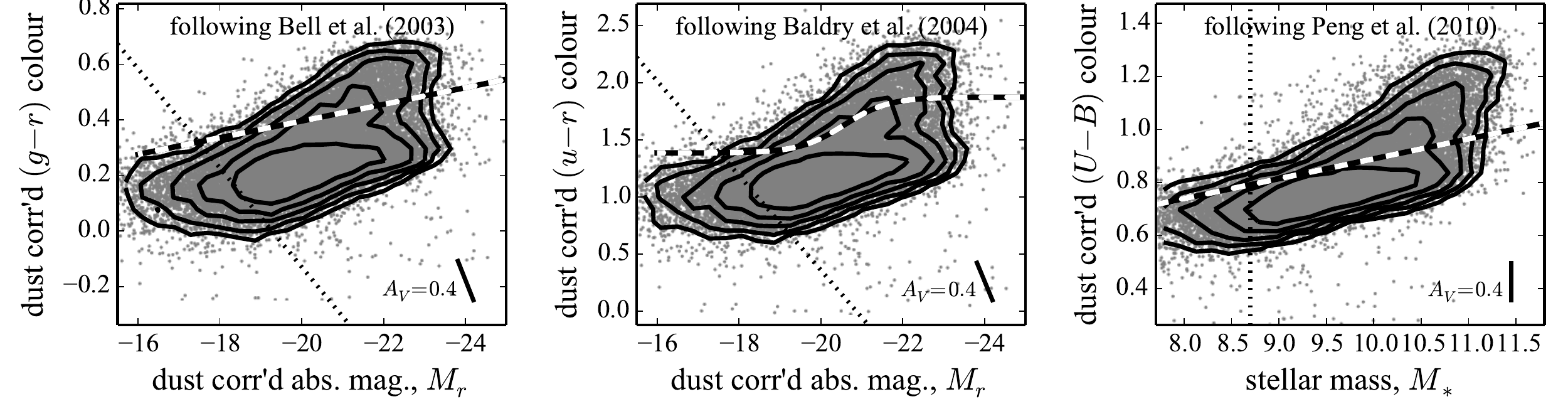}
\includegraphics[width=17.2cm]{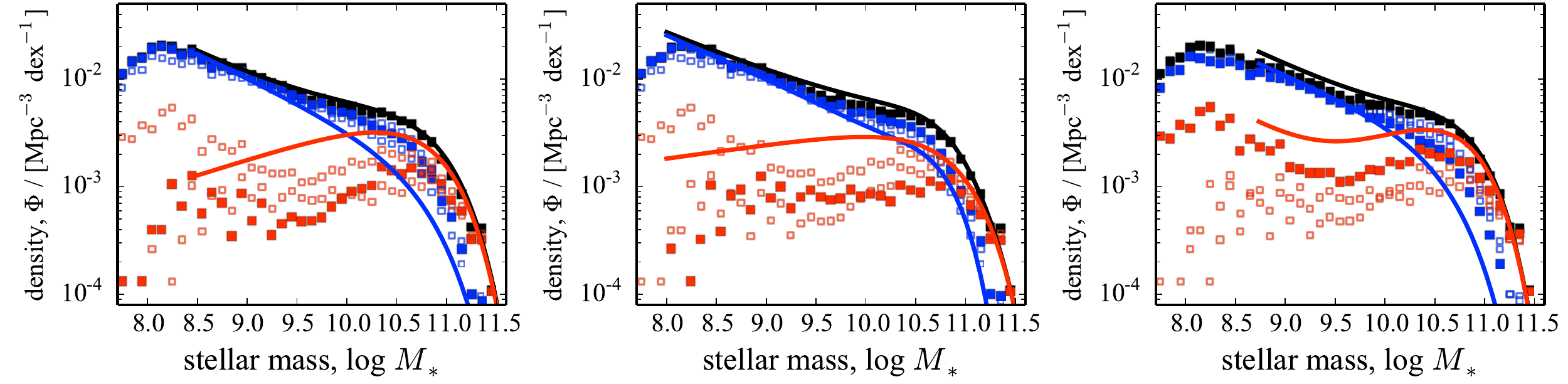} \caption{Adapting
previous analyses to account for dust, and illustrating the critical
importance of how the terms `red' and `blue' are defined.--- In the upper
panels, we have applied dust corrections to the data, to show intrinsic
stellar colours and magnitudes. We have also rescaled the \citet{Bell2003},
\citet{Baldry2004}, and \citet{PengLilly} selection lines, so as to retain
those `red and dead' galaxies with low (but non-zero) dust extinction. The
lower panels show the mass functions that follow from these cuts. Compared to
\figref{others}, these selections lead to a very different picture of the
field galaxy population. {\em Note that we are not suggesting that the results
shown in this Figure provide a fair or accurate representation of the
bimodality in galaxies' stellar populations.} Instead, the conclusion to be
drawn from this Figure is that the inferred MFs depend entirely on how the
`red' and `blue' samples are selected/defined. Compared to one another, the
red galaxy MFs in these figures differ by a factor of $\sim 2$; compared to
the MFs in \figref{others}, the difference is a factor of $\sim 10$. Lacking
any solid theoretical basis for preferring any one selection line over any
other, what is needed is an objective, data-driven means of identifying and
separating the `red' and `blue' populations. \label{fig:otherdust} }
\end{figure*}

\subsection{Dust is not the (only) issue.}

Given that there appears to be a better separation between the `red' and
`blue' populations in the $\starcol$ CMD shown in Fig.s\ \ref{fig:colours} and
\ref{fig:cmds}, the natural question is whether the discrepancies described in
the previous section can be alleviated or removed by focusing on intrinsic
stellar colours/magnitudes. What happens if we try modifying these analyses to
account for dust obscuration/extinction? 

We address this question in \textbf{\figref{otherdust}}. The main difference
between this Figure and \figref{others} is that we have now shifted to
intrinsic (\ie, dust corrected) stellar colours and luminosities, so as to
more directly probe galaxies' stellar populations.

As discussed in \secref{colours} and shown in \figref{colours}, the lowest
values for the SED-fit dust extinctions is $\dust \approx 0.2$, even for
galaxies with no H$\alpha$ emission. In light of this fact, we need to also
rescale each of the selection lines shown in the upper panels of
\figref{others}. What we have done is to shift each selection line by the
equivalent of $\dust = 0.4$ mag. This `correction' is much larger than the
expected dust obscuration for a canonically `red and dead' galaxy. It should
be thought of as a conservative way to exclude the dustiest galaxies, while
retaining those galaxies with genuinely `red' stellar populations.

The bottom panels of \figref{otherdust} shows the MFs for `red' and `blue'
galaxies, selected in this way; the selections themselves are shown in the
upper panels. It is clear that adopting these selections would lead to a
rather different picture of the makeup of the field galaxy population.

Comparing the range of values for the red MF that come from these different
selections, the differences are at the level of a factor of $\sim 2$. This is
considerably larger than the differences seen in \figref{others}. That is,
modifying these selections to account for dust {\em exacerbates} the tension
between these different authors' results, rather than alleviating it. (This
implies that, even when looking at dust-corrected, intrinsic stellar colours,
there is still substantial overlap between the `blue' and `red' populations.
We will show in \figref{stardist}, below, that this is indeed the case.) No
less worrying is the size of discrepancy between the $\log M_* \lesssim 10$
MFs for the $\starcol$-selected `red' galaxies (\figref{otherdust}) and the
$(g-i)$--selected ones (\figref{others}). Perhaps not surprisingly, the size
of this discrepancy is very sensitive to how much one chooses to shift each
selection line---that is, how much dust to allow for in otherwise `red
sequence' galaxies.

So which of the six analyses we have now trialled is right? Are the results
shown in \figref{otherdust} any more or less reliable or meaningful than those
shown in \figref{others}?

The crux of the problem is that there are no clear theoretical grounds for
preferring any one of these `red'/`blue' cuts over any other. In the absence
of a solid, astrophysically meaningful argument for such a cut, this is
necessarily true---without further information, we have no compelling way to
answer this question. While we might offer some empirical or phenomenological
argument in support of our specific cut, the point is that this decision will
always be arguable; that is, {\em arbitrary}.\footnote{Here, it should be
noted that the \citet{Baldry2004} cut is based on an analysis that is similar
in spirit to the one we will pursue below, based on modelling the observed
colour distributions in different magnitude bins. The \citet{Baldry2004} MFs
are thus devised in a qualitatively different way to \citet{Bell2003} and
\citet{PengLilly}, and part of the justification for their particular cut is
that it leads to similar results as are obtained from a more sophisticated
analysis. } This is a point that we will return to in \secref{classes} and
\secref{howelse}.

For now, we can say with some confidence that these different results can be
taken to approximately bracket the range of allowed values for the red/blue
MFs that come from reasonable choices for hard-cut red/blue selections. The
truth probably lies somewhere between the different results shown in
\figref{others} and \figref{otherdust}. This is not very satisfactory,
however, as it means that the shapes of the red and blue MFs are not even well
constrained \emph{qualitatively}, much less quantitatively.

\subsection{The nature of the distinction between `red' and `blue'}
\label{ch:cuts}

In short: the quantitative {\em and qualitative} discrepancies between the
results shown in the lower panels of \figref{others} and \figref{otherdust}
are entirely due to the different ways that each set of `red' and `blue'
samples has been selected---or, said another way, to the different operational
definitions of the terms `red' and `blue'. For example, the fact that
\citet{PengLilly} see an upturn to the red mass function at low masses---where
\citet{Bell2003} and \citet{Baldry2004} do not, despite their using
essentially similar datasets, and even pushing to lower masses---is a direct
consequence of the fact that the \citet{PengLilly} selection line is
relatively bluer than earlier authors.

It is therefore worth reflecting on the two implicit assumptions that underpin
the use of a hard cut to separate `red' from `blue' galaxies, and thus the
results shown in \figref{others} and in \figref{otherdust}. First, it is
presupposed that `blueness' and `redness' are physically meaningful
designations, inasmuch as they encapsulate some fundamental distinction
between the origins or natures of two distinct kinds of galaxies. The second,
and more problematic, assumption is that there is something
special---something astrophysically meaningful---about the particular boundary
used to separate the two galaxy classes.

The methodological appeal of such an approach is that it is well-defined,
inasmuch as the hard cut can be written {\em explicitly} and {\em exactly},
which makes such analyses easily reproducible. In the early days of the SDSS,
the astronomical motivation was also clear. \citet{Strateva2001},
\citet{Blanton2003a}, and many others had shown that there is substantial (but
not total) overlap between a `red sequence' sample and an `early type' sample
selected on the basis of \Sersic\ index. In this way, `redness' and `blueness'
were thought of as indirect proxies for structure, and thus for
morphology.\footnote{See, \eg, \citet{VanDerWel2008} for an excellent
demonstration of how morphology and structure are distinct astrophysical
properties.} As was common at the time, \citet{Bell2003}, \citet{Baldry2004}
and others presented their mass function determinations for `red' and `blue'
galaxies in terms of the Hubble early- and late-type classifications.
\citet{PengLilly}, on the other hand, have phrased their results in terms of
`star-forming' and `quiescent' galaxies. (This is also the explicit goal of,
for example, some colour--colour selections, which are discussed further in
\secref{howelse}.)

Here again, we caution against this conflation of terminology when
interpreting these results. While `early type' samples selected on the basis
of colour, spectral type, morphology, and structure are often treated as if
they are interchangeable, it is now becoming clear that {\em they are not}.
This point, and its importance, have most recently been forcefully made by
\citet{Schawinski2014}, who consider the CMDs for morphologically classified
`early-' and `late-types'.

Further, the use of a hard cut overlooks the empirical fact of scatter around
each of the distinct CMRs for the `red' and `blue' populations, however they
are defined. Any number of authors have shown that, at fixed magnitude or
mass, the distribution of galaxies' (optical) colours can be well described as
the sum of two Gaussians, and that the separation and widths of these two
Gaussian distributions are such that there is considerable overlap between the
two \citep[see, \eg,][]{Balogh2004, Baldry2004, Bell2004, Williams2009,
Wolf2009, Nicol2011, Coppa2011}. Considering these two distributions as
arising from two distinct populations, the implication is that the use of a
hard red/blue cut will yield samples that are both incomplete, and
contaminated \citep[see also, \eg,][]{Driver2006}. We will return to this
issue towards the end of this paper, in \secref{others}.

\subsection{All galaxies are red, but some are redder than others.} \label{ch:overlap}

In light of the above, we will not take quite so simple a view. We will assume
that there {\em is} some meaningful astrophysical distinction to be made
between the two populations: that there is some unknown astrophysical process
that acts to determine whether any given galaxy is a member of either the
`blue' or the `red' population. That is, {\em we will assume that there are
two distinct CMRs.} But we will also allow that some `hidden' parameter (or
parameters) mean that, at fixed mass, there are a range of colours among the
members of each of the two populations, to the extent that these two distinct
populations are observed to overlap in the CMD. That is, {\em we will assume
that there is some intrinsic scatter around each of the two CMRs.}

Adopting this (non-controversial) view of two overlapping populations, the
conceptual difficulty that arises is that some members of the `blue'
population will have quantitatively redder $(g-i)$ or $\starcol$ colours than
some members of the `red' population. Further, two galaxies might have
identical values of $M_*$ and $(g-i)$, but one might `really' belong to the
`blue' population, and the other to the `red' one. Without further
information, it would be impossible to unambiguously determine which is which.

This means that any `red'/`blue' classification of individual galaxies can
only be done probabilistically, in terms of the odds that that galaxy has been
drawn from either the `red' or the `blue' population. While our approach
brings these conceptual quandaries into sharp focus, we stress that similar
criticisms can be levelled at the simple, binary `blue'/`red' distinction used
above: the inferred scatter around the CMRs derived for the hard-cut `blue'
and `red' populations leads to precisely the same conundrum.

Without solid astrophysical justification, the terms `blue' and `red' must be
understood to be defined {\em operationally}, and as such are useful only as
{\em qualitative} descriptors. In acknowledgement of this point, we will from
now on abandon the terms `blue' and `red' as classifiers, and instead use the
more generic idea of a B- and an R-population. Note that these descriptors do
not properly apply to individual {\em galaxies}, but instead to distinct {\em
populations of galaxies}.

Obviously, the designations B and R have been chosen with a nod towards one
being for the bluer population, and the other for the redder one. But we want
to be absolutely clear that these designations are based on {\em
phenomenological} descriptions of the joint colour-magnitude distributions and
should not be taken to be rigorously grounded in astrophysical theory. Any
{\em astrophysical} interpretation of our descriptive B- and R- population
modelling, including those offered in \secref{observations}, must be done with
care. 

Let us stress in particular that we are {\em not explicitly} trying to select
galaxies that are quiescent, quenched, early-type, etc. Our explicit goal here
is {\em only} to distinguish between the generic `developed' and the
`developing' galaxy populations on the basis of their stellar populations. In
this sense, all that the B and R designations are intended to encapsulate are
the distributions of luminosity weighted mean stellar ages, as probed by
either $(g-i)$ or $\starcol$.

Of course, at least for moderate- to high-masses ($\log M_* \gtrsim 9.7$), it
turns out that the galaxies that comprise the R population largely conform to
the prevalent notion of `red and dead' or `quenched' (see \secref{howelse}, as
well as \figref{colours}). This being the case, our results {\em can} be used
to gain insight on the process of quenching, but {\em only insofar as our
operational definition of `red'- or `R'-ness can be taken to mean `quenched'}.

While our approach brings this issue into sharp focus, the same degree of
caution is merited when interpreting the results of past studies of `blue' and
`red' galaxies: bearing in mind the qualitative and quantitative discrepancies
between the results shown in \figref{others} or \figref{otherdust}, which of
the selections shown in these figures can be said to best represent the idea
of `quenched'?

Adopting the working hypothesis of two distinct but overlapping B- and
R-populations in the CMD, the question becomes technical: how best to
distinguish and characterise the two populations on the basis of the observed
CMD. While the designations `B' and `R' must be understood to be qualitative,
inasmuch as they are phenomenological, we want to be able to classify galaxies
{\em quantitatively}. This can be done probabilistically, according to the
chances that they are members of either the B or R population. Further, we
want these classifications to be {\em objective}. In order to achieve these
goals, it is necessary to describe or otherwise account for the actual
underlying colour distributions, including the degree of overlap, and as a
function of mass. This is therefore the task that we have undertaken.

\section{Method --- Objectively Classifying Galaxies in the Colour--Mass
Diagram} \label{ch:method}

This section is devoted to describing and validating our descriptive modelling
of the bivariate or joint $(g-i)$ and $\starcol$ colour--mass distributions
for field galaxies at $z < 0.12$. The most general form of our model is laid
out in \secref{description}, including definitions and descriptions of the 40
parameters that define the model in its most general form. In \secref{mcmc},
we describe the numerical methods that have been used to fit for the free
parameters. In \secref{modelsel}, we describe the process by which we have
selected the best and simplest description of the data from within the more
general family of models that we have considered. (For the more motivated
reader, we present a pedagogical development of the model in
\appref{themodel}, in which we build up our formalism as successive
generalisations of the conventional weighted-$\chi^2$ approach to fitting a
single line.)

The conceptual basis of our descriptive modelling is this: that the observed
data are a sampling of some `true', astrophysical, bivariate colour--mass
distribution. This being the case, our data can be seen to have been drawn
from---generated by---some 2D probability distribution function,
$p(\vec{x}')$, where $\vec{x}' = (x',\,y')$ denotes some generic location in
our 2D data plane. (In this section, we will thus use $x$ and $y$ to further
abbreviate the quantities $\log M_*$ and either $(g-i)$ or $\starcol$.) 

We cannot absolutely know the `true' form of the distribution function
$p(\vec{x}')$. So instead, we aim to construct a parametric description for
what $p(\vec{x}')$ might be, and use the data to constrain the possible and
even likely form of $p(\vec{x}')$. Let us denote the full set of parameters
used or required to describe $p(\vec{x}')$ as $\set{P}$. 

Assume for a moment that we know or can guess the correct form of
$p(\vec{x}'|\set{P})$. Naturally, one does not observe this distribution
directly. Instead, observational errors mean that the observed distribution in
the $(x',\,y')$ plane will be a smeared out version of the true distribution.
Let us also assume that that the observational errors/uncertainties for a
given data point, $\vec{x}\subi = (x\subi, \, y\subi)$, are Gaussian, and so
can be described by the covariance matrix $\mat{S}\subi$ (see Equations
\ref{eq:rho}---\ref{eq:covar}). Using $\gauss_2( \vec{x}\subi, \mat{S}\subi)$
as shorthand for a bivariate Gaussian, the likelihood of observing a
particular datum $i$ is then given by the convolution of the `true',
underlying distribution and the bivariate Gaussian that describes that
measurement, and its associated uncertainty; \ie,
\begin{equation} \begin{split} \label{eq:Elli}
	\Ell\subi(\vec{x}\subi, \mat{S}\subi|\set{P})
	& = \int \mathrm{d} \vec{x}' ~ p( \vec{x}' | \set{P}) 
	        ~ \gauss_2(\vec{x}\subi - \vec{x}', \mat{S}\subi) \\
			& = p( \vec{x}' | \set{P} ) \otimes 
					\gauss_2(\vec{x}\subi, \mat{S}\subi) ~ ;
\end{split} \end{equation}
Note that to satisfy the requirement that a point actually be observed, we
impose to the normalisation conditions that the integral over ($x, y$) space
for $p$ and $\gauss_2$, and hence $\Ell\subi$, be equal to one.

It is crucial to recognise that the value of $p$, and thus the value of
$\Ell\subi$, can only be computed---indeed, are only defined---given an
assumed or trial set of values for each and every of the parameters in
$\set{P}$. In recognition of this fact, these quantities have been written in
\eqref{Elli} and all that follows as $\Ell\subi(\vec{x}\subi, \mat{S}\subi
|\set{P})$ and $p(\vec{x}\subi | \set{P})$.

The crux of the problem is then to construct an appropriate parametric
description of $p(\vec{x}'|\set{P})$. That is our task in this section. At
this stage, the casual or credulous reader whose interest lies only in our
results may wish to skip these technical sections, and move directly to
\secref{dist}, in which we demonstrate the quality of our fits to the observed
bivariate $(g-i)$-- and $\starcol$--$M_*$ distributions.

\subsection{A descriptive model for the distribution of observed data points
in the CMD} \label{ch:description}

In order to accommodate the apparent bimodality in the $(g-i)$ and $\starcol$
CMDs, we split the model for the `true', astrophysical bivariate colour--mass
distribution---\ie, the scalar function $p( \vec{x}' )$---into two distinct B
and R components, which are denoted as $p\blue$ and $p\red$. Each component
has its unique parameter set, denoted as $\set{P}\blue$ and $\set{P}\red$.
Because, in general, these two populations will be observed to overlap, the
probability density at any point $\vec{x}'$ is given by the sum of these two
distributions; \ie,
\begin{equation} \label{eq:good} \begin{split}
	p\good( \vec{x}' | \set{P\good}) =
		(1 -& f\red) \times p\blue( \vec{x}' | \set{P}\blue) \\
			&+ f\red \times  p\red( \vec{x}' | \set{P}\red) ~ .
\end{split}\end{equation}
For now, the `good' subscript can be ignored; its significance will become
clear in a moment. Note that, in line with the probabilistic nature of this
generative model, all of $p\blue$, $p\red$, and $p\good$ should be understood
to be integral normalised to one. The parameter $f\red$ thus sets the relative
normalisation of the B and R components, in terms of the relative number of
R-population galaxies among the global population (given our sample selection
limits).

At fixed mass, we treat the colour distributions of each of the R and B
populations as being Gaussian, and so characterised by three numbers: 1.)\ a
centre, 2.)\ a width, and 3.)\ a normalisation. Each of these three quantities
is allowed to vary parametrically, and independently, as a function of mass,
so that we can constrain: 1.)\ the locii of the B and R CMRs, $\ell\blue$ and
$\ell\red$, 2.)\ the intrinsic scatters around these CMRs, $\zeta\blue$ and
$\zeta\red$, and 3.)\ the mass functions for each population, $\Phi\blue$ and
$\Phi\red$. Using $\gauss_1(y - y_0,\, \sigma_y)$ as short hand for a
(properly normalised) 1D Gaussian with center $y_0$ and width $\sigma_y$, our
model for the bivariate colour--mass distribution for the R-population can be
written as:
\begin{align}\label{eq:model}
    p\red \big(\vec{x}' | \set{P}\red \big) 
    & = \sum_k \bigg( \delta \big(x_k - x'\big) 
                \times \Phi\red \big( x' | \set{P}_{\Phi,}\red \big) \\
    & \times \gauss_1 \big[ y' - \ell\red ( x'| \set{P}_{\ell,}\red ); \,
    \zeta\red (x'| \set{P}_{\zeta,}\red ) \big] \bigg) ~ , \nonumber
\end{align}
with an analogous expression for $p\blue(\vec{x}'|\set{P}\blue)$. Each of
these aspects of the model are described in turn below.

As discussed in detail in \secref{masslike}, we do not actually model the mass
distributions of red and blue galaxies as being continuous. Instead, we model
the mass function using the sum of many Kronecker delta functions whose
amplitudes are modulated by the continuous dual-Schechter mass function,
$\Phi$, defined in \eqref{Phi}. In \eqref{model}, $\Phi$ is thus accompanied
by the Kronecker delta function, $\delta(x' - x_k)$, and the underlying models
$p(\vec{x}')$ can be seen to be constructed as the sum of many discrete
components evaluated at $x' = x_k$. The reason for this decision is to allow
the convolutions in Eq.s (\ref{eq:Elli}), (\ref{eq:model}), and (\ref{eq:bad})
to be done analytically. Our method can be seen as evaluating an approximate
model, which has a discretised mass function, in an exact, analytical way. We
define the $x_k$s as $x_k \cong 8.7 + 0.05 ~ (k + 1/2) : k = 0$, 1, ..., 65;
that is, as a uniformly\footnote{See \secref{masslike} for an explanation for
why this definition of $x_k$ is (very slightly) approximate.} spaced grid in
$x$ with a spacing of 0.05 dex, with grid edges running from our nominal mass
limit of 8.7 up to 12. With this grid spacing, the typical galaxy with a mass
uncertainty $\sigma_x \approx 0.12$ dex has 5 or 6 $x_k$s within its FWHM.

\subsubsection{The mass functions}

The normalised mass functions for the B and R components, $\Phi\blue(x')$ and
$\Phi\red(x')$, respectively, are described using the sum of two \citet{Schechter1976} functions:
\begin{equation} \label{eq:Phi} \begin{split}
	\Phi(x' | \set{P}_\Phi) 
		= (1 - & f_2) \times
		\phi_1(x'|\alpha_1, \log M^\dagger_1) \\
		& + f_2 \times \phi_2(x'|\alpha_2, \log M^\dagger_2) ~ .
\end{split}\end{equation}
Here, the parameter $f_2$ can be understood to govern the relative
normalisations of the $\phi$s by describing the relative number of galaxies
the make up the second of the two Schechter functions, and the shapes of the
two Schechter functions, $\phi(x') \sim (x'/x^\dagger)^{-\alpha} \,
e^{(-x'/x^\dagger)}$, are described by a low-mass power law with logarithmic
slope, $\alpha$, and a characteristic mass, $M^\dagger$, which describes the
`knee' of the mass function. Thus we have up to 5 parameters for each of the B
and R populations, plus the dimensionless parameter $f\red$ defined above, for
a total of 11 parameters to describe the full mass distribution of galaxies,
down to our selection limit.

Again, each of $\Phi$, $\phi_1$, and $\phi_2$ should be understood to be
integral normalised to unity (given our $\log M_* \ge 8.7$ and $z < 0.12$
selections). We must therefore estimate the global normalisation of the mass
function independently of the modelling described in this section. This has
been done after the modelling on the basis of the integrated mass density
among galaxies in our sample; \ie, $\sum \log M_* / V_{{\mathrm{max},i}} =
(1.5944 \pm 0.0010) \times 10^{-2}$ Mpc$^{-3}$. This step introduces a $\sim
0.6$\,\% systematic uncertainty into all of our fit mass functions (but not
the observed ones). With this value fixed, we can compute the values of the
usual characteristic densities, $\phi^\dagger$, in units of Mpc$^{-3}$
dex$^{-1}$, based on the values of all 11 of the MF-defining parameters. When
we give the values of the fit parameters in \figref{mcmc}, we quote the
$\phi^\dagger$ values in place of the $f$s.

\subsubsection{The locii of the colour--mass relations}

Next, the CMRs, $\ell\red(x')$ and $\ell\blue(x')$. We allow the slope of the
B and R CMRs to vary as a function of mass by describing them in the following
way:
\begin{equation}\label{eq:cmrs} \begin{split}
	\ell( x' | \set{P}_\ell) 
        = \big( a ~ x' + c \big)
            + \mathrm{tanh} \left(\frac{x' - x_{\ell,0}}{x_{\ell,s}} \right) 
                \times \big( b ~ x' + d \big) ~ .
    %
\end{split}\end{equation}
Recalling that $\mathrm{tanh}( \ll 0) = -1$, $\mathrm{tanh}(0) = 0$, and
$\mathrm{tanh}( \gg 0)=+1$, this definition can be transparently viewed as the
combination of two linear relations. There is a smooth transition from a
low-mass regime, in which the CMR goes like $( a-b )x' + (c-d)$, to a
high-mass regime where the CMR goes like $(a+b) x' + (c+d)$. We highlight two
special cases: first, if $b=0$, then this parameterisation is equivalent to
the line-plus-tanh parameterisation used by \citet{Baldry2004}; second, if
$d=0$, then we have a smooth transition around the point of intersection
between two lines. The parameter $x_{\ell, 0}$ defines precisely where the
transition takes place, and the parameter $x_{\ell,s} > 0$ governs how
sharp/smooth this transition is. Thus we have six parameters to describe each
of the red and blue CMRs, bringing our running total of fit parameters to 23.

\subsubsection{The scatter around the colour--mass relations}

Finally, there is the scatter around the CMRs, $\zeta\red$ and $\zeta\blue$.
In the most general form of the model, we adopt the same parametric form for
the $\zeta$s as for the $\ell$s; \viz:
\begin{equation}\label{eq:zeta} \begin{split}
	\zeta( x' | \set{P}_\zeta) 
        = \big( p ~ x' + r \big)
            + \mathrm{tanh} \left(\frac{x' - x_{\zeta,0}}{x_{\zeta,s}} \right) 
                \times \big( q ~ x' + s \big) ~ .
\end{split}\end{equation}
This adds another six parameters to describe the scatters around each of the R
and B CMRs, which brings the running total number of parameters in $\set{P}$
to 35.

\subsubsection{Outliers or otherwise `bad' data} \label{ch:outliers}

In order to protect against biasing of our results from outliers, catastrophic
errors or otherwise un- or under-modelled aspects of the observed distribution
in the $(x,y)$ plane, our generative model includes a parametric description
for `bad' data. To this end, we split the model into two components; one for
each of the `good' and `bad' data distributions:
\begin{equation} \begin{split}
	p( \vec{x}' | \set{P} ) =
		(1 -& f\bad) \times p\good( \vec{x}' | \set{P}\good) \\
			&+ f\bad \times  p\bad( \vec{x}' | \set{P}\good, \set{P}\bad) ~ .
\end{split} \end{equation} 
Here, the parameter $f\bad$ describes the fraction of datapoints encompassed
within the `bad' distribution. This is wholly analogous to the use of $f\red$
to parameterise the relative normalisations of the B and R components of the
model. 

In the model, these `bad' data are described by an additional (large) error in
the measured values of $x$ and $y$. In other words, the `bad' component of the
model is simply generated by convolving the `good' component with an
additional 2D Gaussian:
\begin{equation} \label{eq:bad}
	p\bad( \vec{x}' | \set{P}) =
		p\good( \vec{x}' | \set{P\good}) 
			\otimes \gauss_2( \vec{x}', \mat{S}\bad ) ~ .
\end{equation}
The significance of the `good' subscript in \eqref{good} should thus now be
clear.

The defining covariance matrix for this Gaussian, $\mat{S}\bad$ has diagonal
entries $\zeta_{x,}\bad$ and $\zeta_{y,}\bad$; the off-diagonal entries are
zero. (In fact, as we describe below, the fit values of $\zeta_{x,}\bad$ are
$\approx 0$, and we are able to exclude this parameter without compromising
the quality of the fits.) 

The `bad' parameters $f\bad$ and $\zeta\bad$ deserve some further comment.
First, what exactly is meant by `bad'? Before, we have distinguished the B and
R components as having different CMRs as well as different mass function. By
contrast, the `bad' distribution can thus be seen to be just a `poor copy' of
the `good', R--plus--B distribution, having been `smeared' with a large
Gaussian, and with a much lower relative normalisation. Essentially, we are
using these `bad' quantities to parameterise our ignorance of any and all
features in the observed CMDs that are not easily explained by the `good'
model.\footnote{One might ask: can the same be said for any of the other
defining parameters for the model. And the only honest answer would be: yes,
all of them. Our modelling is wholly {\em descriptive}, and in no way {\em
explanatory}: none of the parameters can truly be said to have any real, solid
astrophysical foundation or meaning. That said, the empirical, quantitative
description of the CMRs and mass functions for the (apparently) distinct B/R
or `blue'/`red' populations---their uncertain astrophysical natures and
origins notwithstanding---do provide important empirical constraints for
cosmologically-minded models of galaxy formation and evolution.} This includes
catastrophic errors in the measurements of either $x$ or $y$, but also
includes---at least in principle---any additional components in the true,
astrophysical, joint colour--mass distribution.

Given this, what justification is there for treating the distribution of `bad'
data as Gaussian? In short, there is none. That said, we stress that our
characterisation of the `bad' data is simply in terms of the RMS of `bad' data
points around the `true' CMRs. It is true that the link between the value
of $\zeta\bad$ and the true shape of the distribution of `bad' data in $(x,y)$
space does implicitly assume Gaussianity. But we have no interest in
accurately modelling the shape of this distribution; for the purposes of
objectively identifying and censoring such `bad' data, simply knowing (or,
better, modelling) the RMS scatter is sufficient. 

Further, we are not even really that interested in the precise values of the
parameters $f\bad$ and $\zeta\bad$: what we are interested in using these
parameters to limit the influence of outliers on the fit values of all the
other, more astrophysically meaningful parameters. That being the case, when
we come to reporting our results we will marginalise over the values of both
of these nuisance parameters (see \eqref{margin} below), leaving us only with
the parameters of genuine interest and importance. Readers that remain
concerned about the role of these parameters in our calculation are referred
to our more detailed discussion in \secref{bad}, and also to the excellent
primer on data fitting by \citet{HoggBovyLang}.

\subsubsection{Summary}

In summary, Eq.s (\ref{eq:Elli}--\ref{eq:zeta}) define a model for the
distribution of galaxies in the CMD, which is fully described by up to 40
parameters. (Again, in \secref{modelsel} we will describe the model selection
process by which we have ensured that we reduced this parameter set to ensure
that we have the best and simplest description of the data possible.) There
are five parameters describing each of the red and blue mass functions, so
that $\set{P}_{\Phi,}\red = \{ \alpha_{1,}\red,~ \log M^\dagger_{1,}\red,~
\alpha_{2,}\red,~ \log M^\dagger_{2,}\red,~ f_{2},\red \} \subset
\set{P}\red$, with an analogous five parameters for $\set{P}_{\Phi,
\mathrm{B}}$. Added to these, there is the parameter $f\red \in \set{P}$,
which describes the relative number of red galaxies in our $\log M_* > 8.7$
and $z < 0.12$ sample. There are also six parameters to describe each of the
red and blue CMRs, so that $\set{P}_{\ell,}\red = \{ a\red,~ b\red,~ c\red,~
d\red,~ x_{\ell,0,}\red,~ x_{\ell,s,}\red \} \subset \set{P}\red$, and
similarly for $\set{P}_{\ell,\mathrm{B}}$. And finally there are six
parameters to describe the scatter around each of the red and blue CMRs, so
that $\set{P}_{\zeta,\mathrm{R}} = \{ p\red,~ q\red,~ r\red,~
s\red,~x_{\zeta,0,}\red,~ x_{\zeta,s,}\red \} \subset \set{P}\red$, and
similarly for $\set{P}_{\zeta,\mathrm{B}}$. Then, we have three parameters to
describe outliers or otherwise `bad' data, $\set{P}\bad = \{f\bad,
\zeta_{x,}\bad, \zeta_{y,}\bad \}$. To these should be added the two
parameters, $A_y$ and $b_y$, which are used to rescale the formal
uncertainties in $\starcol$, as discussed in \secref{errors}.

Note that each of these different subsets should be understood to be
formally independent. The mass scale and softening describing the transition
between the high- and low-mass regimes for $\ell$ and $\zeta$ are not assumed
to be related, nor are these transitions in any way formally connected to the
shapes of the mass functions, nor do we place any restrictions on the
relations between parameters for the B and R populations. 

With all of the above definitions, and given a set of trial values for the
parameters in $\set{P}$, we now have the means to compute the value of
$\Ell\subi(\vec{x}\subi|\set{P})$, as defined in \eqref{Elli}. Armed with this
information, it is then straightforward to compute the global likelihood,
$\Ell$ of observing the full dataset $\set{X} = \{\vec{x}\subi\}$, given the
associated uncertainties $\set{S} = \{\mat{S}\subi\}$, as the product of all
the individual $\Ell\subi$s. In practice, it is more convenient to work in
terms of $\ln \Ell\subi$, so that:
\begin{equation} \label{eq:Ell}
	\ln \Ell ( \set{X}, \set{S}, \set{W}|\set{P}) 
		= \sum_i w_i 
			~ \ln \Ell\subi ( \vec{x}\subi, \mat{S}\subi|\set{P}) ~ .
\end{equation}
Here, $\set{W} = \{ w\subi = 1/V\maxi \}$ is the set of 1/$V\max$ weighting
factors that we use to account for incompleteness due to the GAMA apparent
magnitude selections, as defined and discussed in \secref{vmax}.

\subsection{Constraining the values of the model parameters 
	\\ --- \ie, Using the model to fit the data} \label{ch:mcmc}

\begin{figure*} \includegraphics[width=8.4cm]{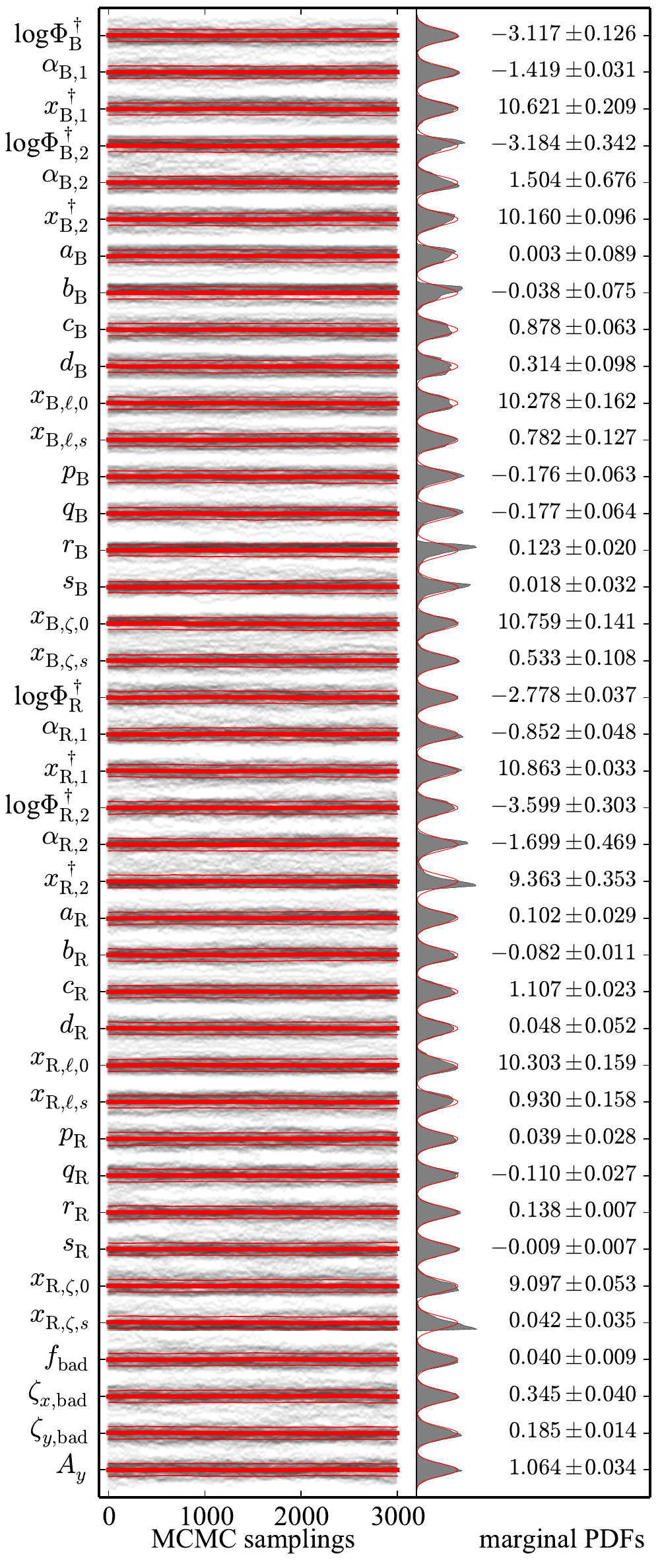}
\includegraphics[width=8.4cm]{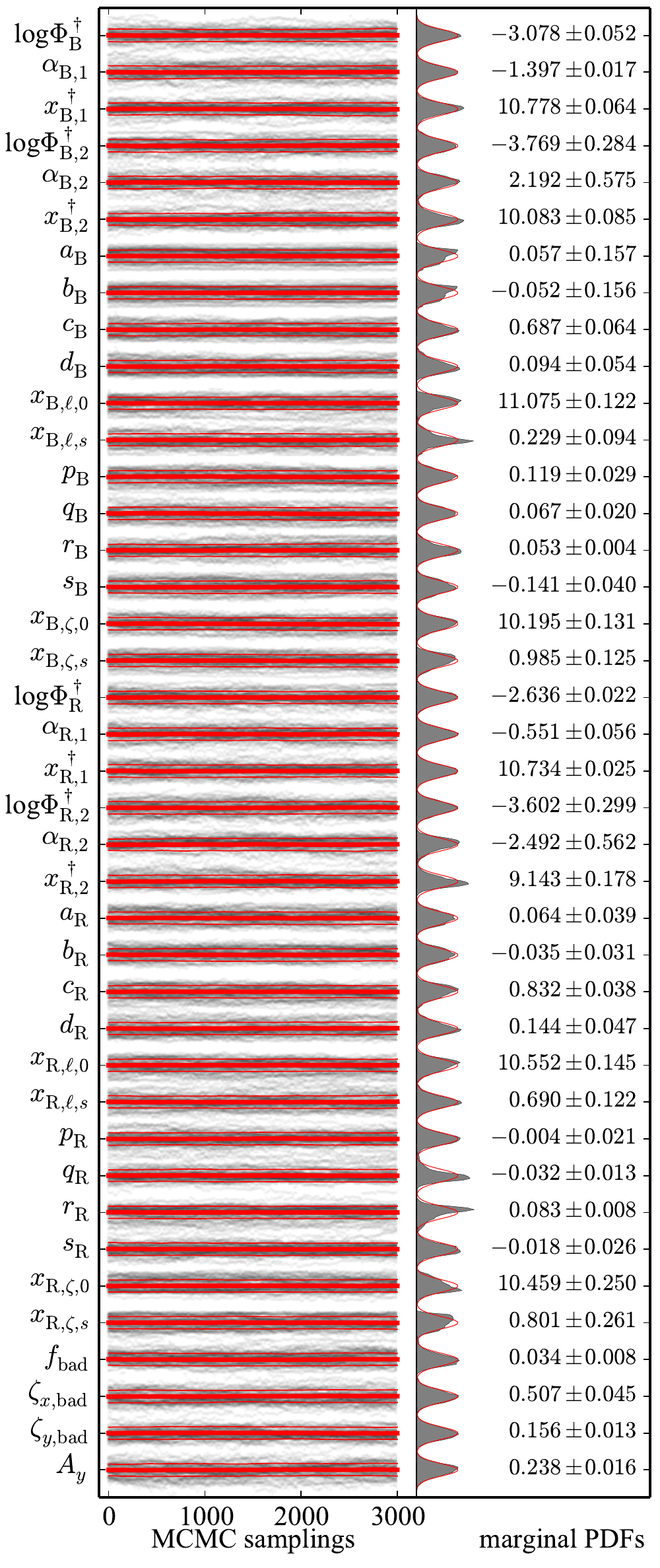} \caption{Diagnostic plots
showing the convergence of our MCMC fits to the $(g-i)$-- and
$\starcol$--$M_*$ CMDs (left and right panel, respectively), and a visual
table of the fit results--- In each panel, the distinct tracks show the values
of each of the parameters in the model at individual MCMC steps, for 200
individual walkers. For each parameter, the tracks have been scaled by the
mean and RMS over all samplings, so that the each track is centred on `the'
fit value, and the width of each track is scaled to match the formal
uncertainty in the value of that parameter. The actual fit values for each
parameter are given with uncertainties at right; machine readable tables of
these results will be published online. The mean and variance in each
parameter is shown as a function of MCMC step number as the heavy and thinner
lines, respectively. That these lines are flat show that the fits have fully
converged. The autocorrelation time for each parameter is in the range 50--150
MCMC steps. We thus have $\gtrsim 4000$ independent samplings of the values of
each parameter. Finally, the grey histograms at right show the distribution of
parameter values over all MCMC samplings; that is, the shape of the
marginalised PDF for each parameter. In all cases, these PDFs can be seen to
be well sampled. Further, in most, but not all cases, the PDFs can be seen to
be well described by a simple Gaussian (the red curves). \label{fig:mcmc} }
\end{figure*}

Given the particular parametric form of our model, and given that we have
observed our specific dataset, what we want to do is to use the data to
constrain the possible values of the parameters in $\set{P}$. In other words,
we want to construct the posterior probability density function (PDF) for each
of the individual parameters, $P_n \in \set{P}$, so that we can evaluate the
probability that the parameter $P_n$ has the value $P_n'$; that is,
$\mathrm{Pr}(P_n = P_n'|\set{X}, \set{S}, \set{W})$

This is done using Bayes' theorem, which can be written as:
\begin{equation} \label{eq:bayes}
	\mathrm{Pr(model|data)} = 
	 \mathrm{ Pr(data|model) } \times \mathrm{ Pr(model) }  ~ .
\end{equation}
Here, $\mathrm{Pr(data|model)} = \Ell(\set{X}, \set{S}, \set{W}|\set{P})$ is
just the scalar likelihood function as defined in \eqref{Ell}, which can only
be computed given a full set of values for the model parameters. In contrast,
$\mathrm{Pr(model|data) = Pr}(\set{P}|\set{X},\set{S},\set{W})$ is the full,
high-dimensional PDF for the values of the parameters $\set{P}$, which is what
we are interested in deriving. 

Bayes' theorem links these two quantities via the {\em prior} distribution
function, $\mathrm{Pr(model)}$, which is an assumed, {\em a priori} statement
of our expectations for the probability of different parameter combinations
with respect to one another. By invoking Bayes' theorem, we are therefore
required to explicitly state our priors on the relative probabilities,
$\mathrm{Pr}(\set{P})$, of different values for each of the individual
parameters $P_n \in \set{P}$. 

The decision of what priors to adopt is by no means trivial, but it is also
inescapable. {\em All fitting algorithms include priors}. One cannot compare
the relative likelihoods of two different trial parameter values without an
implicit or explicit prior, even if that prior is that the two values are,
{\em a priori}, equally likely.

In the absence of any clearly better alternatives, we adopt uniform (or
uninformative) priors on each of the parameters in $\set{P}$. This includes
uniform priors for, for example, the $x^\dagger$s, which is equivalent to
logarithmic priors for the $M^\dagger$s. The exception to this rule is for the
slope parameters for the linear relations that go into both $\ell(x')$ and
$\zeta(x')$. Here we take uniform on the {\em angle} of the relation; that is,
our priors are uniform in, for example, $\mathrm{arctan} ~ a$ and
$\mathrm{arctan} ~ p$.

It is worth noting that with this choice of uniform priors, the prior function
$\mathrm{Pr(\set{P})}$ is constant, and so
$\mathrm{Pr}(\mathrm{model}|\mathrm{data})$ is directly proportional to
$\mathrm{Pr}(\mathrm{data}|\mathrm{model})$. With this decision, the Bayesian
formalism thus all but reverts to that of traditional, frequentist statistics.
In other words, at least in our case, the only difference between the Bayesian
and the more familiar frequentist approach is that, as Bayesians, our priors
are made {\em explicit}.

Formally, the PDF for the single parameter $P_n$ is derived by marginalising
over all of the other parameters in $\set{P}$; ie:
\begin{equation} \label{eq:margin}
	\mathrm{Pr}(P_n = P_n' | \set{X},\set{S}, \set{W})
		\propto \int \mathrm{d} P_m ~ 
				\Ell(\set{X}, \set{S}, \set{W}|\set{P}) ~ \mathrm{Pr}(P_m) , ~
\end{equation}
where $\mathrm{Pr}(P_m)$ is constant for uniform priors, and the integral
should be understood to be evaluated for all $P_m \in \set{P} / \{P_n\}$.
(Here, the symbol `$/$' means the set complement.) In words, this expression
is best understood as a probability weighted integral over all possible
combinations of parameter values, with the condition that the specific
parameter of immediate interest, $P_n$, takes the particular value $P_n'$.

Note that this formalism also works for any quantity $Q(\set{P})$ that can be
deterministically computed from the defining parameters of the model. This
includes, for example, the values of the characteristic densities,
$\Phi^\dagger$, for each of the Schechter functions. This also includes the
values of the individual $\Phi(x_k)$s; \ie, the values of the MFs at any of
the discrete $x_k$s used to define the model. Here, since $\Phi(x_k)$ is not a
member of the defining parameter set $\set{P}$, the `complementary' parameter
set of $P_m$s in \eqref{margin} is the full set $\set{P}$. In this way, we can
derive formal statistical uncertainties on $\Phi(x_k)$, $\ell(x_k)$, or
$\zeta(x_k)$ that fully account for any and all covariances between the 40
parameters in $\set{P}$. We will discuss this point in more detail in
\secref{mfs}.

The fitting of the model thus entails mapping out the scalar likelihood
function $\Ell(\set{X}, \set{S}, \set{W} | \set{P})$ over the 40-dimensional
parameter space defined by $\set{P}$. This is done using the technique of
Monte-Carlo Markov Chain (MCMC) sampling. In essence, MCMC is just a random
walk through the high-dimensional parameter space. The key to MCMC techniques
is that possible steps are considered randomly, but are accepted or rejected
probabilistically. More specifically, the chances of a step being accepted are
defined by the ratios of the PDF --- that is, by the prior-weighted likelihood
function --- at the present and potential future locations in $\set{P}$-space.

In the first instance, this makes MCMC a very robust means of exploring the
parameter space with a view to finding the global maximum of the PDF. In the
second instance---once the algorithm has found itself near to the maximally
likely solution---MCMC sampling represents an extremely convenient means of
sampling the high-dimensional PDF. In this phase of the fitting process, the
key to the utility of MCMC sampling is that it is {\em ergodic}; that is, the
chances of a point in $\set{P}$-space being sampled is directly proportional
to the value of the PDF at that point. As a consequence, the {\em
distribution} of MCMC-sampled points converges to a faithful mapping of the
{\em value} of the PDF in $\set{P}$ space.

This means that the marginalisation integral in \eqref{margin} can be very
easily computed to a high level of accuracy by simply taking a histogram of
sampled values of any of the quantity. Similarly, the joint PDF for any two
(or more) quantities can be computed by taking the two- (or more-)dimensional
histogram over those parameters. Further, the marginalisation integral for the
`most likely' value of any quantity $Q(\set{P})$---properly speaking, the
expectation value for $Q$---can be trivially computed by taking the mean of
all MCMC sampled values for that quantity (\cf\ \eqref{margin}); \ie:
    \begin{equation} \begin{split}
    	\left<Q\right> &\equiv \int \mathrm{d} \set{P} ~ Q(\set{P}) 
    			~ \mathrm{Pr}(\set{P}|\set{X}, \set{S}, \set{W}) \\
    				&\cong \mathrm{mean} \big[ Q(\set{P}_i) \big] ,
    \end{split} \end{equation}
where $\set{P}_i$ represents the individual (post-convergence) MCMC sampled
sets of trial values for the parameters $\set{P}$. Similarly, the uncertainty
on the value of a single parameter can be simply computed as the RMS of MCMC
sampled values (\ie, $\sigma_Q^2 = \left<Q^2\right> - \left<Q\right>^2$), and
the joint, covariant uncertainties on multiple parameters can be computed via
the Pearson correlation coefficient (\ie, as in Eq.\ \ref{eq:rho}).

We have used the publicly available\footnote{Available for download via
\texttt{http://danfm.ca/emcee}} \textsc{python} package \emcee\ \citep{emcee}
to actually perform the MCMC fits presented in this work. Compared to the
standard Metropolis-Hastings MCMC sampling algorithm, the most important
feature of \emcee\ stems from its use of multiple MCMC `walkers' when sampling
the parameter space. The step size for individual walkers are based on the
distribution of the ensemble of all walkers, using an affine-invariant
`stretch move' algorithm, which leads to very efficient sampling, even in the
case of strongly anisotropic PDFs. For this work, a key practical advantage of
using \emcee\ is that it is trivially parallelisable.

The results of this MCMC fitting process are illustrated in
\textbf{\figref{mcmc}}. This figure shows the individual sets of trial values
for each of the 40 parameters in $\set{P}$ for unique MCMC samplings. These
samplings are, in a sense, our results---they represent the high dimensional
PDF for the values of the parameters in $\set{P}$.

For clarity, the tracks for individual parameters in \figref{mcmc} have been
scaled according to the mean and RMS values of the thinned and post-burn MCMC
samplings; that is, according to the Bayesian estimator for the most likely
value, and the uncertainty in that value. These values are given in each panel
of \figref{mcmc}; this Figure thus also serves as a table of the results of
our fits. The fact that each of these lines is horizontal shows that the fits
have in fact converged.

To the extent that the PDF for any given parameter value is truly Gaussian,
these values can be used to fully describe the PDFs. Again, the distribution
of MCMC samplings converges to a faithful mapping of the PDF, with no embedded
assumptions of Gaussianity; these distributions are shown in \figref{mcmc} as
the grey histograms. In all cases, the PDFs can be seen to be well-sampled.
Further, in most---but not all---cases, the PDFs can be seen to be well
described by a simple Gaussian. Again, we stress that these distributions
naturally and fully account for covariances among the values of (many)
different parameters.

\subsection{Model Selection and the Limits of Objectivity} \label{ch:modelsel}

\subsubsection{Model Selection}

How can we be satisfied that, for example, the B population really is (or is
not) better described by a double- rather than a single-Schechter mass
function, or whether or not the blue CMR can be adequately described using
just a simple linear relation? To explore these kinds of issues, we have made
many fits to our dataset, in which we have eliminated one or more of the 40
parameters that go into our most general model. Our tests have been
systematic, but by no means exhaustive. As described below, we have used these
tests to ensure that we are not grossly overfitting the data.

There are a number of Bayesian approaches to the problem of model selection
which we have explored: the Bayes factor, $K$, the closely related Akaike and
Bayesian Information Criteria (AIC and BIC, respectively), and the Deviance
Information Criterion (DIC). While these different approaches are each based
on slightly different assumptions, and are thus strictly valid in slightly
different circumstances, they can all be thought of as being similar in spirit
to a traditional frequentist log-likelihood-ratio test. In asymptotic limits,
AIC $\approx$ BIC $\approx$ DIC $\approx -2 \ln K$. A difference of 2.5 (or
10) in the IC of two different models implies a likelihood ratio of $\approx$
3.5 (or 150), where the model with the lower value for the IC is the preferred
one. However, unlike a simple likelihood ratio test, each of these quantities
includes an explicit or implicit penalty for larger numbers of parameters, so
as to protect against over-fitting of the data.

We have focussed primarily on the AIC and BIC, which are defined with
reference to the maximum of the likelihood function, $\Ell\max$.\footnote{For
each variation of the model, we have found this value using the method of
simulated annealing. This is simply a modification of standard MCMC, in which
$\ln \Ell$ is scaled by a factor $1/T$. A lower $T$ makes steps to lower
values of $\ln \Ell$ harder than they would otherwise, effectively corralling
the MCMC walkers where $\ln \Ell$ is high. It is thus possible to robustly
determine the value of $\Ell\max$ by successively reducing the value of $T$.}
Since extra parameters can only increase the value of $\Ell\max$, the question
is whether or not this improvement is sufficient to merit the inclusion of an
extra parameter. The penalty terms for additional parameters for the AIC and
BIC are $2 k$ and $k \ln n$, respectively, where $k$ is the number of free
parameters in the model, and $n$ is the number of data points. Thus it can be
seen that, all else being equal, the BIC penalises additional parameters more
strongly than the AIC for $n \gtrsim 8$. (In our case, $\ln n \approx 10$, so
the BIC penalty is roughly 5 times larger.)

In other words, the BIC prefers models with fewer parameters. Thus, where the
BIC disfavours a simpler model, this model is definitely too simple, and
should not be used. Conversely, the AIC prefers models with more parameters.
Where the AIC disfavours a model with more parameters, then that the use of
that model is definitely overfitting the data.

In order to ensure that we are not abusing our data, we have used these two
information criteria to explore the consequences of omitting individual
parameters from the model described above. Given the number of parameters that
go into our general model in its most general form, it is impractical to do
this in a properly exhaustive way. Instead, starting from the most general
form of our model, we have considered omitting individual parameters one at a
time to see whether or how our model might be simplified.

First, we have tried successively omitting the parameters $c$, $d$, $r$, and
$s$ (\ie, the parameters that describe the step or bend in the locus of or
scatter around the CMRs). When fitting to the $(g-i)$ CMD, it is possible that
a simpler description of the B CMR is possible: omitting the parameters
$c\blue$ and $d\blue$ improves both information criteria by 8 and by 10,
respectively. While this constitutes positive statistical evidence against the
need for one of these parameters, it is ambiguous which one should be
excluded. When fitting to the $\starcol$ CMD, the results are similar. Based
on the BIC, it is preferable to omit either $r\red$ or $s\red$ (but not both),
and also possibly and $d\blue$ and $b\red$ ($\Delta$BIC = 8 for both).

The ambiguity in these results make perfect sense looking at the table of
results given in \figref{mcmc}. These parameters which may or may not be
necessary are those whose fit values are statistically consistent with being
zero. Their inclusion or exclusion thus makes little if any difference to the
fits, and the decision as to whether or not to include these parameters has no
practical consequences. 

We therefore elect to use the most general description possible for the locii
of and scatters around the CMRs, by fitting for all of these parameters. In
this sense, our results can be thought of as hypothesis testing the need for
each of the $c$s, $d$s, $r$s and $s$s. Where the fit values for any of these
parameters are consistent with zero, then that parameter may be unnecessary
for a good description of the data. These results nonetheless encapsulate
positive information about the forms of locii of and scatters around the B and
R CMRs.

Perhaps more interesting is what happens when we trial alternate descriptions
of the MFs. As well as the general double Schechter parameterisation, we have
trialled a coupled twin-Schechter description, where the two Schechter
functions that describe either the R or B population have the same
characteristic mass (\ie, $x^\dagger_{2} = x^\dagger_{1}$). We have also
trialled using only a single Schechter function (\ie, $f_2 = 0)$.

When fitting to the $(g-i)$ CMD, the BIC definitely disfavours a single
Schechter function description for the B MF ($\Delta$BIC $\gtrsim 30$),
indicating that such a model definitely underfits the data. The BIC also
definitely disfavours the coupled, twin Schechter function description for the
B MF; the data definitely prefer a double Schechter function description of
the B MF, with $x^\dagger{\mathrm{B},2} \ne x^\dagger_{\mathrm{B},1}$. At the
other extreme, the AIC disfavours the most general, double Schechter model
($\Delta$AIC $\lesssim 6$; odds $\approx$ 20:1) for the R MF described. That
said, it is worth noting that both the AIC and BIC (weakly) prefer a single
Schechter description for the R MF. This may not be surprising, given how
weakly constrained the values of $\alpha_{\mathrm{R},2}$, and
$x^\dagger_{\mathrm{R},2}$ are.

When fitting to the $\starcol$ CMD, the BIC disfavours a single Schechter MF
for the R population ($\Delta$BIC $\lesssim 10$). The BIC does not obviously
prefer the more complicated, independent double Schechter descriptions of
either the B or R MFs; the coupled, twin Schechter functions are just as good.
The variation of the model that best balances between the two criteria is the
one that uses coupled, twin Schechter functions to describe both the B and the
R MF.

In light of all of the above, we will continue our analysis using the most
general form for our model, which is fully defined by 40 parameters. This is
despite the fact that both the AIC and BIC prefer a single Schechter
description for the R MF when analysing the $(g-i)$ CMD. (But of course, this
being the case, the fits do not make use of the additional freedom that the
second Schechter component provides, precisely because it is not necessary for
a good description of the data.) In this sense, our fit results can be taken
as limiting the deviations from Schechter-ness in the observed MF for R type
galaxies.

For all of the rest, we can say that we are not definitely overfitting the
data, nor are we definitely underfitting the data. Beyond this point, however,
all we can say is that we have done the best that we know how (and monopolised
64 cores for more than 2 months) to ensure that we are using the best and
simplest model that we can to describe the existing data.

\subsubsection{The Limits of Objectivity} \label{ch:objectivity}

In some of the above, there is some ambiguity, inasmuch as the data do not
provide strong evidence for or against the inclusion of some parameters. This
is particularly true for whether or how the B and R MFs should be described
with a combination of Schechter functions. To the extent that there is
ambiguity, our decisions about whether or not to include these parameters are
subjective, hence {\em arbitrary}.

To the extent that these points are ambiguous, however, they are also {\em
unimportant}. This is true in the sense that these decisions have no strong
impact on the statistical quality of the fits to the data. This is also true
in the sense that the best fit models based on the less-general
parameterisations provide essentially identical results.

What we have done in the above is to explore special cases or restricted
classes of the general form of our model. The question that naturally arises
is whether there are some alternative parameterisation might provide a better
description of the data. For instance, it could be that a Lorentzian or a
Student's $t$ or a skewed Gaussian description of the scatter around the CMRs
yields a better statistical description of the data. Of course, a properly
exhaustive search of all conceivable models is impossible.\footnote{That said,
we have experimented with, for example, using one or two Gaussians to describe
the R MF, and find that this does not work well. We have also experimented
with allowing for a sharp or an exponential cutoff to the R MF at low masses,
and find that the model does not make use of this freedom. For the locii of
and scatters around the CMRs, we have also experimented with using a
polynomial parameterisation in place of our two-line descriptions. This fails
to provide a good description of the locii of the CMRs, nor does it provide a
good description of $\zeta\red$. Remarkably though, using a fifth order (\ie,
six parameter) polynomial description for $\zeta\blue$, we obtain virtually
identical results: the differences in the fit values of $\zeta\blue(x)$ are
nowhere more than 0.01 mag, and the IC is only slightly worse.}

This brings us to the final and most important caveat on our results. While
the formal statistical uncertainties on our CMR and MF fits are impressively
small, the values themselves {\em cannot but} be determined by the decisions
made in constructing the parameterised, descriptive model that has been used.
On the other hand, the same criticism can be levelled at {\em any} model or
modeller---even a fully physically-minded explanatory model is forced to
presuppose the validity of the theoretical framework on which it is based. At
this level, some degree of subjectivity is inescapable.

Once we have set the parametric form of our descriptive model, however, it is
then left to the data to decide on the particular parameter values that
provide the best description, including all the characteristics of the B- and
R- populations. That is, given our choice of parameterisation, it is the data
themselves that define the bimodal distributions. The results of this
modelling thus provide objective B/R classifications {\em insofar as
objectivity is possible}. These results can thus be said to provide an
accurate, reliable, and robust phenomenological description of the observed
data---and this is all that we have set out to obtain.

\section{Results I.\ --- The bivariate colour--mass distributions}
\label{ch:dist}

\subsection{Demonstrating the quality of the fits} \label{ch:quality}

\begin{figure} \hspace{-0.9cm} 
    \includegraphics[height=18.9cm]{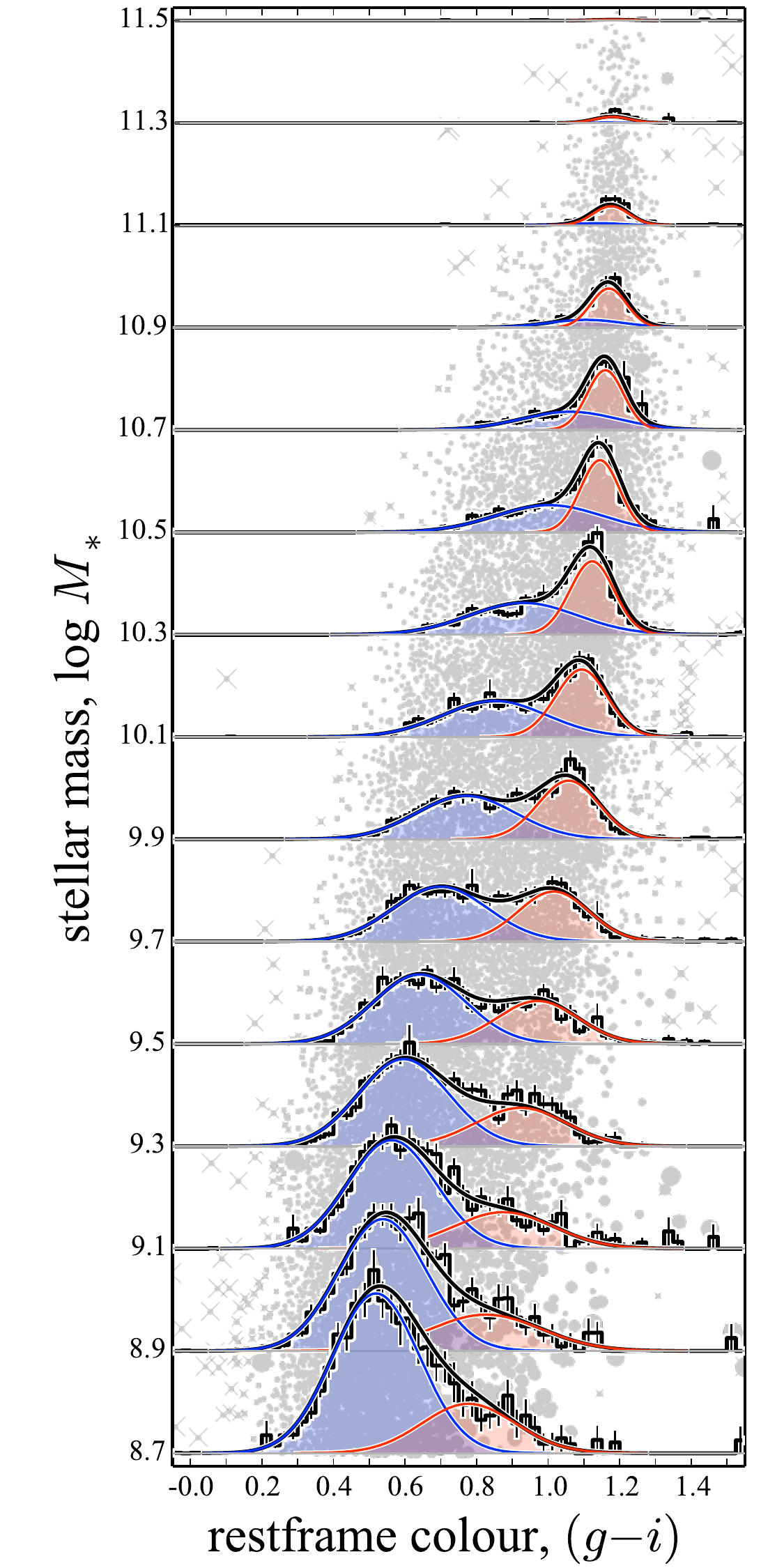}
\caption{Demonstrating the quality of our fits to the joint $(g-i)$--$M_*$
distribution.--- The histograms in this Figure show the observed
$1/V\max$-weighted distribution of restframe $(g-i)$ colours of $z < 0.12$
galaxies, computed in bins of $\log M_*$ centred on $\log M_*$ = 8.8, 9.0,
..., 11.4. The errorbars show the statistical uncertainties on each these
distributions, derived by bootstrap resampling. The smooth curves show the
results of our modelling: the blue and red curves show the fit distributions
for the B- and R-populations; the black curves shows the net B+R
distributions. Underlaid beneath all this, the grey points show the data
themselves. The size of each point is proportional to $1/V\max$. Data
objectively classified as being `bad' are marked with a cross. Note that we
have not binned the data in the course of the fits; the binning in this Figure
is for illustrative purposes only. It is clear that the fit model provides a
good description of the observed data. \label{fig:rfdist} } \end{figure}

\begin{figure}\hspace{-0.9cm}
    \includegraphics[height=18.9cm]{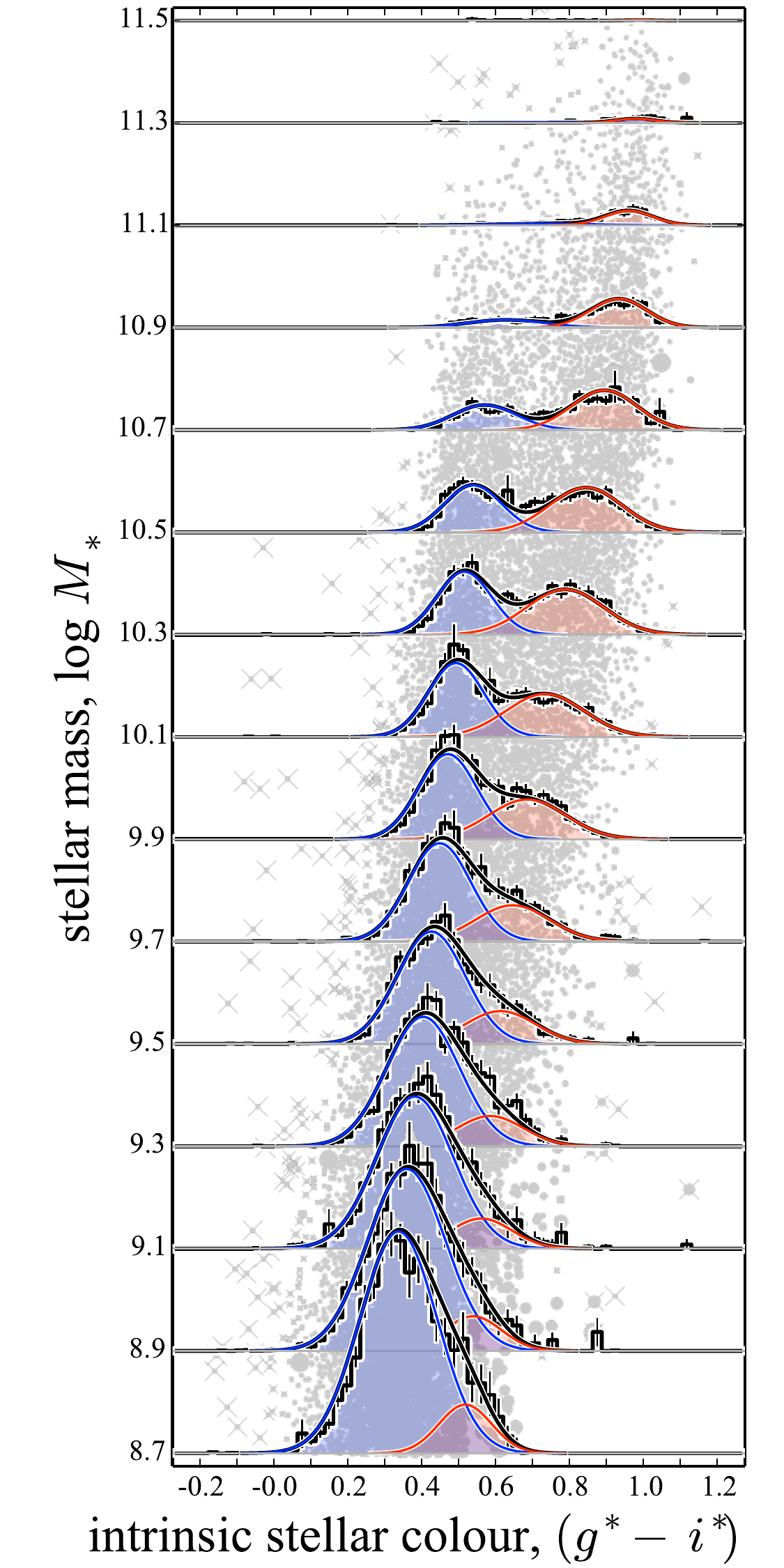}
\caption{Demonstrating the quality of our fits to the joint $\starcol$--$M_*$
distribution.--- The histograms and smooth curves in this Figure show the
observed and fit distributions of intrinsic $\starcol$ colours in bins of
$\log M_*$; all symbols in this Figure and their meanings are analogous to
\figref{rfdist}. In contrast to \figref{rfdist}, the general galaxy population
is better separated into distinct R and B populations on the basis of
$\starcol$ than of $(g-i)$, in the sense that there is less overlap between
these two distributions in this Figure than in \figref{rfdist}. Further, as in
\figref{rfdist}, we see no compelling evidence for the need to include a
third, `green' population. In comparison to \figref{rfdist}, the need for a
multiple-Gaussian description of the {\em observed} colour distributions for
$\log M_* \lesssim 9.3$ is less clear; this point is discussed in detail in
\secref{wrong}. In any case, as in \figref{rfdist}, the fit model can be seen
to provide an excellent description of the observed data. \label{fig:stardist}
} \end{figure}

In order to illustrate our ability to perform such a detailed descriptive
modelling of the data, as well as the quality of the resultant fits, consider
\textbf{\figref{rfdist}} and \textbf{\figref{stardist}}. Our task in this
section is to describe and discuss these Figures. Building on the discussion
begun in \secref{redness}, we will also flag the major issues with, and
caveats on, the interpretation of our results, which should be kept in mind in
all that follows.

\figref{rfdist} and \figref{stardist} show the $(g-i)$ and $\starcol$ CMDs,
respectively. In each panel, we have split our sample up into bins of $\log
M_*$, each with width 0.2 dex. The solid black histograms then show the
incompleteness-corrected colour distributions within each of these mass bins:
the lowest histogram is for the $8.7 < \log M_* < 8.9$ bin, the next is for
the $8.9 < \log M_* < 9.1$ bin, and so on. The normalisation of each of these
histograms is arbitrary, but is the same for all bins in both Figures.

Illustrative statistical uncertainties in the observed colour distributions
are shown by the black error bars. These uncertainties have been estimated
using bootstrap resampling. (We have not attempted to account for
field--to--field variance due to large scale structure.) For $\log M_* \gtrsim
9.5$, these statistical uncertainties are negligible. Further, even for the
lowest $8.7 < \log M_* < 8.9$ bin, the statistics are reasonably good.
Particularly for $9 \lesssim \log M_* \lesssim 11$, then, there should be no
question as to whether or not the data are good enough to allow an independent
double-Gaussian (or some other parametric) fit to any or all of the
individual, mass-binned colour distributions.

The smooth curves in \figref{rfdist} and \figref{stardist} show the results of
our descriptive modelling of the bivariate $(g-i)$ and $\starcol$ colour-mass
distributions. The red and blue curves show the distinct R and B components of
the model; the black curve is the total, R--plus--B distribution. Note that
for these illustrative purposes, we have convolved the modelled
colour-distributions with typical $(g-i)$ or $\starcol$ uncertainties for each
mass bin, so that these curves are directly comparable to histograms for the
observed data.

Before moving on, let us stress that we have {\em not} binned the data in the
course of fitting it: the binning in mass and colour in these figures is for
illustrative purposes only. Further, nowhere in the modelling is it ever
specified---or even relevant---whether any particular galaxy belongs to either
the R or the B populations; there is no binning in this sense either.

If, for example, we were to just fit double Gaussian distributions to
the observed colour distributions for the distinct mass bins that are shown in
\figref{stardist}, then the inferred values for the centres, scatters, and
normalisations of the R and B distributions in each bin would be completely
independent. Given the relatively poor sampling for $\log M_* \lesssim 9.3$
and $\starcol \gtrsim 0.7$ galaxies, we would have no means of robustly
constraining the properties of either of the two Gaussian components of the
$\starcol$ colour distributions for these very low masses---or even whether or
not two Gaussian components should be used.

But this is not what we have done. In effect, we have assumed that the
centres, scatters, and normalisations of the (Gaussian) B and R colour
distributions vary smoothly as a function of mass. In this sense, the
derivation of each and every of the individual modelled B and R colour
distributions shown in \figref{stardist} is based on each and every datapoint
that we have.

Ultimately, all the model considers is the relative probability of finding a
galaxy---irrespective of type---at a particular point in the colour--mass
plane. In other words, we are using a mixture model of two distinct but
overlapping populations to describe the joint colour--mass distribution
function of {\em all} galaxies---we are fitting for the 2D scalar function
that is represented in \figref{rfdist} and \figref{stardist} as the black
curve. In this way, we are able to characterise the CMRs and MFs of the two
populations {\em without ever explicitly considering which galaxies belong to
which population}.

\subsection{What could possibly go wrong? The R population at low masses}
\label{ch:wrong}

While our parametric model does provide a very good description of the data,
{\em a good} description is not necessarily the same as {\em the right} one
(see related discussions in \secref{objectivity} and \secref{fallacies}).
Other parameterisations of the colour distributions (at fixed mass) are
possible, and may lead to different results --- but this is {\em always} true.
The fact that the answer we get depends on how we have devised our analysis is
inescapable. If our assumptions, which are clearly stated and motivated in
\secref{overlap} and \secref{summary}, are shown to be invalid, then our
results go with them. Indeed, we have gone to great lengths to make this point
in \secref{redness}. Without denying these inescapable truisms, the fact
remains that our model does provide a very good description of the data, and
so offers one potential avenue for understanding the data.

At least for $\log M_* \gtrsim 9.7$, the model does perform its intended
function: decomposing the observed data into a mixture of two populations,
which are distinguished and defined by their own distributions of colours,
which we take to be a tracer of the constituent stellar populations.

The skeptical reader's eye may be drawn to the fits at $\log M_* \lesssim 9$,
however, where the suitability of a double Gaussian fit becomes increasingly
problematic. Certainly, in both \figref{rfdist} and \figref{stardist}, below
$\log M_* \sim 9.3$, it becomes difficult for us to claim that we have
robustly separated the general galaxy population into two distinct R- and
B-populations. How then should one interpret our results at these low masses?

Considering this problem from the modelling perspective, we should ask what
aspects of the data drive the fits most strongly. Clearly, it is the shape and
normalisation of the B colour distributions that are the best constrained at
these low masses. Also, from the nature of the fits, it should be clear that
the fits to the B colour distributions are decided primarily by the $(g-i)
\lesssim 0.6$ or $\starcol \lesssim 0.3$ data. Bluewards of the peak of the
observed colour distributions, the data are well described by a Gaussian.
Then, since the colour distributions for each population are assumed to be
symmetric, it is left to the R population to accommodate whatever asymmetries
there are in the observed colour distributions. 

From a more astrophysical perspective, then, what matters is the extent to
which the colour distribution for the B population is expected to be
symmetric. However, this will certainly not be the case for $(g-i)$, where the
asymmetric effect of dust is expected to skew the B colour distribution to the
red. In this case, one would expect the inferred number of R galaxies to
absorb some of the reddest B-type galaxies. Indeed, looking at the lowest two
or three bins of \figref{rfdist}, one can see immediately how such a
description might work. 

Rather than trying to fit for an asymmetric or skewed $(g-i)$ distribution
(which would be considerably computationally more complicated), we can look at
the $\starcol$ distribution, where we have tried to remove dust as a
complicating factor. But looking at the lower three bins in \figref{stardist},
there is no way to decide whether the slight asymmetry of the observed colour
distribution ought to be interpreted as an indication of a separate
population, or instead as nothing more than a slight asymmetry in the
$\starcol$ colour distribution of the B population.

Our conclusion is therefore that at these low masses, the (field) red sequence
dissolves into obscurity---we no longer see clear evidence of two distinct
populations in the $(g-i)$ or $\starcol$ colour distributions for $\log M_*
\lesssim 9.5$. (As we have argued in \secref{complete}, we do not believe that
the apparent dearth of low mass red galaxies is due to incompleteness.)
Instead, we present our inferred MFs for R-type galaxies as {\em an upper
limit} on the number densities of galaxies that have moved (or are moving)
away from the colour distribution that describes most low-mass galaxies; \ie,
the B population.

Again, the fact that the inferred B- and R- populations have substantial
overlap serves to underline the subtleties involved in interpreting our
results in concrete, astrophysical terms. In particular, even at intermediate
masses, it would be unwise to blithely equate the R population with
`quenched'---but recognise how much more problematic it is to apply the term
`quenched' to the hard-cut `red' samples discussed in \secref{redness}.

\subsection{What we have (and have not) done} \label{ch:done}

To sum up: using the parametric model described in \secref{method}, we have
derived a very good description of the observed bivariate distributions
between both $(g-i)$ and $\starcol$ and stellar mass. This analysis is
intended to provide a phenomenological description of the essential
characteristics of the bivariate colour--mass distributions. Such a
description clearly requires (at least) two populations with their own
distinct CMRs and MFs. Our approach enables us to simultaneously and
self-consistently describe the bivariate colour-mass distribution functions of
the two populations; indeed, this is how these populations are defined.

Again, we stress that the designations B and R refer primarily to the two {\em
populations}, rather than to individual galaxies. That is, instead of
characterising the demographics of galaxy samples that are pre-selected to be
`blue' or `red', what we have done is decompose the full population into a
mixture of two distinct, but overlapping, subpopulations; we dub these two
populations `B' and `R'. In this way, we can {\em derive} an operational
definition for the terms `R-type' and `B-type'.

Once this is done, however, it is possible to use these fits to
quantitatively---if probabilistically---classify individual galaxies according
to the chances that the galaxy in question has been drawn from either the B or
R population; \ie, the relative contributions of the B- and R-populations to
the data density at any given point in the CMD. (See \secref{classes}, below.)
But it is important to understand that these classifications {\em follow from}
the fits: during the fitting process itself, the B-ness or R-ness of any
particular galaxy is irrelevant.

We make no pretensions, however, about providing an explanatory model for the
observations. That is, we justify our splitting of the general population into
B and R components on the grounds that {\em these are distinctions that
galaxies somehow care about}. In this way, we derive a phenomenological
description of the bimodal---better, the two population---character of the
galaxy population, in terms of galaxies' stellar populations. 

The underlying physical differences in the origins and natures of the B- and
R-populations, as well as those responsible for the observed ranges of
$\starcol$ colours within each population, remain to be determined (but see
\secref{howelse}), and will be the focus of future works in this series.

\section{Results II.\ --- The Mass Functions and Colour--Mass Relations for
Red and Blue Galaxies} \label{ch:results}

\begin{figure} \includegraphics[width=8.6cm]{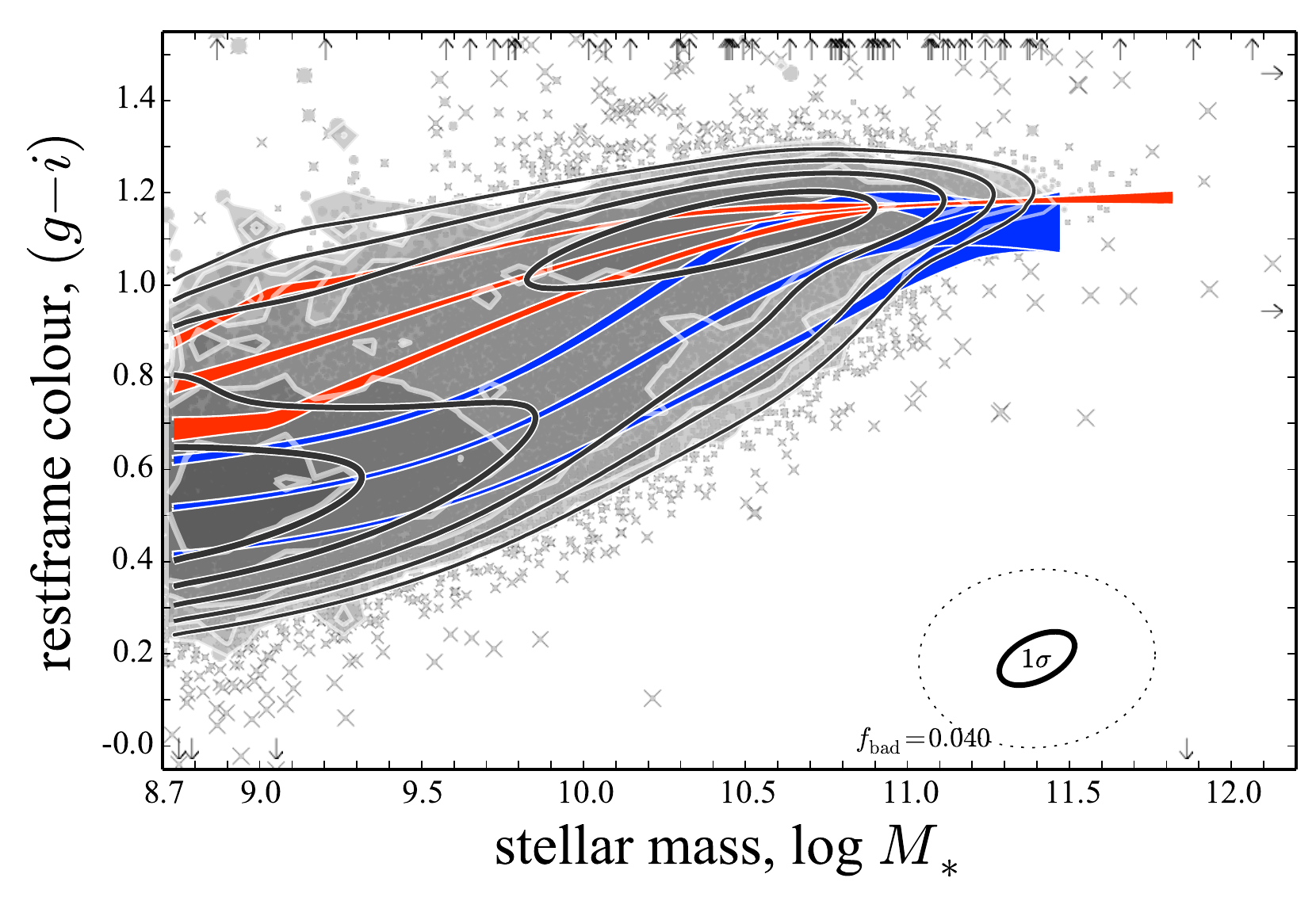}
\caption{Results of our fits to the restframe $(g-i)$ CMD, highlighting the
CMRs for the B and R populations.--- This Figure is discussed at length in
\secref{cmrs}. In this Figure, the points show the data. As in \figref{rfdist}
the size of each point reflects the value of the $1/V\max$ weighting for
incompleteness, and the crosses show those datapoints objectively identified
and censored as `bad'. The small arrows show where galaxies fall outside the
plotted range. The filled greyscale contours show the incompleteness-corrected
data density, in steps of powers of 2. These should be compared to the smooth
line contours, which show the bivariate distribution function from our fits.
For illustration purposes, these have been convolved with the typical
observational errors, shown at the bottom right. Here, the dotted ellipse
shows the inferred errors for `bad' galaxies. The main feature of this plot
are the smooth curves, which show our fits for the locii of, and scatter
around, the $(g-i)$ CMRs for the distinct R- and B-populations. The width of
these lines show the 68\,\% confidence intervals on the values of these
functions, as a function of mass, and including all convariances between model
parameters. This figure should be contrasted with \figref{starcmr}.
\label{fig:rfcmr}} \end{figure}

In \figref{rfdist} and \figref{stardist}, we have illustrated the quality of
our fits to the bivariate $\log M_*$--$(g-i)$ and $\log M_*$--$\starcol$
distributions for our sample of $\log M_* > 8.7$ and $z < 0.12$ galaxies. Our
task in this section is to lay out the actual fit results---\ie, the CMRs and
MFs for R- and B-type galaxies---which describe the two populations. Fit
results are given in a machine readable table as additional online material.

\subsection{Scaling Relations} \label{ch:cmrs}

\begin{figure} \includegraphics[width=8.6cm]{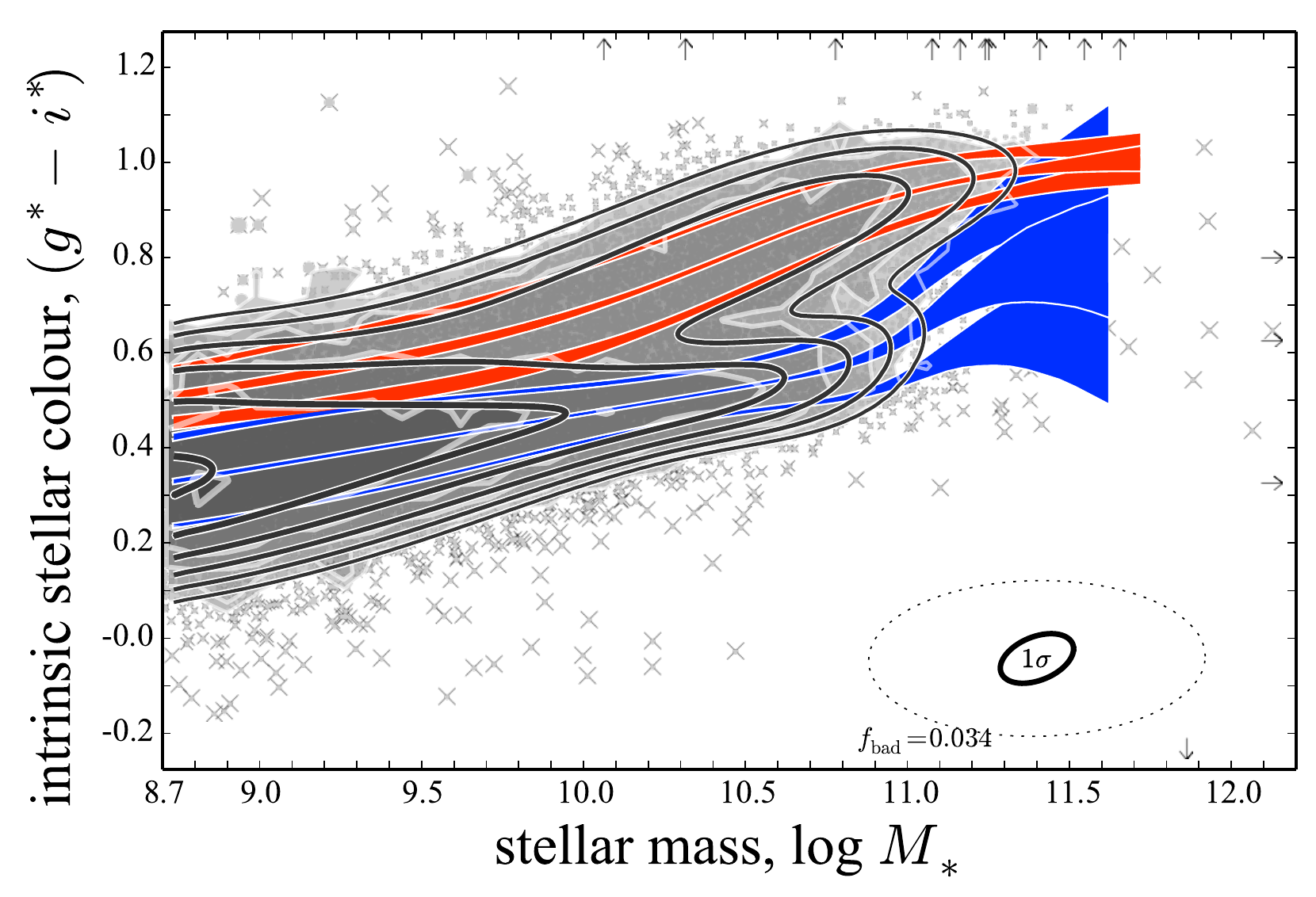}
\caption{Results of our fits to the intrinsic $\starcol$ CMD highlighting the
CMRs for the B and R populations.--- This Figure is discussed at length in
\secref{cmrs}. All symbols and their meanings are analogous to \figref{rfcmr}.
In comparison to the $(g-i)$ CMD, we make the following qualitative
observations. The `blue sequence' in this $\starcol$ CMD is both tighter and
more linear than the $(g-i)$ `blue cloud'. This implies that the upturn in the
`blue' CMR seen in \figref{rfcmr} is caused by a change in the dust properties
of blue galaxies with $9.5 \lesssim \log M_* \lesssim 10.5$, rather than a
change in the stellar populations of these galaxies. Note how the $\log M_*
\gtrsim 10.8$ upturn to the B CMR in this Figure coincides with the
convergence between the $(g-i)$ CMR for the two populations in \figref{rfcmr}.
The rather steeper slope of the R-type CMR the shows how this population is
less homogenous than the B-population: lower mass R-type galaxies have rather
different stellar populations to their higher mass cousins. At the same time,
the relatively tight and smoothly sloping CMR show how across the R-type
population, mass is a relatively good predictor of stellar population,
suggesting a common evolutionary pathway for these galaxies. These points are
discussed further in \secref{observations}. \label{fig:starcmr} } \end{figure}

The results of our fits for the $(g-i)$ and $\starcol$ CMRs are shown in Fig.s
\ref{fig:rfcmr} and \ref{fig:starcmr}, respectively. In both of these Figures,
the grey points show the data themselves. As in Figures \ref{fig:cmds},
\ref{fig:rfdist}, and \ref{fig:stardist}, the size of each point has been
chosen to reflect the value of the $1/V\max$ weighting factor. Further, as in
Figures \ref{fig:rfdist} and \ref{fig:stardist}, each individual datapoint is
marked with a black cross, the size of which has been chosen to reflect the
probability that that datapoint has been drawn from the `bad' distribution.
The marked points have thus been objectively identified as outliers; they make
little to no contribution to the fit CMRs shown. This objective censoring can
be seen to be very effective.

The semi-transparent filled contours in \figref{rfcmr} and \figref{starcmr}
show the observed, $1/V\max$-weighted bivariate data density; these contours
have the same log$_2$ scaling as those in Fig.s
\ref{fig:colours}--\ref{fig:otherdust}. These should be compared to the smooth
line contours, which are interlaid between the data points and the fit CMRs,
and which show the logarithmic probability density contours from the model. As
for \figref{rfdist} and \figref{stardist}, to generate these contours, we have
convolved the model fits with typical uncertainties, so that these contours
are directly comparable to those for the data themselves. These contours thus
reflect the combination of the mass functions and the CMRs for these
populations. These contours are included mostly for illustration; the mass
functions themselves are shown in Fig.s \ref{fig:rfmf} and \ref{fig:starmf},
and are described separately below. The models provide very good descriptions
of the observed bivariate $(g-i)$ and $\starcol$ colour-mass distributions.

In each figure, the smooth lines show the fit locii for the two CMRs, as well
as the fits for the RMS scatter around each CMR. The width of each of these
lines shows, at fixed mass, the 68\,\% confidence intervals for each quantity.
These uncertainties can be seen to behave reasonably: they are very small
where the data concentration is high (\eg, the centres of the red sequences),
and become large where the data concentration is low (\eg, the very high mass
ends of both the red and blue sequences). In connection with our discussion in
\secref{wrong}, note how the uncertainties in the locations of the $+1\sigma$
point of the R $\starcol$ distributions are less than that of the locus of the
CMR, which are in turn smaller than those of the $-1\sigma$ point. This again
shows how the descriptions of each population are constrained principally by
the outer edges of the colour distributions, as well as illustrating how our
Bayesian approach can yield meaningful uncertainties on secondary aspects of
the model (\eg, the CMRs) that nicely and naturally propagate all the relevant
uncertainties in the defining parameters of the model.

In \figref{rfcmr}, there is an upturn to the B CMR that begins around $\log
M_* \sim 9.7$. As expected from \figref{colours}, the behaviour seen in
\figref{starcmr} is rather different: what is seen in the $(g-i)$ CMD as the
blue cloud is seen in the $\starcol$ CMD as a considerably tighter and more
linear blue sequence. This implies that the slope of the $(g-i)$ CMR in
\figref{rfcmr} for the B population is more a product of increasing dust
obscuration in higher mass galaxies, rather than differences in the colours of
the underlying stellar populations.

At the very end of this $\starcol$ blue sequence, there is the hint of an
upturn to redder $\starcol$ colours for $\log M_* \gtrsim 10.8$, but this is
where the uncertainties become large. Intriguingly, looking at \figref{rfcmr},
the B CMR becomes indistinguishable from the R one in the $(g-i)$ CMD at this
mass range: the locii of the two CMRs converge, and the scatter in the B CMR
becomes small. That is, the B population becomes indistinguishable in the
apparent $(g-i)$ CMD for $\log M_* \gtrsim 10.8$. Further, those $\log M_*
\gtrsim 10.8$ B-type galaxies identified in the intrinsic $\starcol$ CMD have
rather different stellar populations to the rest of the blue sequence.

Turning to the R population, the most remarkable aspect of the $(g-i)$ CMD
(\figref{rfcmr}) is how the $\log M_* \gtrsim 10.5$ R CMR flattens and tapers
to have essentially no intrinsic scatter. These are the galaxies that one
might expect to be truly `red and dead'. Below this mass, the $(g-i)$ CMR for
the R population bends to bluer colours for lower masses. The $\starcol$ CMR
shows slightly different behaviour: here, the stellar populations of R-type
galaxies become very gradually redder across the range $9.5 \lesssim \log M_*
\lesssim 10.8$.

Taking the results shown in \figref{starcmr} at face value, the simplest
interpretation would be that the R-type galaxies are moving towards the `dead
sequence' only slowly, and in such a way that creates or preserves a
relatively tight relation between a galaxy's mass and its stellar population.
In this scenario, the higher mass galaxies would appear to have progressed
further in this long migration. We will expand further on the simple
observations above in \secref{observations}.

\subsection{Objective classification, following from the fits} \label{ch:classes}

\begin{figure} \includegraphics[width=8.6cm]{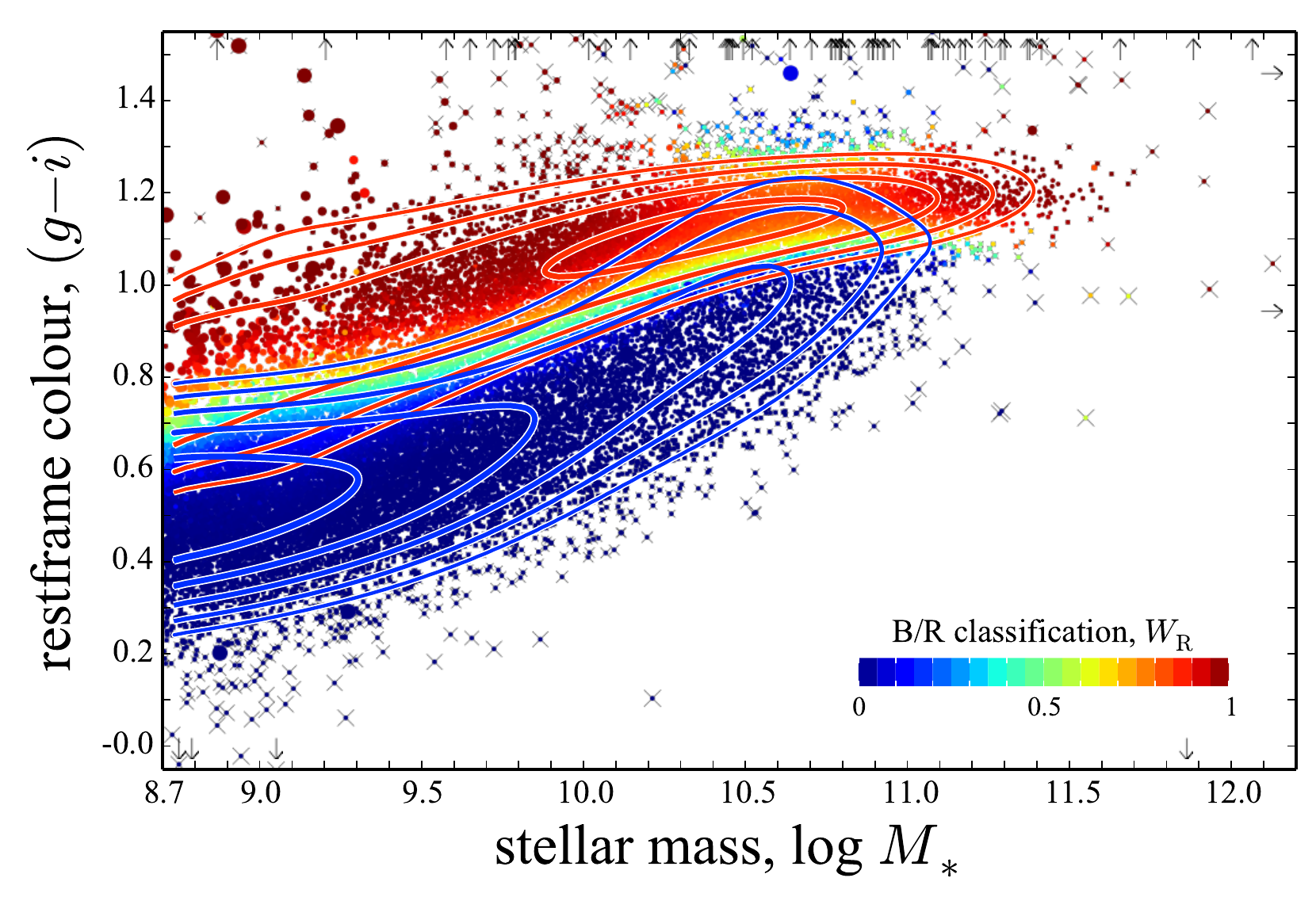}
\caption{Illustrating our objective B/R classifications, based on the
restframe $(g-i)$ colour--mass diagram.--- This Figure is discussed in detail
in \secref{classes}. There are two main features to this Figure. First, the
smooth line contours show the bivariate distributions for the B and R
components of the models, using the same logarithmic scale as in earlier
plots. This information can be used to classify individual galaxies according
to the probability that they have been drawn from one or the other population.
Individual points in this Figure are colour-coded according to these
classifications, $W\red$. Given the empirical fact of scatter in both the blue
cloud and the red sequence (see \figref{rfdist}), and thus the overlap between
the two populations, the individual classifications can be ambiguous,
particularly at high masses. For example, a galaxy observed right on the locus
of the red sequence still has a $\sim 20$--25\,\% chance of having come from
the bluer B-population. \label{fig:rffrac}} \end{figure}

\begin{figure} \includegraphics[width=8.6cm]{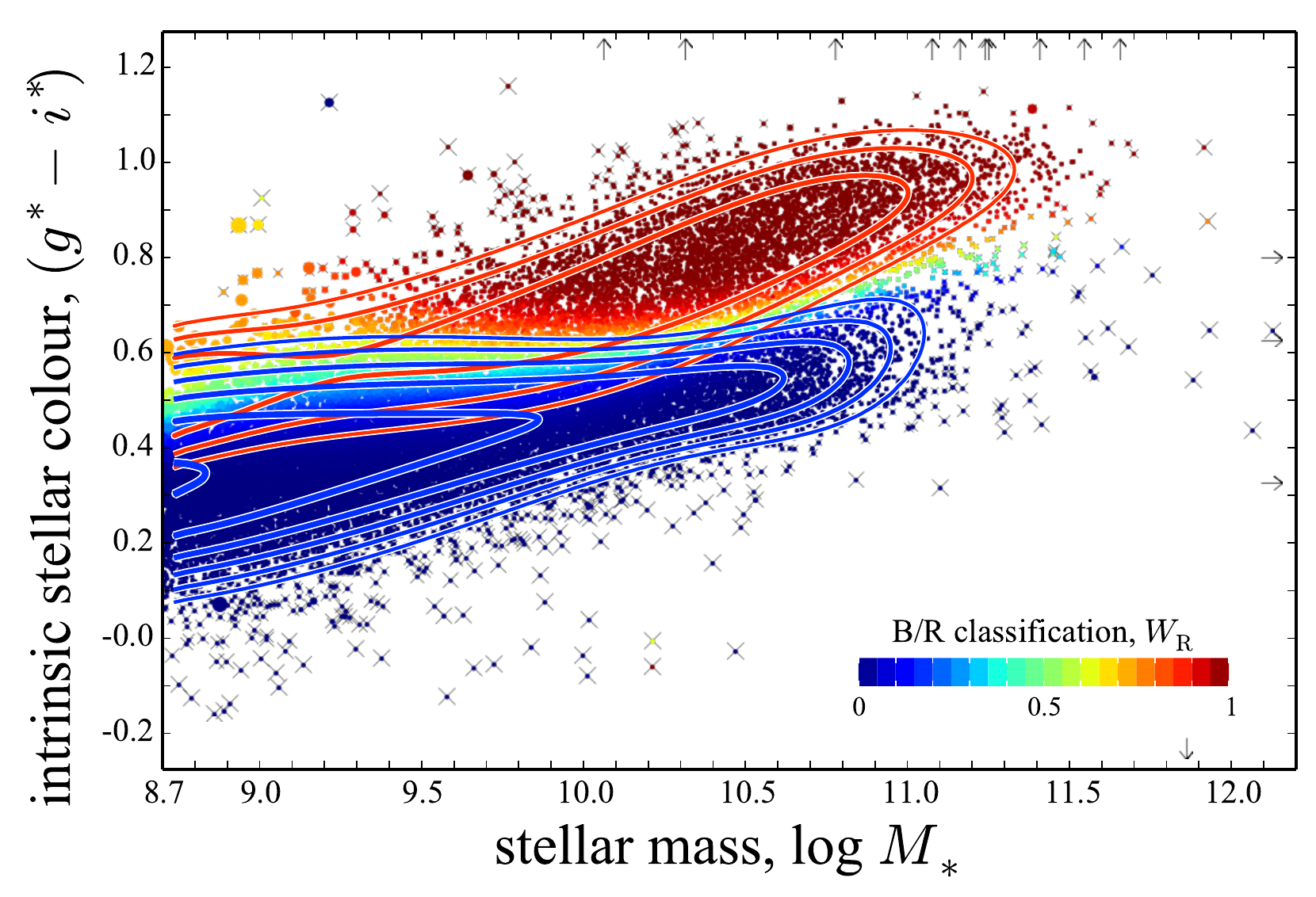}
\caption{Illustrating our objective B/R classifications, based on the
intrinsic $\starcol$ colour--mass diagram.--- All symbols and their meanings
are analogous to \figref{rffrac}. Note how, particularly at higher masses, the
R-population is more unambiguously distinguished in the $\starcol$ CMD than
the $(g-i)$ CMD shown in \figref{rffrac}; only a small fraction of galaxies
have ambiguous classifications (\ie, intermediate values of $W\red$). As
discussed in \secref{classes} and \secref{mfs}, this means that our empirical
determinations of the MFs for B/R classified galaxies are not strongly
dependent on the quality of our fits. At lower masses, and particularly around
$\log M_* \sim 9.5$, the populations are inferred to overlap to the extent
that up to $\sim 20$\,\% of galaxies in the blue sequence are members of the R
population. The overlap between the two populations in this Figure, as well as
\figref{rffrac} demonstrate the importance of accounting for scatter around
the CMR when characterising the two populations. \label{fig:starfrac} }
\end{figure}

In \figref{rffrac} and \figref{starfrac}, we re-present the results of our
fits to the $(g-i)$ and $\starcol$ CMDs in a different way, in order to
illustrate how these fits can be used to develop an objective, quantitative
B/R classification scheme for individual galaxies. In essence, the idea is to
give each galaxy a score, which encapsulates the relative probability, based
on our fits, that that galaxy has been drawn from the R-, B-, or even the
`bad' population.

The R-type score, $W\red\commai$ is given by the relative number of R-type
galaxies expected to be found at the location $\vec{x}\subi$, after convolving
the fit models with the observational uncertainties encapsulated within the
covariance matrix $\mat{S}\subi$. Formally, and using the notation and
definitions introduced in \secref{description}, these values are computed as:
\begin{equation} \label{eq:classes}
     W\red\commai( \vec{x}\subi, \mat{S}\subi | \set{P}_\mathrm{fit}) 
     \equiv \Ell\red\commai(\vec{x}\subi, \mat{S}\subi | \set{P}_\mathrm{fit})
     / \Ell\subi( \vec{x}\subi, \mat{S}\subi | \set{P}_\mathrm{fit} ) ~ .
\end{equation}
Here, $\set{P}_\mathrm{fit}$ is the set of fit-for values for the parameters
$\set{P}$, and $\Ell\red\commai( \vec{x}\subi, \mat{S}\subi |
\set{P}_\mathrm{fit} ) \equiv p\red(x\subi | \set{P}_\mathrm{fit}) \otimes
\gauss_2( \vec{x}\subi, \mat{S}\subi )$. And of course $W\blue$ and $W\bad$
can be defined/computed in an analogous way, so that $W\blue + W\red + W\bad =
1$.

We illustrate how these classifications work in \figref{rffrac} and
\figref{starfrac}. In these Figures, the data have been colour coded according
to their particular values of $W\red$. Note that, because these
classifications depend on the measurement uncertainties as well as the
measured values themselves, there can be some variation in the $W\red$s for
galaxies with very similar colours and masses.

Note how, particularly at higher masses ($\log M_* \gtrsim 10$), there is
considerably less ambiguity in the classifications based on the $\starcol$
CMD, in comparison to those based on the $(g-i)$ CMD. As can be seen in
\figref{rffrac}, the B and R populations overlap in the $(g-i)$ CMD to the
extent that, even along the locus of the R CMR, $\sim 10$-15\,\% of galaxies
come from the B population (see also \figref{rfdist}). Note, too, how the
situation is reversed for $\log M_* \lesssim 10$. The point to be made here is
that, where there is substantial overlap between the two populations, it is
not possible to unambiguously determine whether a particular galaxy is a
member of the R or the B population without some additional information.

It is a legitimate question to ask what additional information could or should
be used to refine these classifications. While it may be tempting to
incorporate morphological or structural information---\Sersic\ index, $n$, for
example--- into the `red' selections, we note that this would result in a
sample of `red and high-$n$' galaxies, rather than a sample of `red' galaxies.
\citep[See][for a detailed discussion of this problem.]{Kelvin2014} In this
sense, there is the very real danger that inclusion of additional parameters
makes things more confusing, rather than less. We will explore this issue
further in future papers in this series.

In this context, we note that these objective, quantitative classifications
can also be cast as {\em weights}, which can be useful in studying the
properties of R- and B-type galaxy samples. \figref{rffrac} and
\figref{starfrac} can thus be alternatively understood as illustrating a
`soft' red/blue selection scheme, which is a smooth function of mass and
colour. This scheme naturally accounts for the scatter around the R and B
CMRs, as well as the R/B fraction, as a function of mass. It also fully
accounts for photometric scatter, due to error/uncertainties in the
measurements themselves. 

This Figure thus helps show how our `B-' and `R-type' designations are
qualitatively different to simple `blue' and `red' colour selections. As
discussed in detail in \secref{others}, below, this is the crucial point that
explains why our determinations of the MFs for the B and R populations differ
strongly from those for `blue' and `red' galaxies given by previous authors.

\subsection{The mass functions} \label{ch:mfs}

\begin{figure} \includegraphics[width=8.6cm]{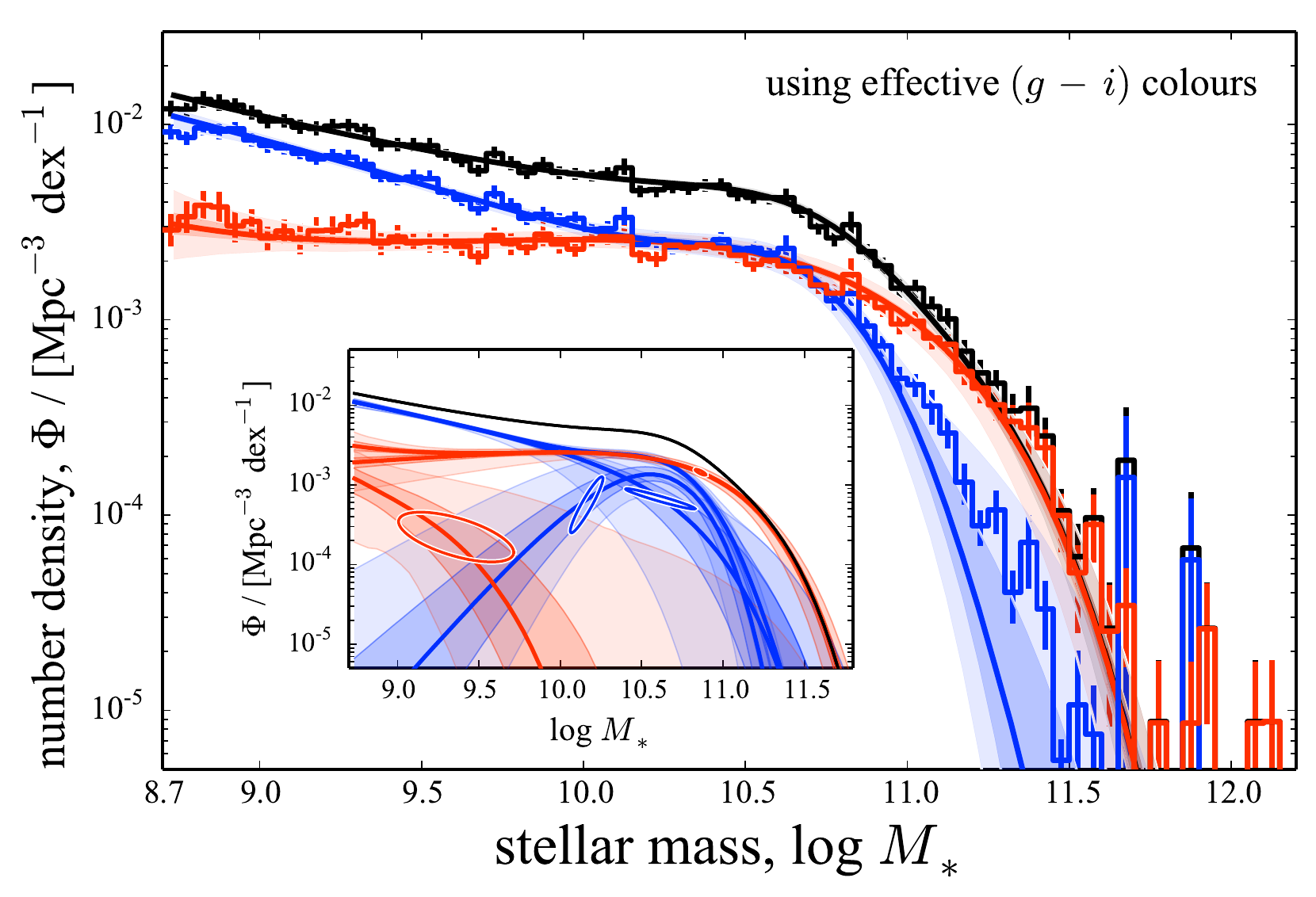} \caption{The
mass functions for the B and R galaxy populations, as derived from our fits to
the restframe $(g-i)$ CMD.--- The smooth blue and red curves show the fit mass
functions for the B and R populations; the black curve shows the net
B--plus--R mass function. The inset panel shows the two separate Schechter
components that go into each of the B and R MFs. The solid lines in this inset
show `the' fit MFs; the shaded regions show the 68 and 99\,\% confidence
intervals for each component. The uncertainties on the total B and R MFs are
shown in the same way in the main panel. Although the individual Schechter
components are partially degenerate, the overall MFs are very well
constrained. Except for the highest masses, the uncertainties are comparable
to the width of the lines used to show the fits, and so are difficult to see.
For comparison to the fit results, the blue and red histograms show the
empirical MFs, where individual galaxies have been weighted by their values of
$W\red$ or $W\blue$. {\em The curves are not fit to the histograms}; instead,
the histograms are derived using the quantitative classifications that come
from the fits. \label{fig:rfmf}} \end{figure}

\begin{figure} \includegraphics[width=8.6cm]{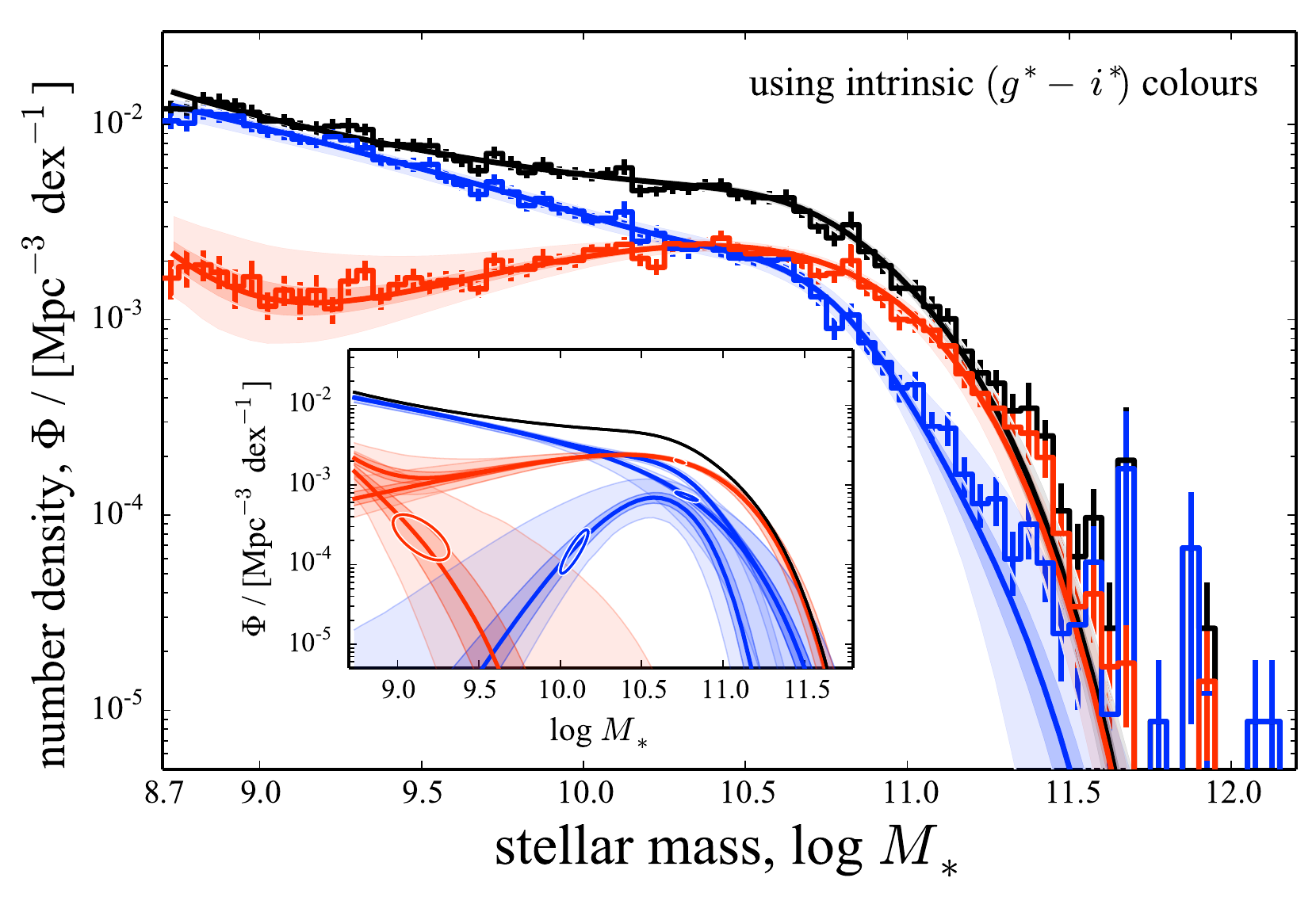} \caption{The
mass functions for B and R galaxy populations, as derived from our fits to the
intrinsic $\starcol$ CMD.--- In contrast to \figref{rfmf}, this Figure shows
the mass functions for the R and B populations as inferred from the
$\starcol$, rather than the $(g-i)$, CMD; otherwise, all symbols and their
meanings are the same as in \figref{rfmf}. Note that we do not consider the
apparent low mass upturn in the R MF to be robust, for the reasons given in
\secref{wrong}, and \secref{mfs} (see also \figref{stardist}). Further, there
is not conclusive statistical evidence that a second Schechter component to
the R MF is required \secref{modelsel}. The quantitative discrepancies between
the results shown in this Figure and in \figref{rfmf} for $\log M_* \lesssim
10$ can be understood looking at the distributions shown in \figref{rfdist}
and \figref{stardist}. The implication from this Figure and \figref{rfmf} is
that systematic uncertainties in the low-mass values of the R MF are at least
a factor of 2. It is worth stressing, however, that the two sets of results
agree at the $\lesssim 2 \sigma$ level for all $\log M_* \gtrsim 10$. Compared
to previous results, we find that the B-population---\ie, galaxies with young
stellar populations---extends to much higher masses, and we find considerably
fewer `old' galaxies at lower masses. (see also
\figref{punchline}).\label{fig:starmf} } \end{figure}

The inferred MFs for the R and B populations, based on the $(g-i)$ and
$\starcol$ CMDs, are shown in \figref{rfmf} and \figref{starmf}, respectively.
In the main panels, the solid lines show our fits for the mass functions when
fitting the model as described in \secref{method}. The heavier and lighter
shaded regions around each line show the 68\,\% and 99\,\% confidence
intervals for each mass function. Note that these are only visible at the
highest masses, and at the low-mass end of the R MF. Elsewhere, the formal
statistical uncertainties in the fits are smaller than the width of the lines.

In both cases, the results shown in these figures involve implicit
marginalisation/averaging over all possible permutations of counting each
individual galaxy as being either B- or R-type. It is true that, in general,
we cannot necessarily say with absolute confidence whether any particular
galaxy is R-type or B-type. As discussed in \secref{quality}, however, {\em
this has virtually no impact on the statistical accuracy of our determination
of the mass functions}.

We can quantify this last statement by marginalising over the uncertainties in
the B/R classifications {\em explicitly}. We have done this by recomputing the
values shown by the histograms (discussed further below) for different
MCMC-samplings of the model parameter space, and so over the range of
statistically allowed classifications for each independent galaxy. The
1$\sigma$ error on the R MF is typically 2--5\,\%, and is everywhere $<
8$\,\%; for the B MF, the error is at the 1--2\,\% level for $\log M_* <
10.5$, rising to $\sim 15$\,\% by $\log M_* \approx 11.3$. For comparison, we
can also quantify the statistical uncertainties on the MF determinations shown
in Figures \ref{fig:rfmf} and \ref{fig:starmf}, via bootstrap resampling. Even
using relatively wide mass bins of 0.10 dex, the statistical uncertainties
from sampling are everywhere at least twice as large as those from
uncertainties in the classifications, and are more typically larger by a
factor of $\gtrsim 4$.

Recall from \secref{description} that, within the modelling, each of the B and
R MFs is parametrically described as the sum of two Schechter functions. This
is illustrated in the insets to each of Figures \ref{fig:rfmf} and
\ref{fig:starmf}, which show the individual Schechter components of the B and
R MFs. Note that while the values of these individual components are generally
not well constrained, the uncertainties or degeneracies in the separate
components are largely irrelevant to the results. The uncertainties in the
primary and secondary Schechter functions are so strongly covariant that they
leave the overall MF essentially unchanged. In this sense, the parameters that
describe the individual Schechter components can be treated as `nuisance
parameters' to be marginalised over in order to determine the values of the
MFs for the two populations at any given mass (see also footnote 10.)

While the secondary Schechter component for the R MF is constrained to be
negligibly small for all $\log M_* \gtrsim 9.3$, both fits show an excess
above a simple Schechter function for $\log M_* \lesssim 9$. This is more
pronounced in \figref{starmf} than in \figref{rfmf}. 

We do not consider this (mildly pathological) behaviour robust, for a number
of reasons. First, as discussed in \secref{modelsel}, there is not a clear
statistical need for a second Schechter function to describe the R MF. Second,
as discussed in \secref{wrong}, based on the binned colour distributions in
\figref{stardist}, it is not clear whether or not our two-population model
really provides a good description of the data at these low masses. Third, we
note that this behaviour is not clearly seen in the red histograms, which are
relatively flat across the range $8.7 < \log M_* \lesssim 9.5$.

This last point requires some explanation. The blue and red histograms shown
have been derived from the data, by weighting the contribution of each datum
by both $1/V\max$, and also by the relative likelihood of each galaxy having
been drawn from the B or R population---that is, by either $W\blue$ or
$W\red$, defined as per \eqref{classes}. In this sense, rather than the curves
being fit to the histograms, the histograms are in fact {\em derived based on
the fits}. In general, there is very good agreement between the fit MFs and
these weighted mass functions for the data. But it is important to recognise
that this agreement is not strictly by construction.

The difference between the two sets of the results is subtle, but important.
As a simple example to illustrate this point, imagine if we had only used a
single Schechter function to describe both the B and the R populations. Given
that the B MF is demonstrably non-Schechter (the two components are not
completely degenerate), we would not be using `the right' description of the B
MF. But given how slight the deviations are from a pure Schechter function, we
would not see very large differences in the values of the $W\red$s or
$W\blue$s. In this case, the histograms in these figures would change hardly
at all: they would still show the same non-Schechter features as are seen in
\figref{rfmf} or \figref{starmf}. (In fact, when we do this experiment, the
values of the histograms change by $\lesssim 5$\,\%.) In this way, the close
agreement between the curves and the histograms thus provides a useful {\em
consistency check}, inasmuch as it shows that the fit CMRs and MFs do provide
a reasonable {\em and self-consistent} description of the data.

In fact, for this reason, we favour these empirical MF measurements over the
fits; it is these histograms that should be taken as `the' results of our
analysis. (The exception is at the highest masses, where the histograms are
prone to the Eddington bias in the normal way.) The fits assume that we have
used {\em the right} parameterisation for our model. Instead, the weightings
that go into the histograms rely only our having chosen {\em a good}
parameterisation---or at least a parameterisation that is {\em good enough} to
derive reasonable classifications. That is, the histograms have a weaker
dependence on the precise parameterisation used to construct the model.

\section{Results III.\ --- The essential characteristics of the developed and
developing populations, as seen in the colour--mass diagram}
\label{ch:observations}

First and foremost, let us stress what we do {\em not} observe. Virtually
nowhere are the colour distributions of galaxies {\em at fixed mass} observed
to be `bimodal' in the strict definition of the word. Almost none of the
distributions in either \figref{rfdist} or \figref{stardist} have two distinct
peaks. With the exception of the limited range $\log M_* \sim 10.3$, and even
then only in the $\starcol$ CMD, there is substantial overlap between the two
galaxy populations.

While our R population fits do extract a distinct and relatively narrow `red
sequence', particularly over the mass range $10 \lesssim \log M_* \lesssim
11$, we do not see this continuing down to lower masses in the either the
$(g-i)$ or the $\starcol$ CMD. There are essentially no galaxies with $(g-i)$
or $\starcol$ colours that are consistent with `red and dead' with stellar
masses $\lesssim 10$.

Instead, the $\starcol$ CMR for the R population becomes progressively bluer
towards lower masses. Indeed, it becomes highly problematic to distinguish two
separate populations, in either the $(g-i)$ or the $\starcol$ CMD, for masses
below $\log M_* \sim 9.3$ (see \secref{wrong}). In other words, the `red
sequence' dissolves into obscurity for $\log M_* \lesssim 9.7$.

Further, we see no clear evidence for a low-mass upturn to the MF for the R
population of the kind discussed by \citet{PengLilly} (see in particular
\figref{rfdist}). We do not believe that these results can be easily explained
by mass incompleteness, for the reasons given in \secref{vmax} and
\secref{vmaxcheck}. Nor do we believe that these results can be easily
explained by outliers or otherwise `bad' data, for the reasons given in
\secref{dist} and \secref{cmrs}.

That said, at least over the range $9.7 \lesssim \log M_* \lesssim 11$, the R
and B populations can be seen in both \figref{stardist} and \figref{starcmr}
to be remarkably well separated in the $\starcol$ CMD. For $\log M_* \lesssim
11$, the fits can be seen to provide an excellent description of what has been
dubbed the `blue cloud' in the effective, restframe CMD, and what we see as a
more uniform `blue sequence' in the dust-corrected, intrinsic stellar CMD.

As a corollary to this observation, we note that, based on either
\figref{rfdist} or \figref{stardist}, there is no obvious need for the
inclusion of a distinct `green', intermediate or transition population. The
data are extremely well described by the double Gaussian model.

Given all of the above, as we describe the basic properties of the bivariate
$(g-i)$-- and $\starcol$--$M_*$ distributions in this Section, we will
relax---but not completely abandon---our self-imposed prohibition against
using the terms `blue' and `red' in connection with our B- and R-population
fits. However, we will limit ourselves to using the terms `blue' and `red' to
those regimes where the B and R population fits can be directly related to the
empirically and astrophysically sensible blue and red sequences described
above (with all the appropriate caveats).

With this as introduction, we make the following qualitative observations
about the bivariate colour--mass distributions from our fits to the R and B
populations.

1.)\ The MF for the R-type population is relatively constant ($\Phi\red
\approx 1$--$2\times 10^{-3}$ Mpc$^{-3}$ dex$^{-1}$) for $\log M_* \lesssim
10.5$. There is the possibility of a slight upturn to the R MF \citep[albeit
at a much lower level than that reported by][see \secref{peng}]{PengLilly},
but we do not consider this result to be robust. The very smooth decline in
the {\em relative} numbers of R-type galaxies---\ie the R-type
fraction---towards lower masses suggests that mass is not {\em the} critical
parameter for determining which population a galaxy is a member of: mass is
not a good predictor of B/R-ness. That is, even though more massive galaxies
are more likely to have `old' stellar populations, quenching cannot be
(uniquely) associated with mass.

2.)\ With regards to the non-Schechter features in the overall galaxy MF,
there are very slight but statistically significant deviations from a simple
Schechter function in the B MF. Specifically, the fits suggest a slight
deficit of galaxies with $\log M_* \sim 10.0$--10.3, which coincides with the
apparent `dip' in the overall MF. Below this mass, the upturn in the total MF
is clearly associated with B-type, rather than R-type galaxies.

3.)\ At the very highest masses, the $(g-i)$ and $\starcol$ colours of the R
population are consistent with `red and dead' stellar populations. For $\log
M_* \lesssim 10.5$, however, the slope of the $\starcol$ CMR implies that
R-type galaxies at these intermediate-to-low masses have relatively younger
luminosity-weighted mean stellar ages and higher dust contents than their
higher mass cousins. Taken at face value, this would imply that the $\log M_*
\lesssim 10.5$ R population has not yet evolved into fully fledged `red and
dead' galaxies. Certainly we can say that, even within the R population, there
are very few $\log M_* \lesssim 10$ galaxies with genuinely `red and dead'
stellar populations (see also \secref{howelse}).

4.)\ At high masses ($\log M_* \gtrsim 10.3$), the scatter around the CMR for
the high mass R population tapers down to become small: $\lesssim 0.03$ mag in
either $(g-i)$ or $\starcol$. This behaviour can be understood in terms of
mergers among `red and dead' galaxies \citep[see][]{Skelton2009}: by the
central limit theorem, the mixing of stellar populations in the individual
merger products leads to convergence towards the mean colour for the
population as a whole. It is interesting that the flattening and focussing of
the R CMR apparently begin at around $\log M_* \approx 10.3$, whereas major
mergers are thought to be most prevalent at slightly higher masses
\citep[$\sim \log M^\dagger \approx 10.7$); see \eg][]{Robotham2014}.

5.)\ At intermediate-to-low masses ($\log M_* \lesssim 10.5$), the scatter in
the $\starcol$ CMR for the R-population is relatively small and relatively
uniform. (Note that this qualitative statement is at worst weakly dependent on
how well we understand the observational errors on the values of $\starcol$
for individual galaxies.) Coupled with the apparent dearth of genuinely `red
and dead' galaxies with $\log M_* \lesssim 10.5$, this implies that the
ongoing evolution of R-type galaxies must proceed in such a way as to create
or preserve the relation between stellar mass and stellar population. This
would seem to go against the idea that (at least at these masses and
redshifts, and outside of the richest clusters) galaxies move on to the `dead
sequence' rapidly and stochastically.

6.)\ In the $\starcol$ CMD, the CMR for the B population (\ie, the blue
sequence) is relatively shallow and very nearly linear for $\log M_* \lesssim
10$. This implies that the slope seen in the $(g-i)$ CMD reflects greater dust
attenuation in higher mass galaxies, as expected from \figref{colours}.
Further, the relatively shallow slope of the $\starcol$ CMR implies an
approximate self-similarity in the stellar populations of B-type galaxies with
$M_* \lesssim 10$. The relatively small and constant scatter around the
relation in this mass range strengthens this idea. Together, these results
suggest, albeit weakly, that the process of star formation---or at least
stellar assembly---proceeds in a roughly self-similar fashion among the
moderate-to-low mass B population.

7.)\ When we fit to the $(g-i)$ CMD, the CMRs for the B and the R population
appear to converge for $\log M_* \gtrsim 10.8$. This mass range coincides with
where there may be a slight upturn to the B CMR in the $\starcol$ CMD, and
with the knee in the overall field galaxy mass function. While it must be
stressed that this is where the statistics for B-type galaxies becomes poor,
this hints that these very massive B-type galaxies may be qualitatively
different from their lower mass cousins, in that they have rather redder
stellar populations, while still having significant amounts of dust. One
possible interpretation is that these B-type galaxies with $M_* \gtrsim
M^\dagger \approx 10.8$ are well on their way to joining the R population. An
alternative is that some recent event (\eg, a merger event) has briefly
rejuvenated the stellar populations and dust content of these massive
galaxies, perturbing them out of the main R population. Either way, the
implication is that the apparent self-similarity among B-type galaxies breaks
down at the highest masses.

\section{Discussion} \label{ch:discuss}

\subsection{Comparisons with other means of separating `red' from `blue'
galaxies} \label{ch:others}

\begin{figure*} \includegraphics[width=17.6cm]{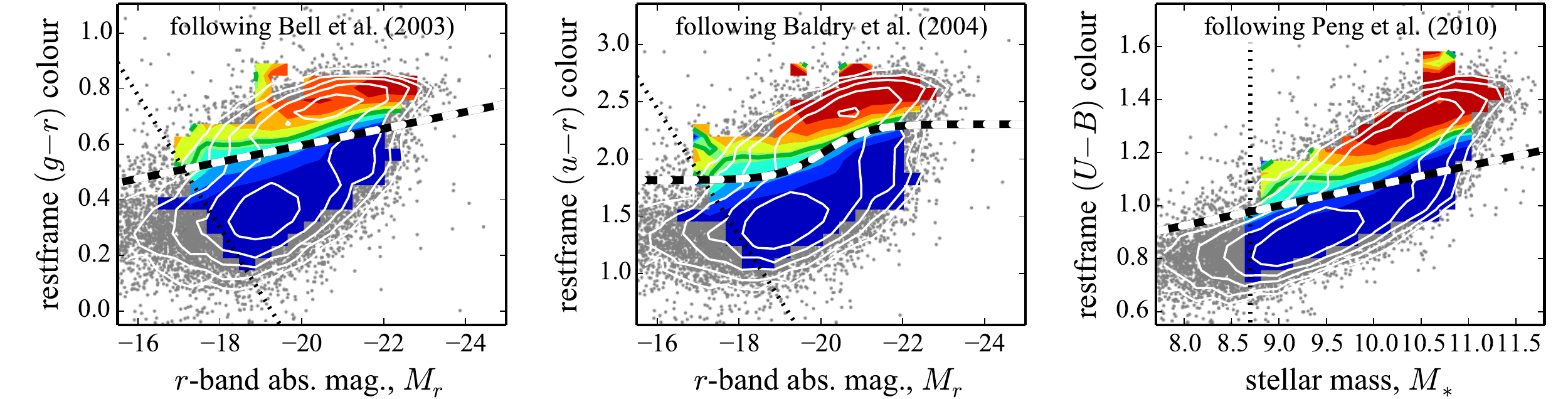}
\includegraphics[width=17.6cm]{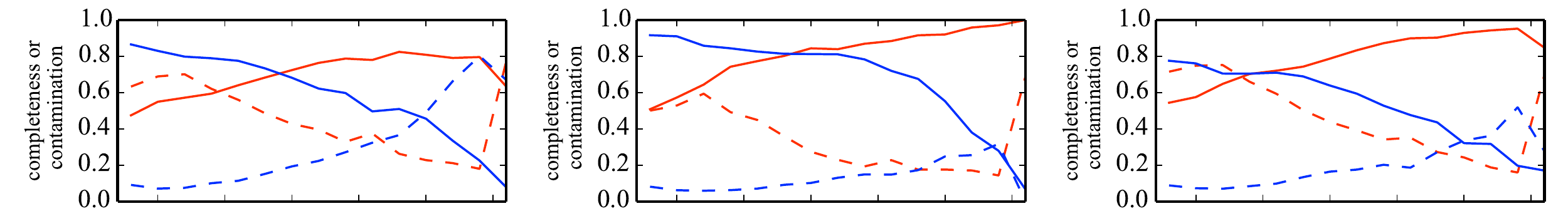}
\includegraphics[width=17.6cm]{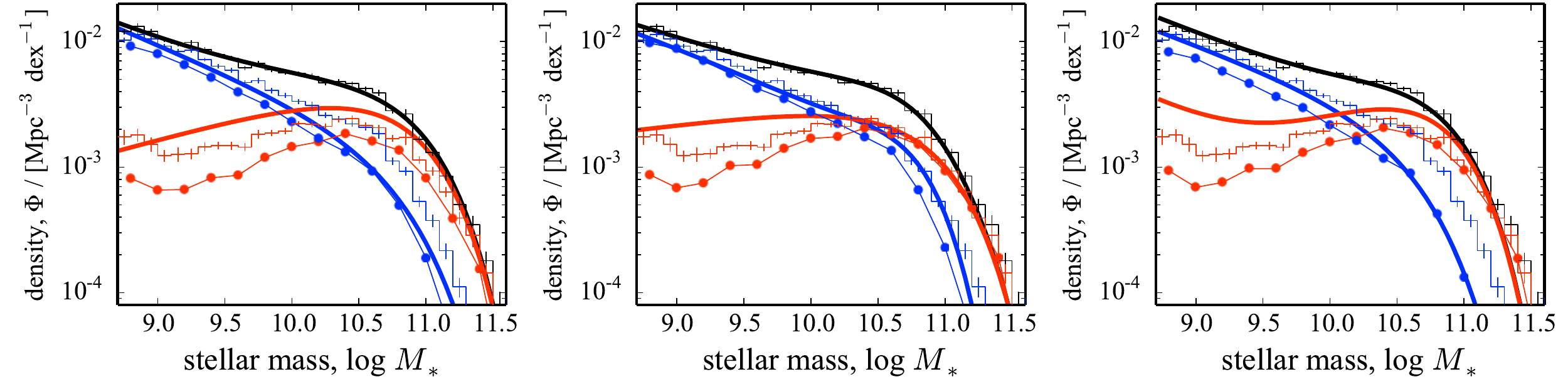}
\caption{Comparisons between our (objective) B/R classifications and the
(largely arbitary) blue/red selections used in previous studies, and how these
differences lead to very different mass functions.--- This Figure is described
and discussed at length in \secref{others}. The message to take from this
Figure is that it is not possible to extract a clean sample of B- or R-type
galaxies using a hard cut in an optical colour--magnitude or colour--mass
diagram. Using the three selections shown, any `red' sample is typically
`contaminated' by B-type galaxies at the $\gtrsim 25$\,\% level; something like
1/3 to 1/2 of all B-type galaxies would be selected as `red'. Further, in
connection with the results of \citet{PengLilly}, we raise the possibility
that the apparent upturn in their `red' MF might be simply explained as an
overly aggressive `red' cut. More than half of all $\log M_* < 9.5$ galaxies
counted as `red' by \citet{PengLilly} are members of the bluer `B-type' galaxy
population. \label{fig:punchline}} \end{figure*}

We have now derived phenomenological descriptions of the B and R populations
(\secref{results}; Fig.s \ref{fig:rfcmr}--\ref{fig:starmf}), and have used
this information to develop an objective, quantitative B/R classification
scheme (\secref{classes}; \figref{rffrac} and \figref{starfrac}). Our final
task is to compare our results to the existing results introduced in
\secref{redness}. Our discussion is based on \textbf{\figref{punchline}}. The
basic point that we are trying to elucidate with this Figure is how our
B/R-type classifications compare to the hard blue/red selection cuts employed
or advocated by \citet{Bell2003}, \citet{Baldry2004}, and \citet{PengLilly},
and how these different operational definitions for `blue' and `red' lead to
very different quantitative results and qualitative conclusions.

Let us begin with a general description of what is shown in the different
panels. As in \figref{others}, the top panels reproduce the colour-magnitude
and colour-mass diagrams used by these authors to define their blue and red
samples. In these panels, the filled coloured contours show how the fraction
of R-type galaxies (as inferred from our fits to the $\starcol$ CMD) varies
across the different colour-magnitude and colour-mass diagrams. These values
have been obtained by 1/$V\max$-weighted averaging of the values of
$W\red\commai \equiv \Ell\red\commai/\Ell\subi$---that is, the objective
B/R-type classifier defined and discussed in \secref{classes}---in bins of
colour and magnitude/mass. In essence, if we were to observe many galaxies
with the same or similar colour and magnitude/mass, the contours thus show
what fraction of these galaxies we would expect to be members of the R
population. Each of these cuts falls slightly blueward of the point where the
R-type fraction is 50\,\%, which is shown by the heavy green line.

Unsurprisingly, there is a rather broad range of colours where there is
substantial overlap between the B- and R-populations ($0.2 \lesssim W\red
\lesssim 0.8$) in each of these diagrams. As we alluded in \secref{cuts}, this
means that the different `blue' and `red' samples selected/defined by these
different hard cuts will comprise a mix of both R- and B-type galaxies, in
different proportions. As a means of selecting R- and B-type galaxies, these
red and blue samples will all be both incomplete and contaminated. Further,
the relative proportions of R and B galaxies in each sample---\ie, the degree
of completeness/contamination---will be a strong function of mass, and will be
sensitive to the precise cut used.

This is quantified in the middle row of \figref{punchline}. We can define a
kind of quasi-completeness, $\mathcal{C}\redsel$, for each of the red
selections by looking at the relative numbers of R-type galaxies that satisfy
each of the different red selections. This is simply derived as
$\mathcal{C}\redsel = \sum ( W\red\commai / V\max\commai ) / \sum
(1/V\max\commai)$, where the sum is over all galaxies satisfying the red
selection, and similarly for $\mathcal{C}\bluesel$. (Note that here, we are
using the subscripts `blue' and `red' to denote the hard-cut samples, as
distinct from our overlapping `B' and `R' populations.) Similarly, we can
define a quasi-reliability, $\mathcal{R}\redsel$, for each red selection as
$\mathcal{R}\redsel = \sum ( W\blue\commai / V\max\commai ) / \sum (
1/V\max\commai )$, and similarly for $\mathcal{R}\bluesel$. The completeness
of each of the three different red and blue selections, so defined, are shown
as a function of mass by the solid lines in the middle panels. We also show
the degree of contamination, ($1-\mathcal{R}$), as the dashed lines.

This same information is re-presented in a different form in the lower panels
of \figref{others}, which also serves to illustrate and elucidate how these
effects lead to qualitatively and quantitatively different determinations of
the red and blue, or R- and B-type, MFs. In each panel, the histograms with
errors show our determinations of the R- and B-type MFs, $\Phi\red$ and
$\Phi\blue$, reproduced from \figref{starmf}; these are the same in each
panel. Then, the solid lines show the fit MFs, $\Phi\redsel$ and
$\Phi\bluesel$, from each of \citet{Bell2003}, \citet{Baldry2004}, and
\citet{PengLilly}. Note that for the purposes of this Figure, we have
renormalised the literature MFs to match the integrated number density for the
GAMA sample, in order to focus on differences in the {\em shapes} of the B/R
and the blue/red MFs.

In each panel of \figref{others}, the points show the absolute numbers of B-
or R-type galaxies that are `correctly' selected as being blue or red; in
other words, these are the galaxies that we can all agree on. These values are
equivalent to $(\mathcal{C}\bluesel \, \Phi\bluesel)$ and $(\mathcal{C}\redsel
\, \Phi\redsel)$. The difference between the points and the histograms thus
reflects the effect of `incompleteness' in the blue/red samples selected using
a hard cut. Similarly, the difference between the points and the curves show
the impact of `contamination' in the hard-cut samples.

Having now described the content of \figref{punchline} in general terms, let
us now turn to discussing the results of each of these three works in the
context of our more sophisticated analysis.

\subsubsection{Comparing our analysis to \citet{Bell2003}} \label{ch:bell}

Looking at the left panels of \figref{others}, it can be seen that our B MF
agrees well with the blue MF from \citet{Bell2003} for $\log M_* \lesssim
9.8$. But this agreement is at least partly coincidental. At these masses,
compared to our B-classifications, the \citet{Bell2003} blue cuts are
80--90\,\% complete, and 80--90\,\% reliable. The red cuts, on the other hand,
are only 50-60\,\% complete, and only 40--60\,\% reliable. It turns out that
these two effects offset one another: the extra 10-20\,\% of B-type galaxies
that are counted as red almost exactly balances the 50\,\% of R-type galaxies
that are missed. In the context of our results, \citet{Bell2003} thus get the
`right' answer for the $\log M_* \lesssim 9.8$ red/blue MFs, but not
necessarily for the `right' reasons: even where our MFs are similar, they are
counting very different galaxies.

At higher masses, there are large discrepancies between our MFs and those of
\citet{Bell2003}. Many of what we call `B-type' galaxies are counted by
\citet{Bell2003} as being `red': the quasi-completeness of the
\citet{Bell2003} selection drops rather smoothly from $\sim$ 75\,\% at $\log
M_* \sim 10$ to $\lesssim 50$\,\% for $\log M_* \gtrsim 11.3$. This means that,
even though the red selection is $\gtrsim 85$\,\% reliable, it is also
contaminated by B-type galaxies at the 20--30\,\% level. The obvious culprit
here is dust, and specifically the dust-induced upturn in the $(g-i)$ CMR for
the B population, as can be seen in \figref{rfcmr} and \figref{rffrac}.

The net effect of these effects is to inflate the red MF by a factor of
$(\mathcal{C}\redsel / \mathcal{R}\redsel) \sim 1.5$, and to depress the blue
MF by a factor of $(\mathcal{C}\bluesel / \mathcal{R}\bluesel) \lesssim 0.6$.
In light of all this, the fact that the \citet{Bell2003} red/blue MFs can each
be well described by a single Schechter function is somewhat coincidental.

\subsubsection{Comparing our analysis to \citet{Baldry2004}} \label{ch:baldry}

The \citet{Baldry2004} determinations of the red and blue MFs are based on
similar assumptions to the ones that we have made. These authors have used an
{\em ad hoc} iterative procedure to fit simultaneously for the centres of and
scatters around the blue/red CMRs, as well as the MFs. Compared to our
analysis, the major differences are: 1.)\ that their fits are to the
$(u-r)$--$M_r$ colour--magnitude diagram, rather than the $\starcol$--$\log
M_*$ CMD; 2.)\ that in their analysis, they bin the data first by magnitude,
and then by colour, and fit to these binned distributions; and 3.)\ that they
then rescale their values for each magnitude bin to stellar mass, using a
simple relation between $(u-r)$ and $M_*/L_r$, whereas we are explicitly
working with $M_*$ estimates from SPS modelling of optical SEDs.

Given the general similarities in, and important differences between, the two
approaches, it is extremely encouraging that there is such good agreement
between the B and R MFs that we derive, and those from \cite{Baldry2004}---at
least at the high-mass end. There are still rather large differences between
the red MF from \citet{Baldry2004} and the R MF that we derive from the
$\starcol$ CMD, but we note that the results based on the $(g-i)$ CMD are in
rather better agreement.

The cut shown in \figref{punchline} has been derived by \citet{Baldry2004} on
the basis of their fits. This cut is designed to maximise the product
($\mathcal{R}\bluesel \mathcal{C}\bluesel \mathcal{R}\redsel
\mathcal{C}\redsel$) in $(u-r)$--$M_r$ space. In this sense, it is designed to
be an optimal hard-cut blue/red selection line (given the specific tanh
parameterisation used). From \figref{punchline}, it can be seen that this
optimal selection is nevertheless imperfect. Using the \citet{Baldry2004}
selection line, $\mathcal{C}\redsel$ drops from $\sim 80$\,\% completeness at
$\log M_* \sim 10$ to $\lesssim 65$\,\% for $\log M_* \gtrsim 10.8$; by the
same token, $\mathcal{R}\bluesel$ is $\lesssim 75$\,\% for all masses.
\citep[These numbers are entirely consistent with the caveats given by
][]{Baldry2004}. This again serves to unambiguously demonstrate the
difficultly of using a hard cut in an optical colour--magnitude or
colour--mass diagram to meaningfully select red/blue galaxy samples.

\subsubsection{Comparing our analysis to \citet{PengLilly}} \label{ch:peng}

Finally, let us turn our attention to the comparison between our results and
those of \citet{PengLilly}. This comparison is particularly interesting and
important, given the elegant `semi-empirical' model for the quenching of star
formation within galaxies that these authors have advanced based on their
results.

The \citet{PengLilly} model predicts a single Schechter function for
star-forming galaxies, and a two-component Schechter function for
quenched/passive galaxies. The model also makes the specific prediction that
the secondary component of the MF for quenched galaxies should have the same
shape as, but a lower normalisation than, the MF for star-forming galaxies.
They show that their observed blue/red MFs, based on data from SDSS and
zCOSMOS, and selected using the cut in the $(U-B)$-$M_B$ colour-magnitude
diagram, can be understood in this way.

From \figref{punchline}, it is clear that the \citet{PengLilly} cut is rather
bluer than others, such that their blue sample is only $\sim 80$\,\% complete
for B-type galaxies, even at the lowest masses. In other words, fully $20$\,\%
of what we classify as `B-type' galaxies would be counted by \citet{PengLilly}
as `red'. By the same token, their red sample is heavily contaminated: more
than half of those galaxies selected as red/quenched by \citet{PengLilly} are
classified by us as being B-type.

This is absolutely crucial in the context of the \citet{PengLilly} model,
which makes the specific prediction that the secondary component of the MF for
quenched galaxies should have the same shape as, but a lower normalisation
than, the MF for star-forming galaxies. It should now be clear how exactly
this kind of behaviour can be produced by using too-blue a cut: once the
red-selected sample becomes highly contaminated, it becomes simply a shadow of
the blue MF. This effect does not rely on, but will be amplified by, the
effects of photometric scatter discussed in \secref{state}.

We thus highlight the possibility that the low mass upturn in the red MF seen
by \citet{PengLilly} and others simply reflects a high degree of
`contamination' of the `red' sample by members of the `blue' population. This
happens where the relative numbers of red galaxies is low, and the hard
red/blue selection limit enters the red wings of the blue distribution. This
interpretation also explains why the red upturn has a similar shape to the
blue MF: because the galaxies responsible for this upturn {\em are} actually
members of the blue population.

Taken at face value, our results would therefore appear to be in conflict with
the quenching model advanced by \citet{PengLilly}. However, this statement is
only true to the extent that our R-type galaxies can be interpreted, in
astrophysical terms, as being genuinely `quenched'. But still, looking at
\figref{punchline} and back to \figref{stardist}, the same criticism could be
levelled at \citet{PengLilly}.

In light of the above, the strongest point that we are prepared to advocate at
this stage is only that the relative shapes and absolute values of the B and R
(or blue and red) MFs depends very sensitively on how the different
populations are defined/selected/classified \citep[see also, \eg][among
others]{Driver2006,Kelvin2014}. Further, we stress the dangers of
over-interpretting the physical significance of any hard red/blue selection,
given the basic fact of overlap between the two R and B populations. This is
precisely why we have set out to derive objective and phenomenological B/R
classifications, which are empirically derived from our statistical
description of the observed data.

\subsection{Comparisons with other approaches of distinguishing galaxies with
`young' and `old' stellar populations} \label{ch:howelse}

\begin{figure*} \includegraphics[width=16.8cm]{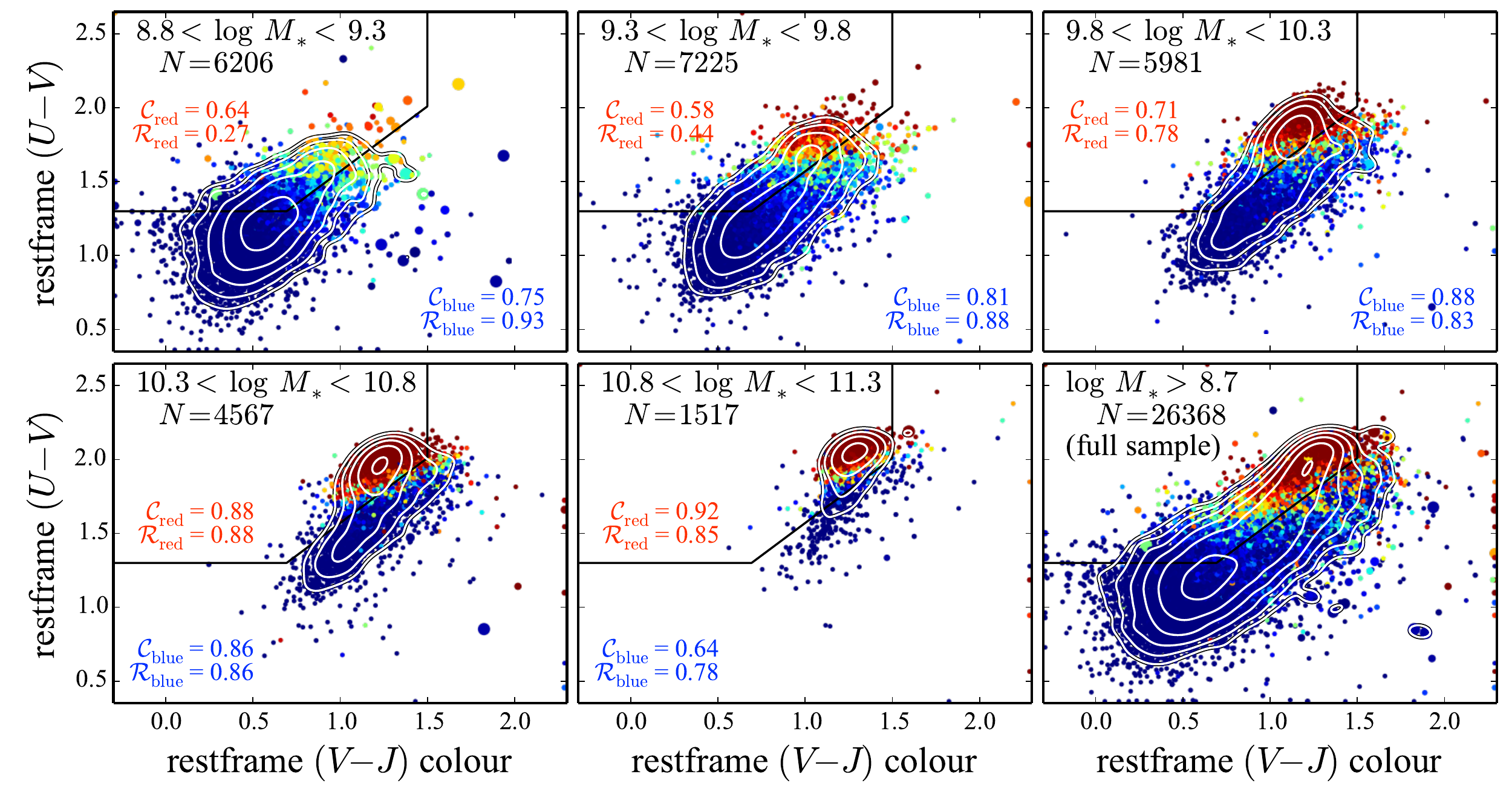}
\caption{The optical-to-NIR SED shapes of B- and R-type galaxies.--- Each
panel shows the restframe {\em UVJ} colour-colour distributions of galaxies in
different mass bins. In each panel, individual galaxies are colour-coded
according to their B/R-type classification, as in \figref{starfrac}. The white
contours show the logarithmic data density. As would be expected if the R-type
galaxies were mostly `red and dead', they fall in a relatively tight clump.
Further, the R-type clump is offset from the red tip of the B-type galaxy
distribution. Note that the derivation of these restframe colours is entirely
independent of the SED fits used to derive the values of $\starcol$, whence
the B/R classifications. The situation of R-type galaxies in the `old' part of
the diagram (shown by the black box) is not by construction. While it is
reassuring that these two complementary approaches qualitatively agree, in
contrast to `hard' colour selections, we stress that our B/R-classification
scheme is {\em objective} (derived from the data, rather than imposed on it),
and {\em quantitative} (explicitly accounting for overlap between two
populations). In this context: consider how many B-type galaxies with $\log
M_* < 9.3$ would satisfy a {\em UVJ} `red' selection, despite clearly being
associated with the main blue clump. Also: consider how the mass function of
these {\em UVJ} `red' galaxies will behave at low masses. \label{fig:nircols}}
\end{figure*}

The basic problem that we have sought to overcome in our analysis is that the
{\em optical} colour distributions of the B and R populations overlap. We are
certainly not the first to apprehend the difficulties in using optical colours
to distinguish `developed' from `developing' galaxies. Our solution to this
problem has been to devise a mixture modelling approach to account for this
fact. Most authors, however, have sought to circumvent these problems, rather
than attempt to confront them head on.

Many others have argued that an optical--NIR colour--colour diagram \citep[see
\eg,][]{Labbe2006, Williams2009, Papovich2012} can be used to isolate
`quiescent' galaxies. The idea is that the dust and age/metallicity vectors
are no longer parallel in a such a diagram, and so `red and dead' galaxies and
`dusty star-formers' are more easily separated. This kind of colour selection
thus acts as a very simple (two-colour) SED-based classification.

Taking this idea to its logical conclusion, other authors have explicitly used
SPS modelling of SEDs to develop a young/old classification scheme. For
example, \citet{Drory2009} have selected `passive' galaxies based on the
best-fit SPS template. After the fact, they then show that these `passive'
galaxies are indeed concentrated in the expected region of the NUV--$r$--$J$
colour--colour diagram.

By considering the intrinsic stellar $\starcol$ color, we have used the
information encoded in the multicolour SEDs to, insofar as is possible, break
the degeneracy between dust and age/metallicity. In this way, we are using
$\starcol$ as a continuous and quantitative diagnostic of a galaxy's stellar
population. Then, we have used a phenomenological, descriptive model of the
bivariate $M_*$--$\starcol$ distribution to derive a quantitative,
probabilistic classification scheme. Next, similar to \citet{Drory2009}, we
can elucidate the nature of---and validate our interpretation of---these
classifications by showing how our classifications map onto other common
diagnostic plots.

\subsubsection{The optical--to--NIR colours of B- and R-type galaxies} \label{ch:nircols}

We explicitly show how our B/R-classification scheme maps onto the {\em UVJ}
diagram in {\bf \figref{nircols}}, which shows the galaxies in our sample,
separated into broad mass bins, and colour-coded according to their objective
B/R-type classification, $W\red$. These plots can be compared to the restframe
{\em UVJ} colour--colour diagram used by, for example, \citet{Williams2009} to and
isolate the `dead sequence' of truly passive galaxies at high redshifts.

For this Figure, the restframe {\em UVJ} photometry have been derived using a
heavily refactored version of the \textsc{InterRest} algorithm for
interpolating restframe photometry \citep{Rudnick2003, interrest}. It is worth
stressing that the derivation of these restframe colours is completely
independent from the SED fits used to derive the values of $M_*$, $(g-i)$,
$\starcol$, etc.

It can be seen that, in broad terms, our $\starcol$-based separation between
B- and R-type galaxies behaves as expected in the {\em UVJ} colour--colour
diagram: the R-type galaxies are tightly clustered, and in a location that is
above and/or to the left of the `blue sequence'. This demonstrates that our
optical SED fits can distinguish between the SEDs of `red and dead' galaxies
and `dusty star formers', and thus that our values of $\starcol$ do provide a
meaningful characterisation of galaxies' stellar populations.

The different panels \figref{nircols} show how our B/R-type classifications is
projected onto the {\em UVJ} diagram at different mass ranges. Looking at the
classifications, the point of transition from mostly B-type (blue points) to
mostly R-type (red points) can be seen to correspond qualitative changes in
the 2D data density that, in simple terms, look like different populations.

\begin{figure*} \includegraphics[width=16.5cm]{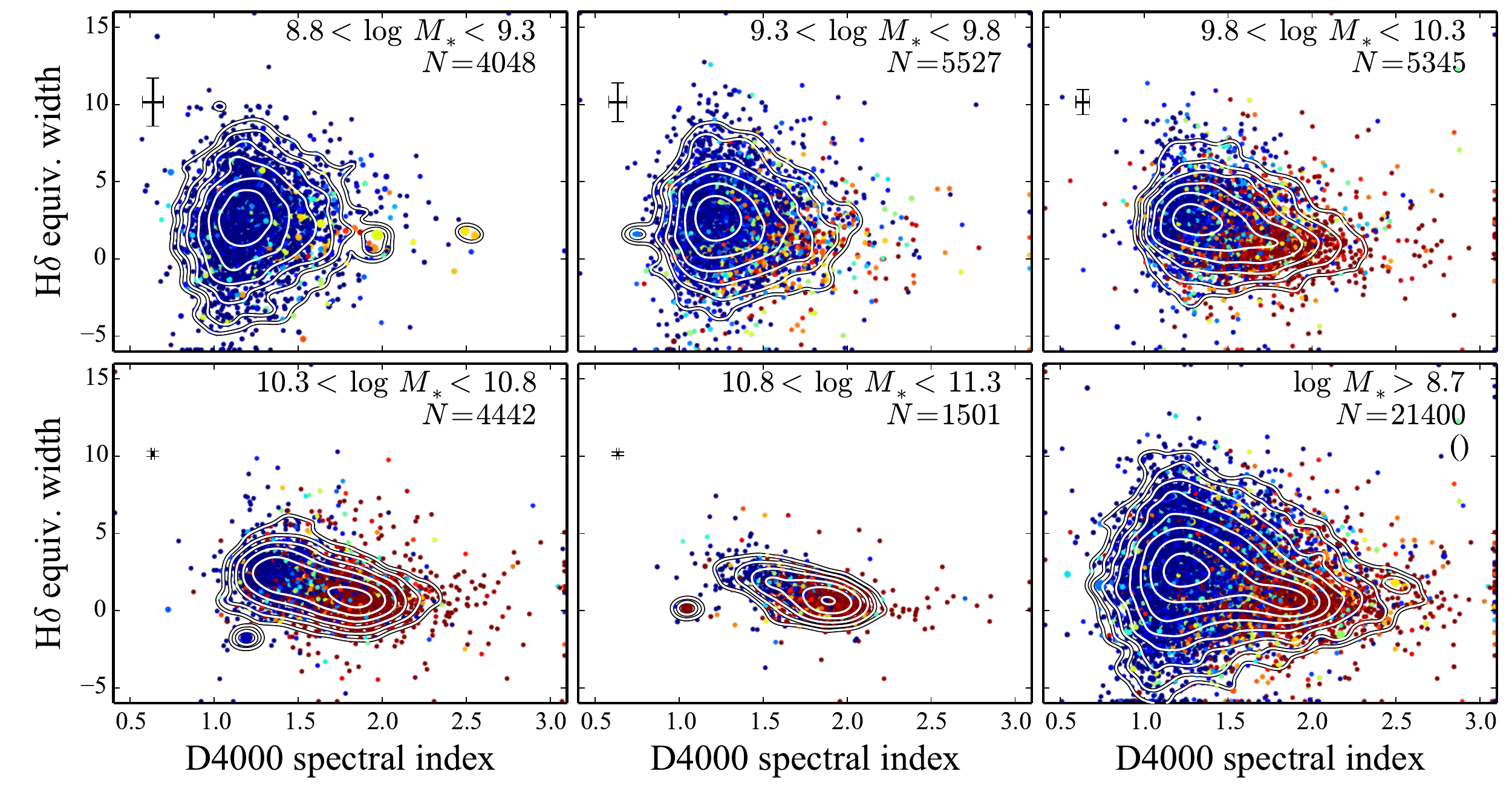}
\caption{The optical spectral shapes of B- and R-type galaxies.--- Each panel
shows the bivariate distribution of 4000 \AA\ break strength, parameterised by
the $D_{4000}$ diagnostic, and the equivalent width of the H$\delta$ line
(positive implies emission). As in \figref{nircols}, each panel is for a
different mass range, and individual points are colour-coded according to
$W_R$. A spectral signal-to-noise cut has been imposed to ensure meaningful
measurements of both $D_{4000}$ and H$\delta$. As for the previous Figure,
these diagnostic plots show how our B/R classifications do meaningfully
encapsulate information about differences in galaxies' stellar populations.
Further, as in \figref{nircols}, the B/R classifications can be seen to
correspond to differences in SED shape that can be broadly understood in terms
of younger/older stellar populations. More specifically, that the light from
R-type galaxies is completely dominated by stars with ages $\gtrsim 1$ Gyr.
\label{fig:HdD4}} \end{figure*}

It can be seen in this Figure that there is some mass-dependent `creep' to how
our B/R classification maps onto the {\em UVJ} diagram. From a
phenomenological point of view, however, the argument could be made that this
is desirable. Certainly, it is clear that the locus of `red' galaxies, however
defined, shifts to redder colours at higher masses. This is just the
colour--mass relation in another guise. Indeed, in the lower mass bins our
B/R-type classifications might provide a better characterisation of the
two-population nature of the data than the standard hard selection box shown.

It is worth explicitly noting that the {\em UVJ} selection box shown would
capture predominately B-type galaxies at the lowest masses: the `reliability'
of a {\em UVJ} selected sample of R-type galaxies would be only $\sim 25$ \%.
In comparison to our determination of the R MF, such a {\em UVJ}-selected
`red' MF would be inflated. Moreover, this would act in such a way that the
inferred `red' MF would have approximately the same low mass slope as the
`blue' one. This is highly significant in the context of the results of
\citet{PengLilly}.

\subsubsection{R-type galaxies have older stellar populations}
\label{ch:HdD4}

\textbf{\figref{HdD4}} shows a similar but different test, this time using
the $D_{4000}$ and H$\delta$ spectral diagnostics. These measurements have
been made using the available GAMA or literature spectra for each galaxy in
our sample. For the purposes of this plot, we have imposed a signal-to-noise
cut across the relevant wavelength range to ensure reasonable data quality.
This effectively introduces a bias against the reddest galaxies, particularly
at low masses. Representative error bars for each bin are shown.

In this diagram, stars with ages $\gtrsim 1$ Gyr will have both $D_{4000}
\gtrsim 1.5$ and H$\delta \lesssim 0$ \citep[see, \eg][among many
others]{Kauffmann2003}. Because these measurements are made over a narrow
wavelength range, they are very weakly sensitive to dust. This diagnostic
diagram shows that the optical spectra of R type galaxies are completely
dominated by old stars, with little to no contribution from stars less than
$\sim 1$ Gyr old. In other words, and modulo the caveat in the previous
paragraph, these results suggest that those galaxies that we have classified
as R-type have seen no significant star formation activity in the past 1 Gyr
or more.

The results presented in this section are intended to demonstrate two
important facts. First, the results shown in each of these two figures
illustrate that---and how---our B-R classifications faithfully and
meaningfully encapsulate differences in galaxies' SED shapes, and hence
stellar populations. In particular, these Figures show how, particularly for
$\log M_* \gtrsim 10$, the R-type population is analogous to commonly used
`red and dead', `passive', or `quenched' selection criteria.

Second, and more perhaps more importantly, we have presented these diagnostic
diagrams as a means to evaluate the meaning of our {\em phenomenological and
empirically derived} B/R classification scheme. Earlier, in \secref{redness},
we have said that our results can be used to gain insight into the process of
quenching, but only to the extent that our operational definitions of `red' or
`R'-ness can be taken to mean `quenched'. In this regard, we can now offer a
more astrophysical characterisation of what distinguishes R- and B-type
galaxies: namely, that the light from R-type galaxies is completely dominated
by stars with ages $\gtrsim 1$ Gyr.

\subsubsection{Understanding the nature of B- and R-type galaxies}

In this work, we have used the intrinsic, dust-corrected $\starcol$ colour as
a diagnostic parameter for a galaxy's stellar population. As mentioned in
\secref{colours}, this quantity is a very good proxy for luminosity-weighted
mean stellar age. We have then used our parametric, descriptive
mixture-modelling of the $\starcol$ CMD to construct an objective B/R-type
classification scheme, and so derive operational definitions for these
designations. In this section, we have used complementary diagnostics to show
that the stellar populations of B- and R-type galaxies do indeed differ, and
in particular that R-type galaxies really do host older stars. This
demonstrates that the $\starcol$ parameter is indeed a useful diagnostic for
distinguishing galaxies based on their stellar populations.

The line of argument that we have now developed is similar to the one in
\citet{Drory2009}. In particular, we have now explicitly demonstrated, in a
mass dependent way, that galaxies classified as R-type or B-type have
different spectral and SED shapes, and thus that the two populations have
qualitatively different stellar populations. (We also show how the kind of
{\em UVJ} cuts commonly used can give very misleading results at low masses.)

The crucial point of difference between our analysis and those mentioned above
is this: we have not explicitly set out to distinguish `young' and `old' (or
`star forming' and `quenched') stellar populations. Instead, we have started
from a phenomenological description of the optical colour--mass diagram, and
used this to disentangle the two apparently distinct populations. Identifying
the distinguishing features of galaxies in these two populations is thus, for
us, a secondary problem. In this sense, the {\em implication} from
\figref{nircols} and \figref{HdD4} is that massive R-type galaxies have
relatively old stellar populations, and little to no dust. This point is
non-trivial, even if it may seem so at first blush.

\subsection{Fallacious arguments against our methods and results}
\label{ch:fallacies}

\subsubsection{40 parameters is too many. Therefore I do not believe your
results.}

In \secref{modelsel}, we have described the various tests that we have done in
an attempt to devise the best and simplest description of the data possible.
This is not to say, however, that our parameterisation is right. Instead, our
claim is only that we do provide a good description of the data, inasmuch as
our model encapsulates all the qualitative features of the bivariate
colour--mass distributions, as illustrated in Fig.s
\ref{fig:rfdist}--\ref{fig:starmf}.

There are some indications that some parameters could be eliminated from the
model without significantly degrading the statistical quality of the fit: most
notably, the data do not clearly demand a two-Schechter description of the R
MF. While there is some ambiguity in whether or not certain parameters ought
be excluded or included, we can say with confidence that we are not grossly
overfitting the data.

There are some parameters for which the inferred values are consistent with
zero (see \figref{mcmc}). To the extent that the model does not make use of
the additional freedoms that these parameters allow, they would appear
unnecessary for the best, simplest possible description of the data. At the
same time, and for the same reason, excluding these parameters would make no
difference to the fits, or to our results. To the extent that the decision as
to whether or not to include these parameters is arbitrary, it is also
unimportant.

\subsubsection{Your results are entirely determined by how you have
parameterised your model. Therefore I do not believe your results.}

As discussed in \secref{objectivity}, this same criticism could be levelled at
any model or modeller. The decision as to how to model one's results is an
inescapable part of any modelling. 

Furthermore, in the specific context of modelling the bimodality in the CMD,
this concern is misplaced. As discussed in \secref{redness} and
\secref{others}, the use of a hard cut to select `red sequence' and `blue
cloud' galaxies directly determines the inferred shapes of the mass functions.
And as discussed in \secref{cuts}, in the absence of some solid theoretical
justification for the specific cuts used, these cuts must be viewed as being
to some extent arbitrary. This criticism is thus much more pertinent for the
vast majority of existing bimodality studies \citep[the notable exception
being][]{Baldry2004}.

Indeed, as laid out in \secref{redness}, the motivation for our analysis is
precisely to address this concern, insofar as is possible. Our primary
motivation for our mixture model of the bivariate colour-mass distributions is
to develop an objective classification scheme, which can be used to
disentangle the apparently distinct populations.

But this is not to deny the truth that the form of the model influences the
outcome. Indeed, as discussed in \secref{wrong}, this idea leads to an
important caveat on our MF determinations for $\log M_* \lesssim 9.3$, where
the application of our model to the observed colour distributions becomes
problematic (see \figref{rfdist} and \figref{stardist}).

\subsubsection{You cannot place strong constraints on the value of parameter
$X$, and anyway, all of your parameters are strongly covariant, if not
completely degenerate. Therefore I do not believe your results.}

By using a Bayesian MCMC sampling scheme to constrain the values of the
parameters in our model---\ie, to do the actual fitting---we fully determine
and account for the covariances between the model parameters. Further, as
discussed in \secref{mfs}, we are not specifically interested in the actual
values of most of the parameters that define the model. For example, the
values of the parameters $\alpha_{\mathrm{B},2}$ and
$M^\dagger_{\mathrm{B},2}$ are important only insofar as they describe the
shape of the MF for the B population. As shown in \figref{rfmf} and
\figref{starmf}, we do not need to be able to uniquely determine the values of
the $\alpha$s or $\phi^\dagger$s in order to obtain very precise
determinations of the MF.

As discussed in \secref{mfs}, `the results' of our analysis should be taken to
be the histograms shown in \figref{rfmf} and \figref{starmf}. In deriving
these results, the role of our parameterised model is only to provide
objective and quantitative B- and R-type classifications, as discussed in
\secref{classes}, which account for the overlap between the two populations in
the CMDs. As discussed in \secref{mfs}, uncertainties in the values of the
different model parameters, propagated through to uncertainties in the
classifications, produce uncertainties in our MFs that are at most $\sim 10$
\%, and are more typically 1--5\,\%. For any given mass in either \figref{rfmf}
or \figref{starmf}, the contribution to the total error budget associated with
the construction of the model is always $\lesssim 40$\,\%, and more typically
$\lesssim 10$\,\%.

\subsubsection{It makes no sense to say that some `red' galaxies are bluer than many `blue' galaxies. Therefore your results are meaningless.}

As discussed in \secref{overlap}, our analysis is predicated on two
assumptions, neither of which are controversial, First, we are assuming that
there are two populations, which are characterised/distinguished by their own
distinct CMRs. Second, we are assuming that there is some scatter around these
relations, to the point where the two populations are observed to overlap.
Taken together, these two assumptions lead to the situation where, in
principle and in practice, the bluest R-type galaxies may have bluer stellar
colours than the reddest B-type galaxies. This is precisely the reason why we
choose to refer to the populations using the more neutral designations `B' and
`R': to try to avoid some of the confusion that comes from the strong
connotations that have come to be associated with these words. Some care is
therefore required in interpreting our results in astrophysical terms (see the
caveats given in \secref{observations}).

Furthermore, we stress that previous results that have used a hard cut
overlook the empirical fact of scatter in the colour distributions for the
`blue' and `red' populations. Some care is also required in interpreting the
results of past bimodality studies in astrophysical terms. We would therefore
invert this criticism to argue that past studies have selected the red tail of
the B population---galaxies that have young, blue stellar populations---and
called these galaxies `quenched' (see \figref{punchline} and in particular
\figref{nircols}).

\subsubsection{Your so-called model contains no physics, and therefore no
information about the process of galaxy formation.}

Again, we make no pretence that our particular parameterisation is in any way
physically meaningful; only that it yields a good description of the
phenomenology of the CMD. Our results thus offer one potential means of
understanding the data, which is the most that any modeller can do. Included
within this is a phenomenological description of how the general population
can be decomposed into two distinct but overlapping populations. Our hope here
is that our empirical results can be used to guide and inform the future
development of genuine semi-analytical and SPH models of galaxy formation in a
cosmological context.

What we have done in this work is to develop a means for objectively and
quantitatively classifying galaxies according to their stellar populations. In
the broader context of this series of papers, our ultimate goal is to use
these classifications to identify the physical differences between B- and
R-type galaxies, with a view to deriving empirical constraints on the physical
processes that act to determine whether any particular galaxy is B- or R-type.
We therefore defer more astrophysically-minded observational studies of the
galaxy bimodality (or bimodalities?) to future works.

\subsubsection{Your R-type galaxies do not conform to existing notions of `red
and dead' or `quenched', and therefore your results tell me nothing about the
different stellar populations and/or star formation histories of `developed'
and `developing' galaxies.}

In the first instance, our goal has been to separate the general galaxy
population into two subpopulations, on the basis of their constituent stellar
populations, and in a mass dependent way. For our analysis of the effective
$(g-i)$ CMD, dust is a confounding factor. For this reason, we have repeated
our analysis looking at intrinsic, dust corrected $\starcol$ colours.

It is true that degeneracies between dust, metallicity, and SFH in such fits
mean that the uncertainties in the values of $A_V$, $Z$, or $\tau$ can be
large, but we maintain that these data are sufficient to make the qualitative
distinction between dusty and old SED shapes. In support of this claim, we
have shown in \secref{howelse} that those galaxies that we classify as being
either B- or R-type occupy different regions of the H$\delta$-D$_{4000}$
spectral line diagnostic diagram (\figref{HdD4}), as well as the restframe
{\em UVJ} colour-colour diagram (\figref{nircols}).

In other words, the B-/R-type classifications really do encapsulate meaningful
information about galaxies' stellar populations. Further, these
classifications closely correspond to the two-population distributions seen in
both diagnostic plots. Conversely, we have also argued the kinds of `red'
selections that are commonly used are a poor proxy for `quenched'.

We have thus accomplished our primary goal of disentangling and characterising
the two apparently distinct populations seen in the CMD. {\em Some} of our
R-type galaxies conform to the usual picture of `red and dead' galaxies, but
we have shown that these galaxies are just the high mass tip of a more
continuous R population.

With our primary goal accomplished, our phenomenological characterisations of
these two populations can then shed light on their astrophysical nature. This
includes the MF for each population, which now represents a target for
cosmological simulations to aim at. This also includes our characterisations
of the $\starcol$ and $(g-i)$ CMRs for each population, which provide
qualitative constraints on the processes of star formation and of star
formation quenching within galaxies.

\subsubsection{Galaxies are complicated, and focussing on only two or three
parameters glosses over all the important details. You are missing the trees
for the forest.}

The process of galaxy formation is complicated, and there are myriad
well-known and studied galaxy types and classes that we have not considered in
our analysis: \eg, radio loud AGN, starbursts, E+A or post-starburst galaxies,
satellites/centrals, interacting and merging galaxies, etc. Some or all of
these processes may play an important role in determining whether a particular
galaxy is R-type or B-type, or conversely, some of these processes may act
exclusively on or within R- or B-type galaxies. In this way, these processes
are presumably also responsible for producing the observed intrinsic scatter
around each of the two CMRs.

By the same token, galaxies are in general multi-component systems. As a
simple example, most massive galaxies have both a disk and a bulge component,
with separate stellar populations and formation mechanisms. For example,
\citet{Driver2006} have argued that galaxies' {\em global} colours are driven
by the mixture of (blue) disk stars and (red) bulge stars, and that the colour
bimodality is thus best understood in terms of the (bimodal) distribution of
bulge-to-disk mass ratios.  

We have deliberately avoided these kinds of questions in this work; our
intention is only to derive an empirical, phenomenological description of the
apparent dichotomy in the stellar populations of field galaxies---that is, the
nature of the R- and B galaxy populations. In doing so, we have derived an
objective, operational definition for these designations. This is a necessary
prerequisite for future studies of the different astrophysical natures of
these two galaxy populations, which we will pursue in future papers.

In light of the above, it is remarkable that this game can be played at all.
Here again, we stress there are no clear signs of an intermediate or
transition `green' population. That the bivariate colour--mass distribution
for the field galaxy population is extremely well described by a simple
two-population model implies that our phenomenological distinction between B-
and R-type galaxies does indeed encapsulate some important astrophysical
differences in the formation histories or evolutionary states of these two
populations---even if we cannot yet articulate what these differences are, or
what the driver for these differences may be.

\section{Summary and Conclusions} \label{ch:summary}

Our particular interest lies in characterising the mass functions (MFs) and
colour-mass relations (CMRs) for the apparently distinct populations seen in
the optical colour-mass diagram (CMD), where the two populations are
distinguished on the basis of galaxies' stellar populations. Our analysis is
based on a sample of $\log M_* > 8.7$ and $z < 0.12$ field galaxies from the
Galaxy And Mass Assembly (GAMA) survey. This sample is properly mass-complete
(volume limited) for $\log M_* \gtrsim 10$; for lower masses, we have used the
standard $1/V\max$ formalism to account for incompleteness (see
\secref{complete}). Note that none of our results or conclusions change if we
limit our analysis to $\log M_* > 9.5$, or to $z < 0.06$.

The immediate motivation for reconsidering this well-studied problem is that,
as discussed in \secref{cuts}, there are quantitative and qualitative
disagreements between the MF determinations that exist in the literature.
First, we have shown that if we analyse our GAMA-sample in the same way as
each of \citet{Bell2003}, \citet{Baldry2004}, and \citet{PengLilly}, we are
able to reproduce each of these authors' SDSS-based results. Then, we argue
that the discrepancies between the results and conclusions of these studies
are due entirely to the different (and most often arbitrary) ways that the
`blue' and `red' galaxy samples have been selected/defined.

Our first and most important conclusion is therefore a qualitative one: that
the largest uncertainty in previous characterisations of the mass functions
for `red' and `blue' galaxies is tied to how these terms are defined. Put
bluntly, the main reason why, say, \citet{PengLilly} see an upturn at the low
end of the `red' MF where, say, \citet{Bell2003} do not is simply because the
\citet{PengLilly} definition of `red' is considerably bluer than the
\citet{Bell2003} one. As a direct consequence, a significant fraction of the
\citet{PengLilly} `red' sample are star forming galaxies with young stellar
populations (see Fig.s \ref{fig:otherdust}, \ref{fig:punchline}, and
\ref{fig:nircols}).

The direct implication is that the power of these results to provide useful
constraints on the process of galaxy evolution is directly limited by the
extent to which the specific `red' and `blue' selections used can be shown to
be astrophysically meaningful. In the absence of convincing arguments in
favour of any one of these selections over the others, important questions
like the mass range over which the galaxy population transitions from mostly
blue to mostly red, or the low mass slopes of the blue and red mass functions,
are left largely unconstrained. In order to address these questions, what is
needed is a well-motivated operational definition for the technical terms
`red' and `blue'. 

To redress this, we have developed a descriptive model for the distribution of
galaxies in CMD, with the specific goal of distinguishing between `developed'
and `developing' galaxies on the basis of their stellar populations. In our
modelling, we treat the observed data distribution as being the sum of two
distinct but overlapping populations. The model also includes a `bad'
component to allow for outliers, catastrophic errors, or otherwise un- or
under-modelled aspects of the observed distributions (see \secref{outliers}
and \secref{cmrs}). The formalism for our modelling, which is developed
pedagogically in \appref{themodel}, is based on the method of (Gaussian)
mixture modelling \citep[see, \eg,][]{HoggBovyLang}.

As outlined in \secref{description}, in its most general form, our descriptive
model is fully defined by 40 parameters. As discussed in \secref{modelsel},
not all of these parameters are strictly necessary for a `good' description of
the data. In particular, it is not clear that the data strictly demand a
second Schechter component to well describe the MF for the R population. As
discussed in \secref{objectivity}, however, where there is ambiguity about
whether or not a parameter is necessary, it will have little to no impact on
the final results. Beyond that, the most that we can say is that we have made
every attempt to ensure that we are not grossly overfitting our data.

In effect, there are two assumptions that underpin our approach. First, we are
assuming that some physical process(es) or hidden parameter(s) act to
determine whether a given galaxy is a member of one or the other population;
that is, {\em we assume that there are two populations, which follow distinct
CMRs}. Then, some secondary process(es) or parameter(s) determines where that
galaxy falls with respect to the main CMR for that population; that is, {\em
we assume that there is some (Gaussian) scatter around each of the two CMRs},
to the point that the two populations are observed to overlap in an optical
CMD (or, indeed, in a {\em UVJ} colour--colour diagram). Neither of these
assumptions ought to be controversial.

Allowing that, at fixed mass, there is some overlap between the `blue' and
`red' colour distributions inescapably implies that some `red' galaxies will
have bluer colours than some `blue' galaxies. In acknowledgement of this
semantic trap, and to avoid some of the confusing connotations associated with
the terms `blue' and `red', we have adopted the more generic designations of
`B' and `R' to describe the two populations seen in the CMD (see
\secref{redness} and \secref{quality}).

That is, rather than considering blue and red {\em galaxies}, we focus on two
galaxy {\em populations}, which we dub B and R. It is not necessarily true
that a particular B- galaxy will be bluer than some other R-galaxy. It is also
not necessarily true that an R-type galaxy can be thought of as being `red and
dead', `early type', `quiescent', etc. It nevertheless remains true that the
distribution of stellar population properties are different for galaxies in
each of the two populations---in particular, at fixed mass, and on average,
galaxies in the B-population have bluer (and so younger) stellar populations
than those in the R-population.

In other words, we are assuming that the phenomenological separation of the
general galaxy population into B- and R-type subpopulations somehow reflects a
qualitative binarity in the formation histories or evolutionary states of
galaxies---but it remains to describe and explain precisely how and why this
is the case. By modelling these populations, we can actually {\em derive from
the data} phenomenological working definitions for the terms `B-' and `R-type'
(see \secref{classes}, \figref{rffrac}, and \figref{starfrac}). What is more,
the classifications are {\em objective}, insofar as objectivity is possible
(see \secref{objectivity}).

Bearing in mind the caveats given above, we go on to describe the basic
characteristics of the B- and R-population in terms of both the effective
$(g-i)$ CMD (\figref{rfcmr}), as well as the intrinsic, dust-corrected colour,
$\starcol$ CMD (\figref{starcmr}). That the members of these two populations
comprise genuinely different stellar populations has been demonstrated in
\secref{howelse}, where we show that R- and B-type galaxies occupy distinct
regions of the {\em UVJ} and $D_{4000}$--H$\delta$ diagnostic diagrams
(\figref{nircols} and \figref{HdD4}).

We find that the intrinsic $\starcol$ CMR for the B population is both
considerably tighter and more linear than the effective $(g-i)$ CMR. This
implies that the both the upturn in the $(g-i)$ CMR at $\log M_* \sim 9.5$ and
the relatively large scatter around the $(g-i)$ CMR at all masses is driven by
the distribution of dust properties, rather than differences in the stellar
populations of these galaxies. At least for $\log M_* \lesssim 10.8$, the
relative flatness of the $\starcol$ CMR for B-type galaxies also implies that,
in terms of stellar colours, the B population is relatively homogenous.

This behaviour changes, however, for $\log M_* \gtrsim 10.8$. At this mass
range, the B population appears to converge with the red sequence in the
$(g-i)$ CMD. This is also where the MF for B galaxies drops off very rapidly.
In other words, there are few if any galaxies with $\log M_* \gtrsim 10.8$
with young (B-type) stellar populations; this represents the top end of the
blue sequence.

For the R population, if nothing else, we can say with confidence that there
are essentially no $\log M_* \lesssim 9$ field galaxies with stellar
populations that are the same as or similar to those of the genuinely `red and
dead' galaxies seen at the highest masses. Instead, at least for $\log M_*
\gtrsim 9.8$, we see a gradual trend whereby less massive galaxies have
progressively bluer stellar colours (\ie, younger luminosity-weighted mean
stellar ages and/or lower mean stellar metallicities) than their more massive
cousins.

The relatively small dispersion in $\starcol$ CMR for the R-population
suggests that, at fixed mass, R-type galaxies have rather similar stellar
populations, and hence similar star formation/stellar assembly histories. At
the same time, the slope of the $\starcol$ CMR for R-type galaxies shows that
there are differences in the evolutionary histories of different members of
the R populations with different present day masses. Said another way, the
evolution of individual R-type galaxies in the field proceeds in such a way as
to create or preserve the correlation between the stellar mass on the one
hand, and stellar population on the other.

These observations---the relative homogeneity of the stellar colours of B-type
galaxies, and the correlation between stellar mass and stellar population for
R-type galaxies---run counter to prevalent notions of the `red sequence' and
the `blue cloud'. They also provide meaningful targets for theoretical models
of galaxy formation and evolution to aim for.

For $\log M_* \lesssim 9.5$, it becomes increasingly difficult to meaningfully
and robustly distinguish two separate B and R populations from the joint
$(g-i)$ and $\starcol$ colour-mass distributions. Indeed, looking {\em only}
at the observed colour distributions in bins of mass (\figref{rfdist} and
\figref{stardist}), there is little to no clear evidence for a distinct R
population at $\log M_* \lesssim 9$.

Our claim is therefore that, below $\log M* \sim 9.3$, we are no longer able
to unambiguously identify a distinct R-component to the general field
population: at least for field galaxies, below $\log M_* \sim 9.3$, the R
population dissolves into obscurity. At these low masses, our characterisation
of the R population may be better interpreted as describing the degree of
asymmetry in the observed colour distributions, and our MF for the R
population may be taken as an {\em upper limit} on the number of galaxies that
lie outside the main colour distribution for `normal' B-type galaxies (see
\secref{wrong}).

We note in passing that the data do not clearly demand an intermediate `green'
population: the data are very well described by a two-population model (see
\figref{rfdist} and \figref{stardist}). Taking this remarkable observation at
face value, this might imply that movement between populations is quick
\citep[see, \eg,][]{Bell2004}. Alternatively, it could imply that movement
{\em within} a population (for example, by satellite accretion/minor mergers)
is much faster than movement {\em between} populations: a galaxy may be
`passed' from one population to the other, and then have its colour rapidly
re-randomised according to the `normal' colour distribution for its new
population.

We also note that optically-identified AGN reside exclusively within the blue
sequence defined by the B population (\figref{colours}). That is, in terms of
their stellar populations, AGN hosts do not clearly differ from `ordinary'
star forming galaxies. Certainly they do not represent a transition population
that is intermediate between the B and R populations.

The MFs that we derive for the R and B populations are rather different to
those for `red' and `blue' galaxies presented by other authors (see
\figref{punchline}). The reasons for these differences are discussed first in
\secref{state}, and then again in close detail in \secref{others}. 

In particular, we find considerably more `B' galaxies with $\log M_* \gtrsim
10$, and a much more abrupt drop-off in the B MF at $\log M_* \gtrsim 10.8$.
Whereas others put the crossover mass, where the two MFs intersect, at $\log
M_* \approx 10.0$--10.3, we find it to be closer to 10.5. Further, our MF for
R-type galaxies is also considerably lower at low masses than those for `red'
galaxies from the literature. {\em If} these red or R-type MFs can be used to
probe the process of quenching, our results would imply that quenching is less
prevalent---or equivalently, that massive galaxies continue forming new stars
for longer---than has been previously thought.

But this leads to the final and most important caveat: as discussed at length
in \secref{redness} and \secref{quality}, there is real danger in
inappropriately reifying the terms `B-' and `R-type'---or, equally, the terms
`blue' or `red'. These terms must be understood as qualitative,
phenomenological designations for galaxy {\em populations}. In particular, one
should not conflate the terms `R' or `red' with the term `quenched'. This is
especially true at low masses, where the stellar populations of `red' or
R-type galaxies are rather different to those found at higher masses.

As discussed throughout this paper, it is necessarily and inescapably true
that our results depend on the choices made in the construction of our
model---we can only answer the question that we have asked. In particular, the
decision to use Gaussians to describe the colour distributions at fixed mass
is primarily one of convenience. Our justification for these decisions is
ultimately empirical---the proof is in the pudding. We have thus shown that
our descriptive modelling provides one potential way of understanding the
data. The fact nevertheless remains that using a different operational
definition for the terms `red' and `blue' will lead to quantitatively and
perhaps qualitatively different results.

We do not pretend to be immune to these difficulties. Instead, our goal is to
highlight the importance of these issues, and to make it clear that the same
criticisms can and should be levelled at any study of the `red' and `blue'
galaxy populations. But then again, we have also validated our results by
showing how the phenomenological B/R classifications that come from our model
do indeed select galaxies with qualitatively different stellar populations.

This point is particularly significant in connection with the elegant
semi-empirical model for quenching presented by \citet{PengLilly}, which
hinges on the presence or absence of an upturn to the MF for quenched galaxies
at $\log M_* \lesssim 9.5$. Looking at the distributions shown in
\figref{rfdist} or \figref{stardist}, (or even the {\em UVJ} diagrams in
\figref{nircols}) it should be immediately obvious how problematic it is to
meaningfully distinguish a separate `red' or `quenched' or even `R'-type
population at these low masses (see \secref{wrong}). Further, it must be
recognised how a sample of `old' or `quenched' galaxies selected using a hard
cut in any of these diagrams---including the {\em UVJ} diagram---will be
dominated at low masses by spillover from the red tail of the distribution of
otherwise normal, young, star-forming, B-type galaxies. The critical issue is
therefore whether or in what sense these galaxies can be thought of as a
distinct, coherent population of quenched galaxies.

This paper is the first in a series in which we explore the different aspects
or manifestations of the bimodality (or bimodalities) in galaxy properties. In
Papers II and III, we will perform a similar analysis to explore bimodalities
in terms of line emission properties, and in terms of \Sersic -fit structural
parameters. Taken together, these three Papers will provide a basis for
robustly and objectively classifying galaxies according to the stellar
populations, present day star formation and/or AGN activity, and structure. In
future works, we will go on to use these results to explore the natures of,
and interrelations between, these bimodalities.

\section*{Acknowledgements}

It is our genuine pleasure to thank the referee, Eric Bell, for his thorough
and thoughtful feedback, which greatly improved the clarity and strength of
the arguments presented. Part of this work was made possible through use of
the edward HPC cluster, maintained by Information Technology Services Research
at the University of Melbourne. PN acknowledges the support of the Royal
Society through the award of a University Research Fellowship and the European
Research Council, through receipt of a Starting Grant (DEGAS-259586). MG is
supported by a European Research Council Starting Grant (DEGAS-259586). GAMA
is a joint European-Australasian project based around a spectroscopic campaign
using the Anglo-Australian Telescope. The GAMA input catalogue is based on
data taken from the Sloan Digital Sky Survey and the UKIRT Infrared Deep Sky
Survey. Complementary imaging of the GAMA regions is being obtained by a
number of independent survey programs including GALEX MIS, VST KiDS, VISTA
VIKING, WISE, Herschel-ATLAS, GMRT and ASKAP providing UV to radio coverage.
GAMA is funded by the STFC (UK), the ARC (Australia), the AAO, and the
participating institutions. The GAMA website is
\texttt{http://www.gama-survey.org/}.

\setlength{\bibhang}{2.0em}
\setlength\labelwidth{0.0em}

\appendix

\section{Developing an Objective Red/Blue Classification Scheme} \label{app:themodel}

\subsection{Introductory Statement of the Problem}

Our overarching goal is to derive an empirical, phenomenological description
of the `bimodality' in the galaxy population, as seen in the $(g-i)$ and
$\starcol$ CMDs. This apparently simple project is made problematic by the
fact that the apparently distinct `red' and `blue' populations are seen to
overlap in both of these CMDs. Accordingly, we want to avoid imposing some
arbitrary `hard' cut to distinguish `red' from `blue'; instead, we want to
develop the means to simultaneously and flexibly describing these two distinct
but overlapping populations. The solution to this modelling problem---the
method of mixture modelling---can alternatively be viewed as the development
of an {\em empirical} and {\em objective} `red'/`blue' classification scheme,
based on the likelihood that a given data point has been drawn from one or the
other population.

To reduce the problem to the simplest possible terms, what we want to do is to
construct a parametric model that describes the distribution of data points in
the CMDs. This model will need to have two components---one for each of the
`red' and `blue' populations---each of the form $p(x,~y)$, where $x = \log
M_*$ and $y = (g-i)$ or $\starcol$. The model itself will be fully described
by a set of parameters, $\set{P}$. The observed data are then considered as
having been randomly drawn from---generated by---this modelled density
distribution, and observed subject to the appropriate measurement errors or
uncertainties. 

The generative model is thus explicitly intended to describe the (scalar)
likelihood, $\Ell\subi$, of observing any given (vector) datapoint,
$\vec{x}\subi = (x_i, y_i)$, and the associated uncertainties, which are
assumed to be Gaussian, and are for now generically represented as $\sigma_i$.
It is crucial to recognise that the generative model is only
calculable---indeed, is only defined---given or assuming a particular set of
trial values for each and every of the parameters in $\set{P}$. That is,
$\Ell\subi$ describes the likelihood of observing the datapoint $\vec{x}\subi$
with formal observational uncertainties $\sigma_i$, given or assuming a
specific set of values for $\set{P}$. To reflect this fact, the likelihood
function is represented as $\Ell\subi(\vec{x}\subi, \sigma\subi|\set{P})$.

The primary goal is thus to use the full observed dataset,
$\set{X}=\{\vec{x}\subi\}$, along with the associated set of observational
uncertainties, $\set{S}=\{\sigma\subi\}$, to constrain the `true' values of
the parameters in $\set{P}$---that is, we aim to generate the posterior
probability distribution function (PDF) for the values of the parameters
$\set{P}$, given that we have observed our data, $\prob(\set{P}|\set{X},
\set{S})$.

Let us begin our discussion by considering only one population, and taking the
simplest possible relation between $x$ and $y$: a perfect line,
$\ell(\vec{x}|m, c) \equiv (m ~x + c - y) = 0$. In this case, the model
parameter set $\set{P}$ just comprises the slope and normalisation of the
line; \ie, $\set{P}=\{m,c\}$.

The traditional---but overly simple and somewhat na\"ive---approach to this
kind of fitting problem is to use the method of weighted $\chi^2$
minimisation, which is described immediately below, to fit independently fit
for a linear relation between $y$ and $x$. Among extragalactic astronomers,
this approach is also frequently referred to as `maximum likelihood'. However,
as is well known and accepted in many other disciplines (including particle
physics and cosmology), and as we shall endeavour to make clear in the next
few sections, the weighted $\chi^2$ fitting formalism is only one instance of
the more general class of maximum likelihood fits that are possible. Moreover,
the very restrictive set of assumptions that underpin the weighted $\chi^2$
formalism means that it is only appropriate to use in very specific---and, in
astronomy, very rare---situations.

In any case, following the weighted $\chi^2$ minimisation formalism, including
the assumption of Gaussian errors, the values of the $\Ell_i$s are computed
as:
\begin{equation} \label{eq:chi2} \begin{split}
	\Ell_i( x_i, y_i, \sigma_{y,i}|m, c) 
		& = \gauss_1( \ell\subi, \sigma_{y,i}^2 ) \\
		& \equiv \frac{1}{\sqrt{2 \pi \sigma_{y,i}^2}}
		 \exp \left[ {-1 \over 2} \frac{\ell_i^2}{\sigma_{y,i}^2} \right] ~ , 
\end{split} \end{equation}
where we have abbreviated $\ell(\vec{x}\subi| m, c) = (m \, x\subi + c -
y\subi$) as $\ell\subi$, and $\sigma_{y,i}$ is the observational uncertainty
associated with the measurement of $y_i$. We have also used this Equation to
introduce $\gauss_1(y', \sigma^2)$ as our shorthand for a 1D Gaussian
distribution, centred on $y' = 0$ and with variance $\sigma^2$, and evaluated
at the location $y'$. Note that in what follows, $\gauss_1$ should always be
understood to be integral normalised to unity.

The global likelihood, $\Ell$, of observing the full dataset is then given by
the product of all the individual $\Ell\subi$s. In practice, it is more
convenient to work in terms of logarithms, so that: 
\begin{equation} \label{eq:maxlike} 
    \ln \Ell \big( \set{X}, \set{S} | \set{P} \big) 
    = \sum_i \ln \Ell_i \big(x_i, y_i, \sigma_{y,i} | m, c \big) ~ .
\end{equation}
Within the traditional weighted $\chi^2$ formalism,
$\ell\subi^2/\sigma_{y,i}^2$ is written as $w_i \chi\subi^2$, where
$\chi\subi^2 = \ell\subi^2$ and $w\subi \propto 1/\sigma_{y,i}^2$. Note that a
`least squares' fit corresponds to the case where all the $w\subi$s, and hence
all of the $\sigma_{y,i}$s, have the same, constant value. With these
definitions, $\ln \Ell$ can be seen to be equal to $-\frac{1}{2} \sum w\subi
\chi\subi^2$ minus a constant.

The parameter values that minimise the sum of $\chi\subi^2$ can thus be seen
to also maximise $\Ell$. This is the statistical justification for the
otherwise {\em purely geometric} rationale that underpins the $\chi^2$
formalism. A weighted-$\chi^2$ minimisation is just a special case of a
maximum likelihood fit.

The `miracle' \citep{HoggBovyLang} of the weighted $\chi^2$ formalism is that
maximisation condition $\partial \ln \Ell / \partial m = 0$ can be solved
analytically. Since we have defined $w\subi \, \chi\subi^2 \propto (\ell\subi
/ \sigma\subi)^2 = (m \, x\subi + c - y\subi)^2 / \sigma_{y,i}^2$, each of the
individual $\ln \Ell\subi$s, and thus the summed $\ln \Ell$, can be seen to be
quadratic in $m$. Scaling the $w\subi$s so that $\sum w\subi = 1$, the result
is:
\begin{equation}
	m_{\mathrm{min\, \chi^2}} = 
		\frac{ \sum \left( w\subi x\subi y\subi \right)
		- \left( \sum w\subi x\subi \right) \left( \sum w\subi y\subi \right)}
			{\sum \left( w\subi x\subi^2 \right) 
				- \left( \sum w\subi x\subi \right)^2 } ~ .
\end{equation}
Note how similar this expression is to Cov$(x\subi,~y\subi)/$Var$(x\subi)$.
For our purposes---fitting the red and blue CMRs---this procedure would be
done separately and independently for the red and blue subpopulations, using
some prior distinction to separate the two. 

There are a number of important and implicit assumptions involved in writing
\eqref{chi2} and \eqref{maxlike}, each of which make the traditional $\chi^2$
approach unsuitable for our purposes. 1.)\ Any and all uncertainties in
measured values of the $x_i$s are ignored---$\sigma_{x,i}$ does not appear in
\eqref{chi2}. There are two facets to the embedded assumption 2.)\ that the
distribution of datapoints is well described by a perfect line. First, 2a.)\
all values of $x$ are considered equally likely; no attempt is made to account
for the underlying distribution function for $x$ (\ie, mass). Second, 2b.)\
the relation between $x$ and $y$ is assumed to be infinitely narrow; no
allowance is made for there being an intrinsic (astrophysical) scatter in the
relation between $x$ and $y$. 3.)\ No allowance is made for outliers,
catastrophic measurement errors, or otherwise `bad' data; the results of the
fit are, in both principle and practice, sensitive to data that contain little
or no useful information. And finally, 4.)\ any two or more populations must
be fit independently using an {\em a priori} distinction; this approach cannot
deal with multiple overlapping populations.

Our task in this Appendix is therefore to develop a better descriptive model
for the data density distribution in the CMD that can overcome the
considerable limitations of the traditional approach, and which can be used to
objectively distinguish between red and blue galaxies. Our treatment and
discussion of the problem, including that immediately above, is heavily
influenced by the excellent pedagogic work of \citet{HoggBovyLang}.

\subsection{Allowing for Covariant Errors in Both $x$ and $y$}

In general, and certainly in our case, there are significant observational
uncertainties on the values of the $x_i$s as well as the $y_i$s. Further, the
measurement errors in $x$ and $y$ are, in general, correlated, in the sense
that if a galaxy's colour is overestimated, then so too will its mass-to-light
ratio, and thus its total stellar mass. In this case, and sticking with the
assumption of Gaussian errors, what we have generically referred to as
$\sigma_i$ is most simply represented by a covariance matrix, $\mat{S}\subi$,
as defined in \eqref{covar}.

Consider observing an instance drawn from our generative model that `really'
lies at the position $\vec{x}' = (x', y')$, which lies somewhere along the
line $\ell(\vec{x}') = 0$. So long as the measurement errors can be treated as
being Gaussian, then the probability of observing this point at the position
$\vec{x}\subi$ is given by:
\begin{align} \label{eq:lineprob}
	& p(\vec{x}\subi|\mat{S}\subi, \vec{x}') 
	= \gauss_2( \vec{x}\subi - \vec{x}', \mat{S}\subi) \\
	& ~~~~ \equiv \frac{ 1 }{ 2 \pi \,  |\mat{S}\subi|^{1/2} }
	\exp \left[ - \frac{1}{2} \left( \vec{x}\subi - \vec{x}' \right)^T ~
		\mat{S}\subi^{-1} ~ \left( \vec{x}\subi - \vec{x}' \right)
		\right] ~ , \nonumber
\end{align}
Where we have now also introduced $\gauss_2(\vec{x}, \mat{S})$ as a shorthand
for the (normalised) 2D Gaussian centred on the point $\vec{x} =
0$, and with covariance given by the matrix $\mat{S}$. This expression can
thus be understood as describing the contribution to the expected, observed
data density at the location $\vec{x}\subi$ in the observed CMD, owing to
the `true', underlying distribution at the location $\vec{x}'$, when observed
with measurement errors described by $\mat{S}\subi$.

In order to derive an expectation for the net observed data density at the
location $\vec{x}\subi$---or, in other words, the overall likelihood of
observing the datapoint $\vec{x}\subi$---it is therefore necessary to
integrate over all possible values of $\vec{x}'$. If the underlying
distribution is truly a perfect, uniformly populated and infinitely thin line,
then this line integral takes the following form:
\begin{align} \label{eq:convol}
	\Ell\subi(\vec{x}\subi, \mat{S}\subi| \set{P}) 
		&= \oint_{\ell} \mathrm{d} \vec{x}' ~  
			p(\vec{x}\subi| \mat{S}\subi, \vec{x'}) \nonumber \\
		&= \int \mathrm{d} \vec{x}' ~ \delta[ \ell(\vec{x}'|\set{P}) ] ~
			\gauss_2(\vec{x}\subi - \vec{x}', \mat{S}\subi) ~ ,
\end{align}
In the second line, we have re-written the original line integral to highlight
the fact that it can alternatively be understood as a convolution between the
underlying data distribution and the measurement error ellipse; \ie,
$\Ell\subi = \delta[\ell(\vec{x}\subi)] \otimes \gauss_2( \vec{x}\subi,
\mat{S}\subi)$. Here, we have used the Kronecker delta function,
$\delta[\cdot]$, to enforce the condition that $\ell(\vec{x}') = 0$ for points
on the line; \ie, $\delta[\ell(\vec{x})]$ is taken to represent the
intrinsic distribution in $(x, y)$ space. We have also used \eqref{lineprob}
to re-express $p(\vec{x}\subi|\mat{S}\subi, \vec{x}')$ as
$\gauss_2(\vec{x}\subi - \vec{x}', \mat{S}\subi)$.\footnote{There is a
subtlety here, in that we have implicitly equated
$p(\vec{x}\subi|\mat{S}\subi, \set{P})$ and $p(\vec{x}\subi, \mat{S}\subi|
\set{P})$. The difference between these two quantities boils down to the
distinction between {\em error} and {\em uncertainty}. When a measurement is
made, there is (almost) always some {\em error}; that is, a difference between
the true and measured values. This quantity is, by definition, unknowable. At
the same time, whenever a measurement is made, it will (or should) always come
with an associated {\em uncertainty}, which reflects the allowed range of
`true' values that are consistent with the measured values. This is something
that can (and should) be estimated. In this sense, the uncertainties represent
our priors on the error distribution, to be marginalised over. The assumption
underpinning the sleight of hand by which we have made
$p(\vec{x}\subi|\mat{S}\subi, \set{P}) = p(\vec{x}\subi, \mat{S}\subi|
\set{P})$ is therefore that the probability distribution for the measurement
{\em errors} is faithfully described by our formal measurement {\em
uncertainties}. In particular, the assumption is that the errors are random,
not systematic, and that the error distribution function is Gaussian in form.}

Sticking with the assumption that the underlying distribution is a perfect
line, this line or convolution integral can be done analytically. The result
is simply a 1D Gaussian:
\begin{equation} \begin{split} \label{eq:covarlike}
	\Ell\subi(\vec{x}\subi, \mat{S}\subi| \set{P}) =
\gauss_1( s_{\perp, i}, \sigma_{\perp, i}^2 ) ~ !
\end{split} \end{equation}
Here, $s_{\perp,i}(\vec{x}\subi|\set{P})$ and
$\sigma_{\perp,i}(\mat{S}\subi|\set{P})$ are the (scalar) projections of the
vector $\vec{x}\subi$ and the error ellipse described by $\mat{S}\subi$,
respectively, onto the normal vector for the line, $\hat{n}$. In other words,
if the slope of the line is $m$, then $\hat{n}$ is a unit vector in the
direction $(-m, 1)$, and $s_{\perp,i}$ and $1/\sigma_{\perp,i}^2$ are defined
as $(\hat{n} \cdot \vec{x}\subi)$ and ($\hat{ n }^T \cdot \mat{S}\subi^{-1}
\cdot \hat{ n }$), respectively.\footnote{The projection vector $\hat{n}$ can
just as easily be thought of in terms of the angle of the line $\theta =
\arctan m$, in which case $\hat{n}$ becomes $(-\sin \theta, \cos \theta)$.}

While this result may at first appear to be surprisingly simple, with a
moment's reflection it becomes immediately intuitive. By a symmetry argument,
the probability density around the line {\em must} depend only on the
perpendicular distance from it. (Think of the electrical field above a charged
wire or plate.) In the simple case of a distribution with $y = 0$, the scatter
in the $x$ direction is immaterial --- for each point from the `true’
distribution that is scattered to the right, another will be scattered from
the left to take its place. Any covariance between $\sigma_{x}$ and
$\sigma_y$, which just represents a shearing of $\sigma_x$ along the $y$ axis,
is similarly immaterial. \footnote{Note, however, that this
argument only holds to the extent that the line is uniformly populated; we
will return to this issue in \secref{masslike}.}

At least in the case that the errors in $x$ and $y$ are uncorrelated, the
solution of $\partial \ln \Ell/\partial m = 0$ is still analytic. (This is
sometimes referred to as Deming regression, particularly among chemists and in
the medical sciences.) It is also possible to derive an analytic solution in
the more general case of correlated errors. However, we would argue that, in
this day and age, it is just as easy to solve the problem numerically, which
can be almost trivially done using any number of established optimisation
algorithms. This is doubly true if one wants to quantify the uncertainties in
the fit parameters (as one should). Furthermore, once one adopts a Bayesian
perspective (as we strongly advocate), the `maximum likelihood' solution
becomes all but meaningless. Instead, the Bayesian strives to constrain the
values of the model parameters {\em given the observed data}; that is, to
derive the posterior probability distribution function (PDF) for the full
range of allowed parameter values.

If the above discussion has been somewhat long and laboured, it is to make the
following point very clear: even within the framework of traditional maximum
likelihood or $\chi^2$ minimisation fitting, {\em it is very easy to account
for completely general Gaussian uncertainties.} The only change to the
formalism that is required is to shift from the offset and scatter in the $y$
direction to those perpendicular to the linear relation; that is, to use the
definition of $\Ell\subi$ given in \eqref{covarlike} in place of that in
\eqref{chi2}. Almost without exception, if a more general description of the
observational errors or uncertainties is available or can be assumed, there is
no good reason not to use this information.

\subsection{Allowing for Intrinsic Scatter in the Underlying Relations}
\label{ch:scatlike}

The next assumption we intend to relax is that the underlying data
distribution comes from an infinitely thin line. Instead, we will allow that
there is some intrinsic, astrophysical scatter around the blue and red
CMRs, which we intend to fit for. In the absence of any other better
motivated alternatives, we will make the simplest and most convenient
assumption that this intrinsic scatter can be treated as being Gaussian. This
requires the introduction of a new parameter into the set $\set{P}$, which we
will denote as $\zeta^2$, and which should be understood as being the
intrinsic variance around the `true' CMR.

Intrinsic scatter can be accommodated by considering it as an additional kind
of `error' on points drawn from a perfect linear relationship. A linear
relationship with Gaussian scatter can be represented as the convolution of
perfect line with a Gaussian; \ie, $p(\vec{x}\subi) = \delta\left[
\ell(\vec{x}\subi) \right] \otimes \gauss_1(y\subi, \zeta^2)$. By the
associativity of convolutions the procedures $(\delta \otimes \gauss ) \otimes
\gauss$ and $\delta \otimes (\gauss \otimes \gauss)$ are equivalent; by the
commutativity of convolutions, it does not matter which of the two Gaussians
in this schema is represents the intrinsic scatter, and which represents the
measurement errors. Then, since the convolution of two
Gaussians with variances $\zeta^2$ and $\sigma^2$ is itself a Gaussian with
variance $(\zeta^2 + \sigma^2)$, it is clear that we are justified in treating
$\zeta$ as being akin to `just another source of error'. Thus we have:
\begin{equation} 
    \Ell\subi = \delta[\ell(\vec{x}\subi)] \otimes \gauss_1( y\subi,
\zeta^2) \otimes \gauss_2( \vec{x}\subi, \mat{S}\subi) ~ .
\end{equation}

There are then two options for how to include $\zeta^2$. If $\zeta^2$ is taken
to be the intrinsic variance {\em perpendicular to the relation}, then it can
be folded into \eqref{covarlike} by simply replacing $\sigma_{\perp, i}^2$
with $\zeta^2 + \sigma_{\perp, i}^2$. The alternative is to define $\zeta^2$
as being the variance {\em in the y direction}. Since this is the more
physically sensible way to envisage astrophysical scatter in the CMRs,
this is how we choose to treat $\zeta^2$; it should thus be thought of as
representing the intrinsic variance in colours at fixed mass. 

Once we have defined $\zeta$ in this way, using the arguments presented in the
previous section, we can now define $\zeta_\perp = \zeta \cos \theta$, where
$\theta = \arctan m$ is the angle of the linear relation. Then,
\eqref{covarlike} becomes:
\begin{equation} \label{eq:scatter} \begin{split} 
	\Ell\subi(\vec{x}\subi, \mat{S}\subi| \set{P}) =
\gauss_1( s_{\perp, i}, \zeta_\perp^2 + \sigma_{\perp, i}^2 ) ~ .
\end{split} \end{equation}

Again, we stress that it is not all that hard to incorporate intrinsic scatter
into the traditional maximum likelihood fitting framework. At this stage,
however, (we believe that) the maximum likelihood problem can no longer be
solved analytically; it requires the use of a computer. It is true that, in
most cases, the inferred values of the linear fit parameters will not be
strongly covariant with the inferred scatter. However, any `goodness of fit'
statistic---and thus uncertainties on these parameters---{\em will} depend on
the true, underlying variance. Furthermore, in many cases---including
ours---the scatter is itself a quantity of astrophysical interest, and it
makes little if any sense to neglect it.

\subsection{Outliers and Catastrophic Errors} \label{ch:bad}

As \citet{HoggBovyLang} stress, traditional weighted $\chi^2$ minimisation
fitting methods are generically sensitive to outliers, or `bad data'. However,
these authors also outline a solution to this problem---the method of
(Gaussian) mixture modelling. The basic idea here is to add a secondary
component to our generative model, so that it is able to generate both `bad'
and `good' data.

Constructing a generative model for such `bad' data obviously requires that we
make some working assumptions about the `true' distribution of `bad' data in
the observed $(x,y)$ plane. Developing a full and an accurate description of
such `bad' data is not something that is easily done; further, it is not
something that we are particularly interested in doing. That said, we echo
\citet{HoggBovyLang} in saying that ``the power of this [method] comes not
from making an {\em accurate} model of the outliers, it comes simply from {\em
modelling} them'' (emphasis in the original). And in the end, as described in
\secref{mcmc}, we will marginalise over these `nuisance parameters' that
describe the `bad' data distribution, leaving us only with the parameters that
we genuinely care about.

Since we have no concrete knowledge of the `bad' distribution, we make the
simplest possible assumption and treat this distribution as Gaussian. We also
consider all data points as having an equal probability of being or becoming
`bad'; our desire to objectively identify `bad' data points requires that we
make no {\em a priori} assumptions as to the `badness' or otherwise of
individual data points. We thus choose to model the `badness' within our data
as an additional source of error in the measured values of $x$ and
$y$.\footnote{Note that this is not exactly what \citet{HoggBovyLang} do in
their primer; instead, they treat the `bad' distribution as being a wholly
independent 2D Gaussian component to the data distribution in $x$--$y$ space,
characterised by five parameters: the mean and variance in both $x$ and $y$,
plus an overall normalisation factor. Our solution is thus less general: we
have fixed the mean values of $x$ and $y$ for the `bad' component to be the
same as for the `good' one. }

In other words, the model for the bad data is constructed by convolving the
`good' distribution with an additional (large) Gaussian error distribution;
this error can then be treated in the same way as in the previous two
sections. Assuming for the moment that a given data point {\em is} bad, then
its likelihood can be written:
\begin{equation} \label{eq:badlike} \begin{split} 
	\Ell_{\mathrm{bad},i}(\vec{x}\subi, \mat{S}\subi| \set{P}) = 
\gauss_1 \left( s_{\perp, i}, \, \zeta_\perp^2 + \sigma_{\perp, i}^2 
	+  \zeta_{\mathrm{bad},\perp}^2 \right) ~ .
\end{split} \end{equation}
In writing this expression, we have defined $1/\zeta^2_{\mathrm{bad},\perp} =
(\hat{n}^T \cdot \mat{S}\bad^{-1} \cdot \hat{n})$, where $\mat{S}\bad$ would
be the error matrix that describes the `badness', in direct analogy to the $\zeta\perp$ introduced in \eqref{scatter}.

Again, $\Ell_{\mathrm{bad},i}$ should be understood as the likelihood of
finding a datapoint at the location $\vec{x}_i$ given or assuming that it is
`bad'. If some fraction $f\bad$ of all datapoints are bad, then the net
likelihood of observing the datapoint $\vec{x}\subi$ becomes:
\begin{equation}  \label{eq:goodbad}
	\Ell\subi(\vec{x}\subi, \mat{S}\subi|\set{P}) 
		 = (1-f\bad) ~ \Ell_{\mathrm{good},i} 
				+ f\bad ~ \Ell_{\mathrm{bad,i}} ,
\end{equation}
where $\Ell_{\mathrm{good},i}$ is the likelihood for `good' data, defined, to
now, as in \eqref{scatter}. Further, the `bad' parameter set $\set{P}\bad =
\{f\bad, \zeta_{\mathrm{bad},\perp}\}$ should now be understood to be included
as a subset of the larger parameter set $\set{P}$.

The above represents our first real departure from traditional frequentist
statistical analysis, inasmuch as we have now introduced the nuisance
parameters $\set{P}\bad$.\footnote{In fact, we have technically become
Bayesian simply by writing \eqref{goodbad}, since its derivation is inherently
Bayesian, as it involves marginalising over all possible combinations and
permutations of good/bad-ness among all datapoints \citep[see Equations
(13--17) of][]{HoggBovyLang}. It is only after this marginalisation that
good/bad-ness can be treated probabilistically, rather than binarily. Also
note that this derivation adopts the prior that all datapoints are equally
likely to be good/bad; this prior is thus embedded in \eqref{goodbad}. } While
these parameters are certainly important in the calculation, we are not
interested in their values {\em per se}. It is through the Bayesian process of
marginalisation that we can push these parameters into the background, and
thus focus on the quantities of genuine interest. The utility of these
parameters is that they allow us to objectively identify and `mask' those
datapoints that cannot reasonably be considered to have been drawn from the
`true' underlying distribution, so as to limit the influence of any and all
`bad' points on the values of the `good' parameters. 

Taking this idea just a little bit further: what we have done is created a
mechanism within the fitting process that operates to accommodate outlying
data within a secondary, `bad' component to the error/uncertainty distribution
function. Further, given set of trial values for the parameters in the full
set $\set{P}$, the `badness' of any individual datapoint can be evaluated be
considering the relative likelihood that that point has been drawn from either
the `bad' or the `good' populations. Specifically, the probability that a
point is `bad' is given by $f\bad \Ell\bad\subi / \Ell\subi$.

This mechanism can equally well be understood in two ways. First, it can be
viewed as using a two-component Gaussian to describe the uncertainties on each
point, with the understanding that we are now {\em fitting} for (part of) the
error/uncertainty distribution function. Alternatively, it can be viewed as
modelling the observed data distribution as being the sum of two components: a
`bad' one, and a `good' one. The `bad' one has both a larger scatter, as given
by $\mat{S}\bad$, and a lower normalisation, given by $f\bad$. By virtue of
the fact that the values of these parameters are determined in the course of
the fit, their inclusion has the effect of flexibly and objectively
determining, on the basis of their `badness', which points ought be
downweighted when fitting for the `good' parameters. In this way, points are
classified according to their `badness' in a way that is both objective, and
empirical; \ie, based on the observed dataset, in its entirety.

\subsection{Simultaneously and Flexibly Describing the Red and Blue Subpopulations} \label{ch:mixture}

Until now, our discussion in this Section has only considered the case of
fitting a single line to an observed dataset. But as we have seen, the general
galaxy population can be decomposed into two distinct but overlapping
populations in the CMD. Our generative model therefore needs to simultaneously
describe these two separate populations. The conceptual basis for how we can
go about doing this has already been laid out in the previous section. In the
same way as we have split the observed $(x,y)$ distribution into `good' and
`bad' components, we now split the `good' distribution into distinct `blue'
and `red' components. We can then consider separately the likelihoods of a
particular galaxy as having been drawn from---or as being a member of---either
the red or the blue subpopulation.

This can be done by redefining $\Ell_i$ as:
\begin{equation} \begin{split}
	\Ell_{\mathrm{good},i}( \vec{x}_i, \mat{S}\subi | \parset ) 
		= & f\bluei ~ 
			\Ell\bluei(\vec{x}_i, \mat{S}\subi|\parset\blue) \\
		& ~~ + f\redi
			~ \Ell\redi(\vec{x}_i, \mat{S}\subi|\parset\red) 
			~ .
\end{split} \end{equation}
Here, we have defined two independent subsets of $\set{P}$ that comprise those
parameters pertaining exclusively to each population; \eg, $\parset\red =
\{m\red, c\red, \zeta\red^2\}$ and similarly for $\parset\blue$. Then,
$\Ell\redi$ and $\Ell\bluei$ are defined analogously to \eqref{covarlike} as
the likelihood of drawing the datapoint $\vec{x}\subi$ from either the red or
blue distributions. Finally, in a similar way to $f\bad$ above, the parameters
$f\bluei$ and $f\redi$ define the relative amplitudes of the blue and red
probability distribution functions at the point $\vec{x}\subi$ in the `true',
astrophysical CMD. Further, it should now be clear the ways in which mixture
modelling and objective classification are two sides of the same coin.

In contrast to $f\bad$, we have deliberately written $f\bluei$ and $f\redi$
with a subscript $i$. To understand the motivation for this decision, consider
the following two examples. If we were to somehow have perfect {\em a priori}
knowledge of which galaxies were blue/red (if such a thing is even possible),
then we could set each individual $f\bluei$/$f\redi$ to either 1 or 0. This
case would be equivalent to simultaneously---but still independently---fitting
for the relations within the two subpopulations. The next level of complexity
would be to fit for the relative fraction of red galaxies by treating
$f\bluei$ and $f\redi$ in the same way as we have $f\bad$---that is, to use a
global parameter $f\red$ to modulate the relative amplitudes of the Gaussians
used to define $\Ell\redi$ and $\Ell\bluei$, so that $f\redi = f\red$ and
$f\bluei = ( 1 - f\red)$. 

Of course, neither of these two simple cases are suitable for our purposes.
Instead, what we want to do is to account for the relative numbers of red and
blue galaxies {\em as a function of mass}; that is, we want to allow the
values of $f\bluei$ and $f\redi$ to vary explicitly with $x$. Our method for
doing so is the subject of the next Section.

\subsection{Modelling the Different Mass Distributions for Red and Blue Galaxies} \label{ch:masslike}

Our next task is to find a way to incorporate a more general, non-uniform
distribution in $x$ values---\ie, the mass function---into our
modelling/fitting algorithm. As the starting point for this Section, let us
restate \eqref{scatter} in the following form:
\begin{equation} \label{eq:restate} \begin{split} 
	& \Ell_{\mathrm{blue/red}, i}( \vec{x}\subi, \mat{S}\subi | \set{P} ) =
		\\ 
		& ~~~~~ \delta\left[ \ell_\mathrm{blue/red}(x\subi) \right] 
		\otimes \gauss_1( y\subi, \zeta_\mathrm{blue/red}^2 )
		\otimes \gauss_2( \vec{x}\subi, \mat{S}\subi ) ~ ,
\end{split} \end{equation}
where, again, $\ell(\vec{x}) = m x + c - y$ is the defining function for a
linear relation. Let also restate equation \eqref{badlike} as:
\begin{equation}
	\Ell_{\mathrm{bad},i}( \vec{x}\subi, \mat{S}\subi | \set{P} )
		= \left( \Ell\bluei + \Ell\redi \right) 
			\otimes \gauss_2( \vec{x}\subi, \mat{S}\bad ) ~ .
\end{equation}
Now, realise that the principal advantage of assuming Gaussian distribution
functions to characterise all the different aspects of our generative model is
that by doing so, each of these convolutions can be done analytically. This is
why the method of mixture modelling is typically phrased in terms of Gaussian
distributions.  

So, how are we to proceed? One way of allowing for $f\bluei$ and $f\redi$ to
vary with $x$ would be to treat the red and blue populations as 2D Gaussian
distributions in the CMD with finite widths along the direction of the CMR
line, by replacing the $\delta\left[ \ell \right] \otimes \gauss_1( y\subi,
\zeta^2 )$ with something like $\gauss_2( \vec{x}\subi - \vec{x}_0, \mat{S})$.
This would be the well established method of Gaussian mixture modelling, and
would have the advantage of keeping the calculation of the $\Ell\subi$s
analytic.

Of course, this is not what we want to do, because it is not astrophysically
sensible. The \citet{Schechter1976} function:
 \begin{align}
	\phi(x'|\alpha, x^\dagger, & \phi_0) \, \mathrm{d} x' \\ \nonumber
		= ~ & \phi_0 \, 
		 \left( 10^{x' - x^\dagger} \right)^{\alpha+1} 
			\exp\left( -10^{x' - x^\dagger} \right) \, \ln 10 ~ \mathrm{d} x' ~ .
\end{align}
has been found to provide a very good description for the mass distribution
function of field galaxies---as well as that of many important subpopulations.
Ultimately, this is the distribution function that we want to use as the basis
for our characterisation of the blue/red mass functions. For now, though, let
us continue our discussion in generalised terms, taking $\phi(x)$ as a
generic, parametric, functional description of the distribution of $x$ values.

As the first step towards folding in a more astrophysical and more general
(\ie, non-uniform and non-Gaussian) mass distribution, let us define
$g_\mathrm{blue/red}(\vec{x}' | \set{P}_\mathrm{blue/red}) = \delta\left[
\ell_\mathrm{blue/red}(\vec{x}') \right] \otimes \gauss_1( y',
\zeta_\mathrm{blue/red}^2 )$ to represent the linear parts of our generative
model; \ie, the convolution between the (infinitely thin) locii for the blue
and red CMRs, as given by $\ell(\vec{x'}) = 0$, and some scatter in the
$y$ direction, described by the variance $\zeta^2$. The values of these $g$s
are analytic, as are the convolutions $g \otimes \gauss$, which are necessary
to describe the effect of observational errors/uncertainties.

Now, the idea is to use the $\phi$s to modulate the relative amplitudes of
these Gaussians---both the relative amplitudes of the `blue' and `red'
Gaussians {\em at fixed mass}, and the relative amplitudes of these
distributions {\em as a function of mass}. In this way, the model for the data
distribution in the ($x,y$) plane becomes:
\begin{equation} \label{eq:redblue} \begin{split}
	p\good(\vec{x}'|\set{P}) 
		= (1 -& f\red) \, \phi\blue(x') \, g\blue (\vec{x}')  \\
			& ~ + f\red \, \phi\red(x') \, g\red(\vec{x}') ~ .
\end{split} \end{equation}
It is important at this point to remember the integral normalisation condition
on $p$, and thus also on the two $\phi$s. As written, all of $p$,
$\phi_\mathrm{blue/red}$, and $g_\mathrm{blue/red}$ should be understood to be
integral normalised to unity over the ($x', y'$) domain; in our case, this is
$(( 8.7, \infty ), (-\infty, \infty))$. This means that we cannot fit for the
absolute overall normalisation of $\phi\blue$ or $\phi\red$. That said, we can
fit the relative normalisations of $\phi\blue$ and $\phi\red$, using the
global parameter $f\red$, which we have now effectively defined via
\eqref{redblue}, and which from now on should be considered to be an element
of $\set{P}$.

Now, as per \eqref{restate}, the value of $\Ell\subi$ comes from the
convolution between $p(\vec{x}'|\set{P})$, and the (Gaussian) measurement
error ellipse:
\begin{equation}
	\Ell\subi(\vec{x}\subi, \mat{S}\subi|\set{P})
		= p(\vec{x}\subi|\set{P}) 
			\otimes \gauss_2 ( \vec{x}\subi, \mat{S}\subi) ~.
\end{equation}
But this presents a problem: the convolution $\phi \otimes \gauss$ is no
longer analytic. At this point we are thus forced to make our first formal
approximation. Our solution is to break the continuous mass function $\phi$
into the sum of many discrete $\delta$ functions:
\begin{equation} \label{eq:discrete}
	\phi(x'|x^\dagger, \alpha) \rightarrow 
		\sum_k \delta(x_k - x') \, \phi( x' | x^\dagger, \alpha ) ~ ,
\end{equation}
with the appropriate scaling to satisfy the integral constraint.

Note that we have not had to appeal to any special properties of the Schechter
function to make this approximation; we are free to use any form of $\phi$.
\citep[Indeed, we could even include the many $\phi_k = \phi(x_k)$ as a suite
of independent parameters to be fit for; see, \eg,][]{Blanton2003b} Let us
now turn from the general case, and define the specific parametric form for
the mass functions that we actually use.

We have chosen to model both the blue and red mass functions as the sum of two independent Schechter functions, \ie:
\begin{align}
	\Phi\red(x'|\set{P}\red) = \sum_k \delta(x_k -& x')
		\, \big[ (1 - f_{\mathrm{r},2}) \,
					\phi_{\mathrm{r},1}(x'|x_{\mathrm{r},1}^\dagger, 		
												\alpha_{\mathrm{r},1}) 
\nonumber \\
		+ & f_{\mathrm{r},2} \,
					\phi_{\mathrm{r},2}(x'|x_{\mathrm{r},2}^\dagger, 		
												\alpha_{\mathrm{r},2})
				\big] ~ ,
\end{align}
with an analogous expression to define $\Phi\blue$. Here, the two parameters
$f_{\mathrm{b/r},2}$ govern the relative amplitudes of the two mass functions
$\phi_{\mathrm{b/r},1}$ and $\phi_{\mathrm{b/r},2}$ in the same way that 
$f\red$ does for $\Phi\blue$ and $\Phi\red$.

The approximation given in \eqref{discrete} can be thought of in two
complementary but, at least at this stage, equivalent ways. For any value of
$k$, $\phi_k = \phi(x_k)$ is just a scalar normalisation factor for $g(x_k,
y')$, and thus for the series of convolutions; \ie, $\Ell = \sum \phi_k \times
(\delta_k \otimes g \otimes \gauss)$, which is analytic. With this way of
thinking, we are exactly and analytically solving an approximate model with a
discrete, stepped mass function. Alternatively, the analytic convolutions $g
\otimes \gauss$ could be thought of as being computed first. In this way of
thinking, the shift to a discretised $\phi$ can be seen simply as computing
the convolution $\delta \otimes \phi \otimes g \otimes \gauss $ numerically,
using Euler's method for numerical integration. In the former way of thinking,
we have made a formal approximation in the {\em construction} of the model for
$p(\vec{x}')$; in the latter way of thinking we have made a numerical
approximation in the {\em computation} of the values of $p(\vec{x}')$. For
reasons that will become clear in the following two Sections, we would
advocate the former interpretation over the latter.

With this in mind, our last task for this Section is to explicitly define the
values of $x_k$ that we adopt for the fits. Let us take the $x_k$s to be
evenly spaced in the $x$ dimension, with a spacing given by $\Delta_k$,
ranging from the lower limit for our sample up to some high value. In a sense,
this can be thought of as using a histogram with bin centres $x_k = 8.7 + ( k
+ 1/2) \, \Delta_k : k = 0, 1, ..., N$ in place of the continuous $\Phi$.
However, because the mass function is not generally linear in $x$, the mean
value of $x'$ for this bin will be slightly different from the geometric
centre of the bin; similarly the population of the bin will be slightly
different to the value of $\Phi(x_k)$. This is akin to the so-called
Eddington bias, and is most true in the high mass,
exponential tail. To explicitly account for this, then, we compute:
\begin{align} x_k + \varepsilon_k & = \frac{ \int_{x_k - \Delta_k/2}^{x_k +
\Delta_k/2} \mathrm{d}x' ~ x' \, \Phi(x') } { \int_{x_k - \Delta_k/2}^{x_k +
\Delta_k/2} \mathrm{d}x' ~ x' } ~ , \\ \Phi_k & = \int_{x_k - \Delta_k/2}^{x_k
+ \Delta_k/2} \mathrm{d}x' ~ \Phi(x') ~ , \end{align} using trapezoidal
numerical integration.

This is perhaps an unnecessary flourish: this choice has very little impact on
our results. For this reason, we have glossed over this aspect of our
calculation in \secref{method}, where we describe the full model {\em in
toto}. Given that we have to perform this numerical integration anyway to
enforce the normalisation condition on the $\Phi$s, however, making this
choice has a negligible cost in terms of computational runtime. We have
therefore elected to retain this very minor `correction', if only because we
can.

\subsection{Allowing for more general relations between $x$ and $y$}

With the approximation made in \eqref{discrete}, we were able to relax the
assumption of a uniformly populated line, and in so doing, accommodate a
wholly general (if parametric) description of the $x$ distribution function;
\ie, the mass function. Until now, the significance of assuming a uniform
distribution function for $x$ has been that it made the line/convolution
integral in \eqref{convol} analytic. Freed from the assumption of a uniform
distribution function for $x$, the approximation in \eqref{discrete} also
allows us to relax the assumption of both a purely linear relation between $x$
and $y$, as well as the assumption of a uniform scatter around the CMRs.
Instead, what appears in \eqref{discrete} is nothing more than the centre and
width of a Gaussian distribution at each and every of the discrete $x_k$s;
\ie, $\delta_k \otimes g$, where $g = \delta \left[ \ell(\vec{x}\subi) \right]
\otimes \gauss(y\subi; \zeta^2)$.

In the same way as we have chosen to model the `red' and `blue' mass functions
as the sum of two Schechter functions, we must now state our specific
parametric description of the locus of, and scatter around the `red' and
`blue' CMRs; we do this in the next two Sections.

In order to allow the slope of each of the CMRs to vary with $x$, we elect
to describe them as a combination of two linear relations. The exact
definitions for the descriptive model for $\ell(\vec{x}')$ are given in Eq.s
\ref{eq:cmrs} and \ref{eq:zeta}; the exact functional form $\ell(\vec{x}')$ is
immaterial in this pedagogical discussion. Here, let us simply note that
$\ell\red(\vec{x}')$ is described by five new parameters $\{m_\mathrm{r,lo},
m_\mathrm{r,hi}, x_{\mathrm{r},\ell,0}, x_{\mathrm{r,}\ell,s} \} \subset
\set{P}\red$, with analogous parameters used to describe $\ell\blue$.

Now, using the arguments made in the previous Section, we can relax the
assumption of a perfectly linear relation between $x$ and $y$ simply by
substituting this two-line definition for $\ell$ in place of the linear $\ell$
we have assumed thus far. Using the same formalism as before, $g(\vec{x}')$
remains defined as $ \delta( \ell(\vec{x}') ) \otimes \gauss_1(y', \zeta^2)$,
with no further adjustments required in our formalism.

We note, however, that this is only true because we have independently
characterised the distribution of points in the $x$ dimension as in the
previous section. If $\ell(\vec{x})$ were uniformly populated {\em along the
line}, the equivalent of the line integral given in \eqref{convol} would have
to include a factor of $\mathrm{d}\ell/\mathrm{d}x$ to account for the fact
that there would be more points (per unit $x$) where the slope of $\ell$ is
steeper. This would mean that the line/convolution integral would no longer be
analytic. In this sense, the transformation mapping the $x$-axis (\ie, the
line $y=0$) to the relation described by $\ell$ should be understood in terms
of shearing and shifting, rather than rotation, since the $x$ distribution
function is left unchanged.

\subsection{The next level of sophistication in the intrinsic scatter in $y$}

The final aspect of the model that remains to be developed is the allowance
for the scatter in the CMRs to vary with $x$. Using the same arguments
advanced in the previous two Sections, it should be clear that we can now
simply redefine the (until now) constant $\zeta_\mathrm{blue/red}$ that goes
into the definitions of $g_\mathrm{blue/red}$ with some functional description
$\zeta_\mathrm{blue/red}(\vec{x}')$. The exact description that we have
adopted comes from \citet{Baldry2004}, and is given in \eqref{zeta}.

The final formal step that remains to be taken, then, is to fold this change
into our existing definitions for $g(\vec{x}')$, $p(\vec{x}')$, and
$\Ell(\vec{x}\subi)$. Combining the results of the last three sections, we
have:
\begin{align}
	p\red(\vec{x}'|\set{P}\red)
	& = \sum_k \Phi_k \times \delta( x_k + \varepsilon_k - x' ) \nonumber \\
	& \hspace{1.15cm}		\otimes \delta( \ell(x') )
				\otimes \gauss_1( y', \zeta\blue^2(x') ) \\ \nonumber
	& = \sum_k \Phi_k \times \gauss_1\big( \ell\red( x_k + \varepsilon_k) ), 
								\zeta\red^2( x_k + \varepsilon_k ) \big) ~ .
\end{align}
That is, as previously described, the model for the true $(x, y)$ distribution
of $x = \log M_* > 8.7$ and $z \le 0.12$ galaxies comprises a discrete
distribution of $x$ values; at each of these discrete masses, the distribution
of $y$ values (\ie\ colours) is taken to be a `bimodal' distribution of two
(1-D) Gaussians, the normalisations, centres, and widths of which are all
allowed to vary parametrically as a function of $x$; \ie, mass.

For completeness, then, we have the model probability density distribution for 
`good' data as given by the sum of distinct `blue' and `red' components:
\begin{equation} \begin{split}
	p\good(\vec{x}'|\set{P}\good) 
		= (1 -& f\red) \, p\blue(\vec{x'}|\set{P}\blue) \\
			& ~ + f\red \, p\red(\vec{x'}|\set{P}\red) ~ , 
\end{split} \end{equation}
with an additional `bad' component, which is just a badly smeared version of 
the `good' distribution:
\begin{equation} 
p\bad(\vec{x}'|\set{P}\bad) 
 	= p\good(\vec{x}'|\set{P}\good) \otimes \gauss_2( \vec{x}', \mat{S}\bad)~.
\end{equation}
The generative model for the net, observed ($x,y$) distribution is a then 
mixture of these `good' and `bad' components:
\begin{equation} \begin{split}
	p( \vec{x}' | \set{P} )
	= (1 -& f\bad) \, p\good(\vec{x'}|\set{P}\good) \\
		& ~ + f\bad \, p\bad(\vec{x'}|\set{P}\bad) ~ ,
\end{split} \end{equation}
and the likelihood of observing any given datapoint is determined by 
convolving this model with the observational uncertainties/errors:
\begin{equation}
	\Ell\subi(\vec{x}\subi, \mat{S}\subi|\set{P}) 
		= p(\vec{x}\subi|\set{P}) \otimes G_2( \vec{x}\subi, \mat{S}\subi) ~ .
\end{equation}

\subsection{Accounting for Incompleteness}

The last remaining aspect of the observational dataset that remains to be
accounted for is incompleteness arising from the fact that our sample is
ultimately selected on the basis of apparent magnitudes; specifically,
$r$-band \petrott\ magnitudes, as reported in the SDSS DR6 catalogues.
This has been accounted for using the standard technique of $1/V\max$
weighting as motivated, described, and validated in \secref{vmax}.

The basic idea is as follows. Take the specific example of a galaxy with mass
and colour such that it would only satisfy the GAMA spectroscopic target
selection criteria over, say, 1/3 of the total GAMA survey volume this side of
our analysis redshift interval of $z \le 0.12$. In this case, there would be
(modulo the effects of large-scale clustering) 3 times as many galaxies in the
real $z \le 0.12$ universe as are found in the GAMA catalogues. To account for
this, we should count this single putative galaxy 3 times over. That is, each
galaxy should be given a weighting $w\subi \propto 1/V\max$.

In this way, we arrive at last at our final expression for the overall
likelihood of observing our dataset, assuming some fiducial trial values for
each and every parameter in the set $\set{P}$:
\begin{equation}
	\Ell( \set{X}, \set{S}, \set{W} | \set{P} )
		= \prod_i \Ell\subi(\vec{x}\subi, \mat{S}\subi|\set{P} )^{1/V\maxi}~,
\end{equation}
or, equivalently:
\begin{equation}
	\log \Ell( \set{X}, \set{S}, \set{W} | \set{P} )
		= \sum_i w\subi \, 
			\Ell\subi(\vec{x}\subi, \mat{S}\subi|\set{P} )~.
\end{equation}

\subsection{Summary: A Generative Model for the Observed Distribution of Galaxies in the Colour--Mass Plane} \label{ch:modelsumm}

We have now fully developed our model to predict or describe the observed
distribution of galaxies in the CMD, accounting for distinct blue and red
populations with different mass functions, and allowing both the slope of and
scatter around the blue and red CMRs to vary with mass. We have also accounted
for fully generalised (if Gaussian) covariant errors in the measured values of
$x$ and $y$, as well as allowing for un-modelled or under-modelled aspects of
the ($x,y$) distribution as `bad' data. The origins of this `bad' data may be
be astrophysical, in the sense of some additional component in the $(x,y)$
plane not included in the model, or observational, in the sense of
catastrophic errors in the measurements. Finally, we have also accounted for
incompleteness arising from the magnitude-limited nature of the GAMA sample.

In our pedagogical development of the model, we have attempted to make it
clear that the conceptual framework that underpins our modelling is just an
extension of the traditional (frequentist) weighted $\chi^2$ or `maximum
likelihood' formalism for fitting a perfectly straight, perfectly narrow, and
uniformly populated line. Indeed, this simple and highly idealised situation
can be seen as a special case of our (much) more general model. In this sense,
there is only one `trick' that we have introduced here---the method of
(Gaussian) mixture modelling. This is what allows us to flexibly describe, and
thus objectively identify and quantify, the distinct `blue', `red', and `bad'
components. As stated in \secref{bad}, this method is intrinsically Bayesian,
inasmuch as the formal justification involves implicit marginalisation over
all possible binomial combinations of blue-/red-/bad-ness for each individual
point, with the implicit prior that all points are treated as having an equal
{\em a priori} probability of being blue, or red, or `bad'.

With this caveat, there is nothing preventing us from using this model to
perform a frequentist `maximum likelihood' fit---this would simply involve
identifying the set of parameter values $\set{P}$ that maximises the scalar
likelihood function $\Ell(\set{X},\set{S},\set{W}|\set{P})$. However, at least
with the formalism as we have laid it out, this would be dishonest. To see
this, recognise that $\Ell$ is only defined for a given or assumed set of
values for $\set{P}$. Any comparison between the values of $\Ell$ for
different $\set{P}$s thus inescapably, if implicitly, assumes that the
different values of $\set{P}$s are equally likely, or not. {\em All fitting
algorithms include priors}; the difference between Bayesian and a frequentist
statistics is only that these priors are made explicit in a Bayesian setting.
This is especially important when it comes to deriving formal uncertainties on
the values of the fit parameters---by considering only
$\Ell(\set{X},\set{S},\set{W}|\set{P})$, the frequentist does not have a good
formal basis for making such a calculation, since they cannot simultaneously
assume two distinct sets of values for $\set{P}$. In this sense, a
traditional, frequentist `maximum likelihood' fit is just a Bayesian maximum
{\em a posteriori} (MAP) determination---\ie, the identification of the global
maximum for the posterior PDF---with (unstated) uniform priors.

We have chosen our specific priors with this in mind. In the absence of any
clearly better alternatives, we assume uniform priors for just about all of
the parameters in $\set{P}$. This includes uniform priors on the {\em
logarithm} of the characteristic masses for the Schechter functions,
$x^\dagger = \log M^\dagger$, rather than uniform priors on $M^\dagger$ {\em
per se}), as well as on the values of the $f$s (\cf\ $\log f$) that are used
in place of the $\phi_0$s to modulate the relative amplitudes of the different
Schechter function components. The only exception to this rule is for the
slopes of any and all relations, which are assumed to be linear in $\theta =
\arctan m$, rather than linear in $m$. This ensures that steeper slopes are
not `artificially' down-weighted in preference of flatter ones. In all cases,
we have checked to ensure that the allowed range for the priors is reasonable,
and in particular that these ranges are broad enough to ensure that they do
not `artificially' cut off the PDFs. The only exceptions to this rule are the
sensible and obvious ones; \eg, we require that all of the $f$s be in the
range $[0,1]$, aand that $\zeta$s be positive.

\end{document}